\documentclass[%
 reprint,
superscriptaddress,
 amsmath,amssymb,
 aps,
prb,
]{revtex4-2}

\usepackage{physics}
\usepackage{adjustbox}
\usepackage{color}
\usepackage{subfigure}
\usepackage{ulem}

\usepackage{hyperref}
\hypersetup{hypertex=true,
colorlinks=true,
linkcolor=blue,
anchorcolor=blue,
citecolor=blue}

\def\cup{\! \smile \!}
\renewcommand{\eqref}[1]{Eq.~(\ref{#1})}

\begin{document}

\preprint{APS/123-QED}

\title{Stacking Group Structure of Fermionic Symmetry-Protected Topological Phases}

\date{\today}

\author{Xing-Yu Ren}
\affiliation{Department of Physics, The Chinese University of Hong Kong, Shatin, New Territories, Hong Kong, China}
\author{Shang-Qiang Ning}
\affiliation{Department of Physics, The Chinese University of Hong Kong, Shatin, New Territories, Hong Kong, China}
\author{Yang Qi}
\affiliation{Center for Field Theory and Particle Physics, Department of Physics, Fudan University, Shanghai 200433, China}
\affiliation{State Key Laboratory of Surface Physics, Fudan University, Shanghai 200433, China}
\author{Qing-Rui Wang}
\affiliation{Yau Mathematical Sciences Center, Tsinghua University, Haidian, Beijing 100084, China}
\author{Zheng-Cheng Gu}
\affiliation{Department of Physics, The Chinese University of Hong Kong, Shatin, New Territories, Hong Kong, China}

\begin{abstract}
In the past decade, there has been a systematic investigation of symmetry-protected topological (SPT) phases in interacting fermion systems. Specifically, by utilizing the concept of equivalence classes of finite-depth fermionic symmetric local unitary (FSLU) transformations and the fluctuating decorated symmetry domain wall picture, a large class of fixed-point wave functions have been constructed for fermionic SPT (FSPT) phases. Remarkably, this construction coincides with the Atiyah-Hirzebruch spectral sequence, enabling a complete classification of FSPT phases.
However, unlike bosonic SPT phases, the stacking group structure in fermion systems proves to be much more intricate. The construction of fixed-point wave functions does not explicitly provide this information. In this paper, we employ FSLU transformations to investigate the stacking group structure of FSPT phases. Specifically, we demonstrate how to compute stacking FSPT data from the input FSPT data in each layer, considering both unitary and anti-unitary symmetry, up to 2+1 dimensions.
As concrete examples, we explicitly compute the stacking group structure for crystalline FSPT phases in all 17 wallpaper groups and the mixture of wallpaper groups with onsite time-reversal symmetry using the fermionic crystalline equivalence principle. Importantly, our approach can be readily extended to higher dimensions, offering a versatile method for exploring the stacking group structure of FSPT phases.
\end{abstract}

\maketitle

\tableofcontents

\section{\label{sec1}Introduction}
\subsection{\label{sec1.1}Goal of this paper}

The exploration of topological materials, such as fractional quantum Hall (FQH) states and electronic topological insulators, has brought about a revolutionary shift in our comprehension of quantum phases and their fundamental characteristics. Over the past few decades, the study of topological phases in quantum matter has garnered considerable interest because of their exceptional properties that go beyond the conventional Landau paradigm. Moreover, these phases exhibit great potential for applications in quantum information processing and fault-tolerant quantum computation. An intriguing revelation is that a systematic understanding of topological phases in quantum matter can be achieved through the analysis of entanglement patterns. 
The underlying entanglement structures provide a powerful framework for characterizing and classifying different topological phases.  

Based on the concept of finite-depth symmetric local unitary (LU) transformations~\cite{Chen2010}, gapped quantum states can be categorized as long-range entangled (LRE) or short-range entangled (SRE) states. Topological order, arising from LRE, exhibits stable universal properties even under local perturbations. These states show ground state degeneracy on nontrivial manifolds like torus and possess fractional excitations with unique non-Abelian statistics. When stacking two topological states, the resulting state inherits the elementary excitations from each individual state. However, it is not possible to define an "inverse" topological state for a given state, preventing the collection of all topologically ordered states from forming a group. Interestingly, there exists a special class of topological states called invertible topological states that do form a group. These states lack fractionalized bulk  excitations, and the chiral central charge serves as their only nontrivial characteristic. Examples of invertible topological states include integer quantum Hall states and $p+ip$ topological superconductors.

The interplay between symmetry and entanglement makes the classification of phases even richer. The LRE states with symmetries are called symmetry-enriched topological (SET) phases ~\cite{BBCW,YeZou1,YeZou2}. In the presence of global symmetry, even short-range entangled states can be divided into various subclasses. Among these subclasses, there are symmetry-protected topological (SPT) phases~\cite{gu09,chenScience2012,chen13}, which are special invertible states protected by global symmetry. As SPT phases are SRE, they can be connected to a trivial product state through finite-depth local unitary transformation if the symmetry is broken. Examples of SPT phases include well-known systems like topological insulators~\cite{hasan10,qi11} and Haldane chains~\cite{haldane83}. Like other invertible topological phases, SPT states lack fractionalized bulk excitations, and their nontrivial properties are characterized by (symmetry-protected) gapless edge states.
It is widely recognized that the classification of SPT phases requires a certain generalized cohomology theory. It is not surprising that SPT states associated with a specific symmetry class always form an Abelian group. While the stacking group structure is evident for bosonic SPT phases and free fermion SPT phases, understanding interacting fermion systems, particularly those with anti-unitary symmetry, poses greater challenges. 

It turns out that the construction and classification of interacting fermionic SPT (FSPT) states are more complex compared to BSPT states, which have been well-known for over a decade~\cite{fidkowski10,fidkowski11,qi13,yao13,ryu12,gu14b,you2014,GuWen2014,kapustin14,freed14,fidkowski13,bonderson13,wangc13b,cheng2014,chong14,wangc-science, metlitski14,chen14a,witten15,kitaevsre,metlitski15,cheng15,morimoto15,Tarantino2016,wangcj16,wanggu16,Gaiotto2016,gaiotto16,freed16,morgan16,WangNingChen2017,Kapustin2017,WQG2018,Morgan2018,Kapustin2018,ChenjieFSPT,Tantivasadakarn,Juven2018,Fidkowski2018}. In Refs. \cite{WangGu2018,WangGu2020}, the authors proposed a systematic and fruitful scheme for classifying interacting FSPT states based on the concept of decorated fluctuating domain walls. Mathematically, this construction can be described using the Atiyah-Hirzebruch spectral sequence. Consequently, given a symmetry group, the classification of FSPT states as a set can be computed using the algorithm introduced in Ref. \cite{Ouyang2021}.

Unfortunately, the existing classification results for interacting FSPT phases only provide information about the number of distinct phases. It is crucial to understand the group structure of these phases under the stacking operation. 
In this paper, we will determine the stacking group structure for interacting FSPT phases with internal symmetry in spacetime dimensions ranging from $0+1$D to $2+1$D. The stacking rules will be summarized in the upcoming subsection \ref{sec1.2}. 
Using these stacking rules, we will proceed to calculate several examples, which are summarized in subsection \ref{sec1.3}. It is well known that there exists a one-to-one correspondence between internal BSPT phases protected by a group $G$ (considered as an onsite symmetry) and crystalline BSPT phases protected by the same group $G$ (considered as spatial symmetry) based on the crystalline equivalence principle~\cite{Else2018}. However, the application of this crystalline equivalence principle to fermionic SPT phases is more intricate, and we will explain this intricacy in Section \ref{sec6}. Utilizing the fermionic crystalline equivalence principle, we calculate the stacking group structures of 2+1D internal FSPT states protected by 2D wallpaper groups in Section \ref{sec6}.


\subsection{\label{sec1.2}Summary of main results}

The total symmetry group $G_f$ of a fermionic system can be understood as a central extension of the bosonic symmetry group $G_b$ by the fermion parity subgroup $\mathbb Z_2^f$. This central extension is defined by the 2-cocycle $\omega_2 \in H^2(G_b, \mathbb Z_2)$. To distinguish the unitary or antiunitary nature of a symmetry operator, we use the group homomorphism $s_1 \in H^1(G_b, \mathbb Z_2)$, which maps elements from $G_b$ to $\mathbb Z_2$. Therefore, the symmetry group of a FSPT can be uniquely specified by a triple $(G_b, \omega_2, s_1)$.

The classification of FSPT involves cochains satisfying specific consistency conditions. In lower dimensions, up to Kitaev chain decorations, we encounter a triad of data denoted as $(n_{d-1}, n_d, \nu_{d+1})$, where $d$ is  the spatial dimension. When we stack two FSPT states, say $(n_{d-1}, n_d, \nu_{d+1})$ and $(n_{d-1}', n_d', \nu_{d+1}')$, they combine to form a new FSPT state, which can be exposed by a  fermionic symmetric local unitary (FSLU) transformation. This transformation places all nontrivial data in one layer while leaving the other in a trivial product state. The nontrivial layer is described by a distinct set of parameters, denoted as $(N_{d-1}, N_d, \mathcal{V}_{d+1})$. This process unveils a stacking group structure that governs FSPT phases, characterized by the addition operation:
\begin{align}
(n_{d-1},n_d,\nu_{d+1}) + (n_{d-1}',n_d',\nu_{d+1}')
=(N_{d-1},N_d,\mathcal{V}_{d+1}).
\end{align}
This approach establishes an Abelian group structure that emerges when FSPT phases are stacked together. Importantly, this structure also corresponds to the group structure of 't Hooft anomalies in a fermionic system one dimension lower.

\subsubsection{\label{sec1.2.1}Stacking group structure of 0+1D FSPT}

Given two 0+1D FSPT states $( n_{0} ,\nu _{1})$ and $( n'_{0} ,\nu '_{1})$ with the same symmetry group $( G_{b} ,\omega _{2} ,s_{1})$, we can stack them to get a new 0+1D FSPT state $( N_{0} ,\mathcal{V}_{1})$:
\begin{align}
N_{0} &=n_{0} +n'_{0} \ (\bmod\ 2) ,\label{eq1.1}\\
\mathcal{V}_{1} &=\nu _{1} \nu '_{1} \label{eq1.2}.
\end{align}

\subsubsection{\label{sec1.2.2}Stacking group structure of 1+1D FSPT}

Given two 1+1D FSPT states $( n_{0} ,n_{1} ,\nu _{2})$ and $( n'_{0} ,n'_{1} ,\nu '_{2})$ with the same symmetry group $( G_{b} ,\omega _{2} ,s_{1})$, we can stack them into a new 1+1D FSPT state $( N_{0} ,N_{1} ,\mathcal{V}_{2})$: 
\begin{align}
N_{0} &=n_{0} +n'_{0} \ (\bmod\ 2) ,\label{eq1.3}\\
N_{1} &=n_{1} +n'_{1} + s_{1} \cup n_{0} \cup n'_{0} \ (\bmod\ 2) ,\label{eq1.4}\\
\mathcal{V}_{2} &=\nu _{2} \nu '_{2}( -1)^{s_{1} \cup ( n_{1} +n'_{1}) \cup n_{0} \cup n'_{0} +n_{1} \cup n'_{1}}  .\label{eq1.5}
\end{align}
Note that this is the simplified version after setting $\omega _{2}=0$. For the original results, please see Eqs. (\ref{eq4.20})-(\ref{eq4.22.2}). The reason why we can set $\omega _{2}=0$ is also explained below the original results.

\subsubsection{\label{sec1.2.3}Stacking group structure of 2+1D FSPT}

Given two 2+1D FSPT states $( n_{1} ,n_{2} ,\nu _{3})$ and $( n'_{1} ,n'_{2} ,\nu '_{3})$ with the same symmetry group $( G_{b} ,\omega _{2} ,s_{1})$, we can stack them into a new 2+1D FSPT phase $( N_{1} ,N_{2} ,\mathcal{V}_{3})$: 
\begin{align}
N_{1} & =  n_{1} +n'_{1} \ (\bmod\ 2) , \label{eq1.6}\\
N_{2} & =  n_{2} +n'_{2} + n_{1} \cup n'_{1} +s_{1} \cup ( n_{1} \cup _{1} n'_{1}) \ (\bmod\ 2) , \label{eq1.7}\\
\mathcal{V}_{3} & =  \nu _{3} \nu '_{3}\mathcal{E}_{3} , \label{eq1.9}\\
\mathcal{E}_{3} & =  ( -1)^{\epsilon _{3}} e^{2\pi i\theta _{3}[ n_{1} ,n'_{1}]} \; [\mathrm{see \; Eqs.~(\ref{eq5.13})-(\ref{eq5.15.4})}]. \label{eq1.10}
\end{align}
The expressions of $\epsilon _{3}$ and $\theta _{3}$ in \eqref{eq1.10} are too complicated to show here.

\subsection{\label{sec1.3}Summary of main examples}

Consider the stacking group structure of the interacting crystalline FSPT~\cite{Hemele2017,Meng2018,Zhang2022} protected by space groups or a combination of space groups and internal symmetries. Based on the fermionic crystalline equivalence principle~\cite{Debray2021,Manjunath2023}, there is a one-to-one correspondence of the crystalline FSPT protected by $(G_{b},\omega_{2},s_{1})$ (the bosonic part $G_b$ contains space symmetries) and the internal FSPT protected by $(G_{b}, \omega _{2} ^{\mathrm{eff}}, s_{1} ^{\mathrm{eff}})$ ($G_b$ is viewed as onsite symmetries). The data $\omega _{2} ^{\mathrm{eff}}$ and $s_{1} ^{\mathrm{eff}}$ are defined in \eqref{eq6.1} and \eqref{eq6.2}. We denote the 2d wallpaper groups as $G_{\mathrm{wp}}$. There are four main examples we calculated: 
\begin{enumerate}
    \item \textbf{Spin-1/2 crystalline} FSPT protected by the 2d wallpaper groups, $G_{b}=G_{\mathrm{wp}}$. This corresponds to \textbf{spinless internal} FSPT protected by the same $G_{b}$ groups viewed as internal symmetries. The results are listed in Tab.~\ref{tab7}.
    \item \textbf{Spinless crystalline} FSPT protected by the 2d wallpaper groups, $G_{b}=G_{\mathrm{wp}}$. This corresponds to \textbf{spin-1/2 internal} FSPT protected by the same $G_{b}$ groups viewed as internal symmetries. The results are listed in Tab.~\ref{tab8}.
    \item \textbf{Spin-1/2 crystalline} FSPT protected by the direct product of the 2d wallpaper groups (acting on space) and time-reversal symmetry (acting onsite), $G_{b}=G_{\mathrm{wp}} \times \mathbb{Z}_{2} ^{T}$. This corresponds to \textbf{spinless internal} FSPT protected by the same $G_{b}$ groups viewed as internal symmetries. The results are listed in Tab.~\ref{tab9}.
    \item \textbf{Spinless crystalline} FSPT protected by the direct product of the 2d wallpaper groups (acting on space) and time-reversal symmetry (acting onsite), $G_{b}=G_{\mathrm{wp}} \times \mathbb{Z}_{2} ^{T}$. This corresponds to \textbf{spin-1/2 internal} FSPT protected by the same $G_{b}$ groups viewed as internal symmetries. The results are listed in Tab.~\ref{tab10}.
\end{enumerate}

\begin{table}
\caption{\label{tab7}Stacking group structure of interacting \textbf{spin-1/2 crystalline} FSPT protected by 2d wallpaper groups (viewed as \textbf{space} symmetries). This is mapped to \textbf{spinless internal} FSPT by fermionic crystalline equivalence principle.}
\begin{ruledtabular}
\begin{tabular}{ccccc}
$G_{b}$ & MC & CF & B & Extension \\ \hline
p1 & $\mathbb{Z}_{2} ^{2}$ & $\mathbb{Z}_{2}$ & $\mathbb{Z}_{1}$ & $\mathbb{Z}_{2} ^{3}$\\
p2 & $\mathbb{Z}_{2} ^{3}$ & $\mathbb{Z}_{2} ^{4}$ & $\mathbb{Z}_{2} ^{4}$ & $\mathbb{Z}_{4} \times \mathbb{Z}_{8} ^{3}$\\
p1m1 & $\mathbb{Z}_{2}$ & $\mathbb{Z}_{2} ^{2}$ & $\mathbb{Z}_{2} ^{2}$ & $\mathbb{Z}_{4} \times \mathbb{Z}_{8}$\\
p1g1 & $\mathbb{Z}_{2} ^{2}$ & $\mathbb{Z}_{2}$ & $\mathbb{Z}_{1}$ & $\mathbb{Z}_{2} ^{3}$\\
c1m1 & $\mathbb{Z}_{2}$ & $\mathbb{Z}_{2}$ & $\mathbb{Z}_{2}$ & $\mathbb{Z}_{2} \times \mathbb{Z}_{4}$\\
p2mm & $\mathbb{Z}_{1}$ & $\mathbb{Z}_{1}$ & $\mathbb{Z}_{2} ^{8}$ & $\mathbb{Z}_{2} ^{8}$\\
p2mg & $\mathbb{Z}_{2} ^{2}$ & $\mathbb{Z}_{2} ^{3}$ & $\mathbb{Z}_{2} ^{3}$ & $\mathbb{Z}_{4} \times \mathbb{Z} _{8} ^{2}$\\
p2gg & $\mathbb{Z}_{2} ^{2}$ & $\mathbb{Z}_{2} ^{2}$ & $\mathbb{Z}_{2} ^{2}$ & $\mathbb{Z}_{2} \times \mathbb{Z}_{4} \times \mathbb{Z} _{8}$\\
c2mm & $\mathbb{Z}_{2}$ & $\mathbb{Z}_{2}$ & $\mathbb{Z}_{2} ^{5}$ & $\mathbb{Z}_{2} ^{4} \times \mathbb{Z} _{8}$\\
p4 & $\mathbb{Z}_{2} ^{2}$ & $\mathbb{Z}_{2} ^{3}$ & $\mathbb{Z}_{2} \times \mathbb{Z}_{4} ^{2}$ & $\mathbb{Z}_{2} \times \mathbb{Z} _{8} ^{3}$\\
p4mm & $\mathbb{Z}_{1}$ & $\mathbb{Z}_{1}$ & $\mathbb{Z}_{2} ^{6}$ & $\mathbb{Z}_{2} ^{6}$\\
p4gm & $\mathbb{Z}_{2}$ & $\mathbb{Z}_{2}$ & $\mathbb{Z}_{2} ^{2} \times \mathbb{Z}_{4}$ & $\mathbb{Z}_{2} ^{3} \times \mathbb{Z}_{8}$\\
p3 & $\mathbb{Z}_{1}$ & $\mathbb{Z}_{2}$ & $\mathbb{Z}_{3} ^{3}$ & $\mathbb{Z}_{2} \times \mathbb{Z}_{3} ^{3}$\\
p3m1 & $\mathbb{Z}_{1}$ & $\mathbb{Z}_{2}$ & $\mathbb{Z}_{2}$ & $\mathbb{Z}_{4}$\\
p31m & $\mathbb{Z}_{1}$ & $\mathbb{Z}_{2}$ & $\mathbb{Z}_{6}$ & $\mathbb{Z}_{12}$\\
p6 & $\mathbb{Z}_{2}$ & $\mathbb{Z}_{2} ^{2}$ & $\mathbb{Z}_{6} ^{2}$ & $\mathbb{Z}_{3} \times \mathbb{Z}_{8} \times \mathbb{Z}_{12}$\\
p6mm & $\mathbb{Z}_{1}$ & $\mathbb{Z}_{1}$ & $\mathbb{Z}_{2} ^{4}$ & $\mathbb{Z}_{2} ^{4}$\\
\end{tabular}
\end{ruledtabular}
\end{table}

\begin{table}
\caption{\label{tab8}Stacking group structure of interacting \textbf{spinless crystalline} FSPT protected by 2d wallpaper groups (viewed as \textbf{space} symmetries). This is mapped to \textbf{spin-1/2 internal} FSPT by fermionic crystalline equivalence principle.}
\begin{ruledtabular}
\begin{tabular}{ccccc}
$G_{b}$ & MC & CF & B & Extension \\ \hline
p1 & $\mathbb{Z}_{2} ^{2}$ & $\mathbb{Z}_{2}$ & $\mathbb{Z}_{1}$ & $\mathbb{Z}_{2} ^{3}$\\
p2 & $\mathbb{Z}_{1}$ & $\mathbb{Z}_{2} ^{3}$ & $\mathbb{Z}_{2}$ & $\mathbb{Z}_{2} ^{4}$\\
p1m1 & $\mathbb{Z}_{2} ^{2}$ & $\mathbb{Z}_{2} ^{3}$ & $\mathbb{Z}_{2}$ & $\mathbb{Z}_{2} ^{6}$\\
p1g1 & $\mathbb{Z}_{2} ^{2}$ & $\mathbb{Z}_{2}$ & $\mathbb{Z}_{1}$ & $\mathbb{Z}_{2} ^{3}$\\
c1m1 & $\mathbb{Z}_{2} ^{2}$ & $\mathbb{Z}_{2}$ & $\mathbb{Z}_{2}$ & $\mathbb{Z}_{2} ^{4}$\\
p2mm & $\mathbb{Z}_{1}$ & $\mathbb{Z}_{2} ^{4}$ & $\mathbb{Z}_{2} ^{4}$ & $\mathbb{Z}_{2} ^{8}$\\
p2mg & $\mathbb{Z}_{2}$ & $\mathbb{Z}_{2} ^{3}$ & $\mathbb{Z}_{2}$ & $\mathbb{Z}_{2} ^{5}$\\
p2gg & $\mathbb{Z}_{2}$ & $\mathbb{Z}_{2}$ & $\mathbb{Z}_{2}$ & $\mathbb{Z}_{2} ^{3}$\\
c2mm & $\mathbb{Z}_{1}$ & $\mathbb{Z}_{2} ^{3}$ & $\mathbb{Z}_{2} ^{2}$ & $\mathbb{Z}_{2} ^{4}$\\
p4 & $\mathbb{Z}_{1}$ & $\mathbb{Z}_{2} ^{2}$ & $\mathbb{Z}_{2} \times \mathbb{Z}_{4}$ & $\mathbb{Z}_{2} ^{3} \times \mathbb{Z} _{4}$\\
p4mm & $\mathbb{Z}_{1}$ & $\mathbb{Z}_{2} ^{4}$ & $\mathbb{Z}_{2} ^{3}$ & $\mathbb{Z}_{2} ^{7}$\\
p4gm & $\mathbb{Z}_{1}$ & $\mathbb{Z}_{2} ^{2}$ & $\mathbb{Z}_{2} ^{2}$ & $\mathbb{Z}_{2} ^{4}$\\
p3 & $\mathbb{Z}_{1}$ & $\mathbb{Z}_{2}$ & $\mathbb{Z}_{3} ^{3}$ & $\mathbb{Z}_{2} \times \mathbb{Z}_{3} ^{3}$\\
p3m1 & $\mathbb{Z}_{2}$ & $\mathbb{Z}_{2}$ & $\mathbb{Z}_{2}$ & $\mathbb{Z}_{2} ^{3}$\\
p31m & $\mathbb{Z}_{2}$ & $\mathbb{Z}_{2}$ & $\mathbb{Z}_{6}$ & $\mathbb{Z}_{2} ^{2} \times \mathbb{Z}_{6}$\\
p6 & $\mathbb{Z}_{1}$ & $\mathbb{Z}_{2}$ & $\mathbb{Z}_{3} \times \mathbb{Z}_{6}$ & $\mathbb{Z}_{2} \times \mathbb{Z}_{3} \times \mathbb{Z}_{6}$\\
p6mm & $\mathbb{Z}_{1}$ & $\mathbb{Z}_{2} ^{2}$ & $\mathbb{Z}_{2} ^{2}$ & $\mathbb{Z}_{2} ^{4}$\\
\end{tabular}
\end{ruledtabular}
\end{table}

\begin{table}
\caption{\label{tab9}Stacking group structure of interacting \textbf{spin-1/2 crystalline} FSPT protected by $G_{b} = G_{\mathrm{wp}} \times \mathbb{Z}_{2} ^{T}$, which is a direct product of the 2d wallpaper groups $G_{\mathrm{wp}}$ (viewed as \textbf{space} symmetries) and the time-reversal symmetry $\mathbb{Z}_{2} ^{T}$ (viewed as \textbf{onsite} symmetries). This is mapped to \textbf{spinless internal} FSPT by fermionic crystalline equivalence principle.}
\begin{ruledtabular}
\begin{tabular}{ccccc}
$G_{\mathrm{wp}}$ & MC & CF & B & Extension \\ \hline
p2 & $\mathbb{Z}_{2}$ & $\mathbb{Z}_{2} ^{3}$ & $\mathbb{Z}_{2} ^{4}$ & $\mathbb{Z}_{2} ^{8}$\\
p1m1 & $\mathbb{Z}_{2}$ & $\mathbb{Z}_{2} ^{3}$ & $\mathbb{Z}_{2} ^{4}$ & $\mathbb{Z}_{2} ^{3} \times \mathbb{Z}_{4} \times \mathbb{Z}_{8}$\\
p1g1 & $\mathbb{Z}_{2}$ & $\mathbb{Z}_{2} ^{2}$ & $\mathbb{Z}_{1}$ & $\mathbb{Z}_{2} \times \mathbb{Z}_{4}$\\
c1m1 & $\mathbb{Z}_{2}$ & $\mathbb{Z}_{2} ^{2}$ & $\mathbb{Z}_{2} ^{2}$ & $\mathbb{Z}_{2} ^{2} \times \mathbb{Z}_{8}$\\
p2mm & $\mathbb{Z}_{2}$ & $\mathbb{Z}_{2} ^{4}$ & $\mathbb{Z}_{2} ^{12}$ & $\mathbb{Z}_{2} ^{8} \times \mathbb{Z}_{4} ^{3} \times \mathbb{Z}_{8}$\\
p2mg & $\mathbb{Z}_{2}$ & $\mathbb{Z}_{2} ^{3}$ & $\mathbb{Z}_{2} ^{4}$ & $\mathbb{Z}_{2} ^{5} \times \mathbb{Z} _{8}$\\
p2gg & $\mathbb{Z}_{2}$ & $\mathbb{Z}_{2} ^{2}$ & $\mathbb{Z}_{2} ^{2}$ & $\mathbb{Z}_{2} ^{3} \times \mathbb{Z} _{4}$\\
c2mm & $\mathbb{Z}_{2}$ & $\mathbb{Z}_{2} ^{3}$ & $\mathbb{Z}_{2} ^{7}$ & $\mathbb{Z}_{2} ^{6} \times \mathbb{Z}_{4} \times \mathbb{Z}_{8}$\\
p4 & $\mathbb{Z}_{2}$ & $\mathbb{Z}_{2} ^{2}$ & $\mathbb{Z}_{2} ^{3}$ & $\mathbb{Z}_{2} ^{6}$\\
p4mm & $\mathbb{Z}_{2}$ & $\mathbb{Z}_{2} ^{3}$ & $\mathbb{Z}_{2} ^{9}$ & $\mathbb{Z}_{2} ^{6} \times \mathbb{Z}_{4} ^{2} \times \mathbb{Z}_{8}$\\
p4gm & $\mathbb{Z}_{2}$ & $\mathbb{Z}_{2} ^{2}$ & $\mathbb{Z}_{2} ^{4}$ & $\mathbb{Z}_{2} ^{4} \times \mathbb{Z}_{8}$\\
p3 & $\mathbb{Z}_{2}$ & $\mathbb{Z}_{1}$ & $\mathbb{Z}_{1}$ & $\mathbb{Z}_{2}$\\
p3m1 & $\mathbb{Z}_{2}$ & $\mathbb{Z}_{2}$ & $\mathbb{Z}_{2} ^{2}$ & $\mathbb{Z}_{2} \times \mathbb{Z}_{8}$\\
p31m & $\mathbb{Z}_{2}$ & $\mathbb{Z}_{2}$ & $\mathbb{Z}_{2} ^{2}$ & $\mathbb{Z}_{2} \times \mathbb{Z}_{8}$\\
p6 & $\mathbb{Z}_{2}$ & $\mathbb{Z}_{2}$ & $\mathbb{Z}_{2} ^{2}$ & $\mathbb{Z}_{2} ^{4}$\\
p6mm & $\mathbb{Z}_{2}$ & $\mathbb{Z}_{2} ^{2}$ & $\mathbb{Z}_{2} ^{6}$ & $\mathbb{Z}_{2} ^{4} \times \mathbb{Z}_{4} \times \mathbb{Z}_{8}$\\
\end{tabular}
\end{ruledtabular}
\end{table}

\begin{table}
\caption{\label{tab10}Stacking group structure of interacting \textbf{spinless crystalline} FSPT protected by $G_{b} = G_{\mathrm{wp}} \times \mathbb{Z}_{2} ^{T}$, which is a direct product of the 2d wallpaper groups $G_{\mathrm{wp}}$ (viewed as \textbf{space} symmetries) and the time-reversal symmetry $\mathbb{Z}_{2} ^{T}$ (viewed as \textbf{onsite} symmetries). This is mapped to \textbf{spin-1/2 internal} FSPT by fermionic crystalline equivalence principle.}
\begin{ruledtabular}
\begin{tabular}{ccccc}
$G_{\mathrm{wp}}$ & MC & CF & B & Extension \\ \hline
p2 & $\mathbb{Z}_{1}$ & $\mathbb{Z}_{2} ^{3}$ & $\mathbb{Z}_{2} ^{4}$ & $\mathbb{Z}_{2} ^{7}$\\
p1m1 & $\mathbb{Z}_{2} ^{2}$ & $\mathbb{Z}_{2} ^{5}$ & $\mathbb{Z}_{2} ^{5}$ & $\mathbb{Z}_{2} ^{4} \times \mathbb{Z}_{4} \times \mathbb{Z}_{8} ^{2}$\\
p1g1 & $\mathbb{Z}_{2} ^{2}$ & $\mathbb{Z}_{2} ^{2}$ & $\mathbb{Z}_{2}$ & $\mathbb{Z}_{2} ^{2} \times \mathbb{Z}_{8}$\\
c1m1 & $\mathbb{Z}_{2} ^{2}$ & $\mathbb{Z}_{2} ^{3}$ & $\mathbb{Z}_{2} ^{3}$ & $\mathbb{Z}_{2} ^{2} \times \mathbb{Z}_{8} ^{2}$\\
p2mm & $\mathbb{Z}_{1}$ & $\mathbb{Z}_{2} ^{4}$ & $\mathbb{Z}_{2} ^{11}$ & $\mathbb{Z}_{2} ^{15}$\\
p2mg & $\mathbb{Z}_{2}$ & $\mathbb{Z}_{2} ^{4}$ & $\mathbb{Z}_{2} ^{4}$ & $\mathbb{Z}_{2} ^{6} \times \mathbb{Z} _{8}$\\
p2gg & $\mathbb{Z}_{2}$ & $\mathbb{Z}_{2} ^{2}$ & $\mathbb{Z}_{2} ^{2}$ & $\mathbb{Z}_{2} ^{3} \times \mathbb{Z} _{4}$\\
c2mm & $\mathbb{Z}_{1}$ & $\mathbb{Z}_{2} ^{3}$ & $\mathbb{Z}_{2} ^{6}$ & $\mathbb{Z}_{2} ^{9}$\\
p4 & $\mathbb{Z}_{1}$ & $\mathbb{Z}_{2} ^{3}$ & $\mathbb{Z}_{2} ^{3}$ & $\mathbb{Z}_{2} ^{4} \times \mathbb{Z}_{4}$\\
p4mm & $\mathbb{Z}_{1}$ & $\mathbb{Z}_{2} ^{4}$ & $\mathbb{Z}_{2} ^{8}$ & $\mathbb{Z}_{2} ^{10} \times \mathbb{Z}_{4}$\\
p4gm & $\mathbb{Z}_{1}$ & $\mathbb{Z}_{2} ^{3}$ & $\mathbb{Z}_{2} ^{3}$ & $\mathbb{Z}_{2} ^{4} \times \mathbb{Z}_{4}$\\
p3 & $\mathbb{Z}_{1}$ & $\mathbb{Z}_{2}$ & $\mathbb{Z}_{1}$ & $\mathbb{Z}_{2}$\\
p3m1 & $\mathbb{Z}_{2}$ & $\mathbb{Z}_{2} ^{2}$ & $\mathbb{Z}_{2} ^{2}$ & $\mathbb{Z}_{2} ^{2} \times \mathbb{Z}_{8}$\\
p31m & $\mathbb{Z}_{2}$ & $\mathbb{Z}_{2} ^{2}$ & $\mathbb{Z}_{2} ^{2}$ & $\mathbb{Z}_{2} ^{2} \times \mathbb{Z}_{8}$\\
p6 & $\mathbb{Z}_{1}$ & $\mathbb{Z}_{2}$ & $\mathbb{Z}_{2} ^{2}$ & $\mathbb{Z}_{2} ^{3}$\\
p6mm & $\mathbb{Z}_{1}$ & $\mathbb{Z}_{2} ^{2}$ & $\mathbb{Z}_{2} ^{5}$ & $\mathbb{Z}_{2} ^{7}$\\
\end{tabular}
\end{ruledtabular}
\end{table}

The obtained results are based on mapping the crystalline FSPT to its corresponding internal FSPT using the fermionic crystalline equivalence principle \eqref{eq6.1} and \eqref{eq6.2}. Subsequently, we perform calculations for the internal FSPT using the formulas derived in this paper (summarized in \ref{sec1.2}).


\subsection{\label{sec1.4}Organization}

In the rest of this paper, we will investigate the stacking rules of interacting FSPT in spacetime dimensions ranging from 0+1 to 2+1. We will begin by examining the 0+1D case in Sec.\ref{sec3}, followed by the 1+1D case in Sec.\ref{sec4}, and culminating with the 2+1D case in Sec.\ref{sec5}. In Sec.\ref{sec6}, several examples will be provided.
It is important to note that our construction is based on Refs.~\cite{WangGu2018,WangGu2020}, while our computation of examples is based on Ref.~\cite{Ouyang2021}. Therefore, in Sec.\ref{sec3}-\ref{sec5}, we will review the classification results derived by Ref.~\cite{WangGu2020} in the first few subsections. 

\section{\label{sec3}Stacking group structure of 0+1D FSPT}

\subsection{\label{sec3.1}Classification of 0+1D FSPT}

To warm up, let us consider the 0+1D case. In 0+1D spacetime, the system is just a point in the space, and this point possesses two layers of degrees of freedom:
\begin{enumerate}
    \item $|G_{b} |$ levels of bosonic (spin) states $\ket{g} \ ( g\in G_{b})$ at this point;
    \item $|G_{b} |$ species of complex fermions $c^{\sigma } \ ( \sigma \in G_{b})$ at this point.
\end{enumerate}

An FSPT phase can be classified by a pair $( n_{0} ,\nu _{1})$ given the symmetry group $( G_{b} ,\omega _{2} ,s_{1})$, where $n_{0} \in H^{0}( G_{b} ,\mathbb{Z}_{2})$ represents the existence of complex fermions and $\nu _{1} \in C^{1}_{s_1}( G_{b} ,U( 1))$ represents the bosonic phase. If $n_{0}( g) =0$, there is no complex fermion (where the count of complex fermions is always taken modulo 2); otherwise if $n_{0}( g) =1$, there is a complex fermion $c^{g}$ at this point. The pair $( n_{0} ,\nu _{1})$ must satisfy the conditions~\cite{WangGu2020}
\begin{align}
\mathrm{d} n_{0} &=0\ (\bmod\ 2) ,\label{eq3.1}\\
\mathrm{d}_{s_1} \nu _{1} &=( -1)^{\omega _{2} \cup n_{0}} .\label{eq3.2}
\end{align}
The first equation signifies the conservation of fermion parity, while the second equation denotes projective representation.

\subsection{\label{sec3.2}Stacking of 0+1D FSPT}

Given two 0+1D FSPT phases $( n_{0} ,\nu _{1})$ and $( n'_{0} ,\nu '_{1})$ with the same symmetry group $( G_{b} ,\omega _{2} ,s_{1})$, we can stack these two phases to create a new 0+1D FSPT phase classified by the pair $( N_{0} ,\mathcal{V}_{1})$. Then it's nature to ask what is the relation between the new data $( N_{0} ,\mathcal{V}_{1})$ and the old data $( n_{0} ,\nu _{1})$ and  $( n'_{0} ,\nu '_{1})$? The answer is given by the following equations:
\begin{align}
N_{0} &=n_{0} +n'_{0} \ (\bmod\ 2) ,\label{eq3.3}\\
\mathcal{V}_{1} &=\nu _{1} \nu '_{1} \label{eq3.4}.
\end{align}
The first \eqref{eq3.3} is derived from fermion parity conservation of the complex fermion decoration. Once we have obtained \eqref{eq3.3}, we can naturally derive the stacking rule \eqref{eq3.4} of the bosonic layer from (\ref{eq3.2}): 
\begin{equation}
\mathcal{V}_{1} =( -1)^{\omega _{2} \cup N_{0}} =( -1)^{\omega _{2} \cup ( n_{0} +n'_{0})} =\nu _{1} \nu '_{1} .
\end{equation}
We observe that the two classification data are stacked separately in 0+1D without twist.

\section{\label{sec4}Stacking group structure of 1+1D FSPT}

\subsection{\label{sec4.1}Classification of 1+1D FSPT}

Now we turn to the more intricate 1+1D spacetime. The system consists of three layers of degrees of freedom:
\begin{enumerate}
    \item $|G_{b} |$ levels of bosonic (spin) states $\ket{g_{i}} \ ( g_{i} \in G_{b})$ on each vertex $i$ of the spatial lattice;
    \item $|G_{b} |$ species of complex fermions $c_{ij}^{\sigma } \ ( \sigma \in G_{b})$ at the center of each link $\langle ij\rangle $ of the spatial lattice;
    \item $|G_{b} |$ species of complex fermions (split into Majorana fermions) $a_{i}^{\sigma } =\left( \gamma _{i,A}^{\sigma } +i\gamma _{i,B}^{\sigma }\right) /2\ ( \sigma \in G_{b})$ on each vertex $i$ of the spatial lattice.
\end{enumerate}

An FSPT phase is classified by a triplet $( n_{0} ,n_{1} ,\nu _{2})$, where $n_{0} \in H^{0}( G_{b} ,\mathbb{Z}_{2})$ represents the presence of a Majorana chain, $n_{1} \in C^{1}( G_{b} ,\mathbb{Z}_{2})$ represents the complex fermion decoration, and $\nu _{2} \in C^{2}_{s_1}( G_{b} ,U( 1))$ represents the bosonic phase. It should be noted that in Ref.~\cite{WangGu2020}, a FSPT phase is considered as just a pair $( n_{1} ,\nu _{2})$, without the inclusion of the Majorana chain decoration $n_{0}$ since Kitaev's Majorana chain is a invertible state. In this paper, we include the $n_{0}$ data to ensure generality.
Therefore precisely speaking, we are constructing the 1+1D fermionic invertible topological order with symmetries.

\begin{figure*}
    \centering
    \includegraphics[width=0.65\textwidth]{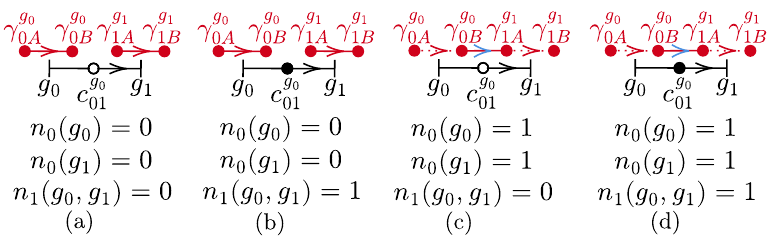}
    \caption{
    The decorations in the complex fermions and Majorana chains are as follows: a black hollow circle represents no complex fermion decoration, indicated by $n_{1}( g_{0} ,g_{1}) =0$; a black solid circle represents one complex fermion decoration, indicated by $n_{1}( g_{0} ,g_{1}) =1$; a red solid circle represents a Majorana fermion; a red solid line with an arrow represents Majorana pairings; a red dashed line with an arrow represents the absence of Majorana pairing between the species labeled by $g_{i} \in G_{b}$ for the site $i$, while all other species $\sigma \neq g_{i}$ exhibit trivial pairing.
    }
    \label{fig1}
\end{figure*}

If the $n_{0}( g_{i}) =0$, then the Majorana operator $\gamma _{i,A}^{\sigma }$ and $\gamma _{i,B}^{\sigma }$ are both in trivial pairing $-i\gamma _{i,A}^{\sigma } \gamma _{i,B}^{\sigma }=1$ for all $\sigma \in G_{b}$. If $n_{0}( g_{i}) \neq 0$, only the Majorana pair $\gamma _{i,A}^{g_{i}}$ and $\gamma _{i,B}^{g_{i}}$ are in non-trivial paring (i.e. paired with adjacent Majorana operators, respectively), the other Majorana pairs $\gamma _{i,A}^{\sigma }$ and $\gamma _{i,B}^{\sigma }$ for $\sigma \neq g_{i}$ are all in trivial paring. If $n_{1}( g_{i} ,g_{j}) =0$, there is no complex fermion at the center of the link $\langle ij\rangle $. Otherwise, if $n_{1}( g_{i} ,g_{j}) =1$, there is a complex fermion $c_{ij}^{g_{i}}$ decorated at this center. Fig.~\ref{fig1} demonstrates the decorations of complex fermions and Majorana chains. In this paper, we always only label the potentially non-trivial Majorana fermions and complex fermions in all the figures, unless otherwise specified. For example, in Fig.~\ref{fig1}(a), the potentially non-trivial Majorana fermions are $\gamma _{0A}^{g_{0}}$ and $\gamma _{0B}^{g_{0}}$ at the site $0$. All the other Majorana fermions $\gamma _{0A}^{\sigma }$ and $\gamma _{0B}^{\sigma }$ for $\sigma \neq g_{0}$ are irrelevant and will not be discussed.

\subsubsection{\label{sec4.1.1}Obstruction function}

The triplet $( n_{0} ,n_{1} ,\nu _{2})$ must satisfy the following conditions: 
\begin{align}
\mathrm{d} n_{0} &=0\ (\bmod\ 2) ,\label{eq4.1}\\
\mathrm{d} n_{1} &=\omega _{2} \cup n_{0} \ (\bmod\ 2) ,\label{eq4.2}\\
\mathrm{d}_{s_1} \nu _{2} &=( -1)^{\omega _{2} \cup n_{1} +\mathrm{d} n_{1} \cup _{1}\mathrm{d} n_{1} +\mathrm{d} n_{1} \cup n_{1} + s_{1} \cup \mathrm{d} n_{1}} .\label{eq4.3}
\end{align}
Note that the above equations are re-derived in our paper, which are slightly different from the equations outlined in equations (8)-(11) of Ref.~\cite{WangGu2020} because we add the Majorana chain decoration while this layer is not considered in the reference.
The first condition \eqref{eq4.1} ensures that the Majorana chain forms a closed loop, without any dangling Majorana modes since the state should be gapped. The second condition \eqref{eq4.2} is derived from the fermion parity conservation of both Majorana fermions and complex fermions. The last condition \eqref{eq4.3} is obtained from the symmetry transformation and consistency condition of $F$ moves. The $F$ move is defined as 
\begin{widetext}
\begin{gather}
\Psi \left( \adjincludegraphics[valign=c,scale=0.8]{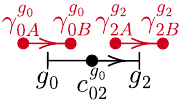} \right) =F( g_{0} ,g_{1} ,g_{2}) \Psi \left( 
\adjincludegraphics[valign=c,scale=0.8]{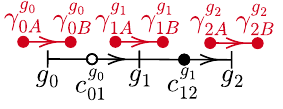} \right) ,\label{eq4.4}\\
F( g_{0} ,g_{1} ,g_{2}) =|G_{b} |^{1/2} \nu _{2}^{1-2s_{1}( g_{0})}( g_{0} ,g_{1} ,g_{2})\left( c_{02}^{g_{0} \dagger }\right)^{n_{1}( 02)}\left( c_{01}^{g_{0}}\right)^{n_{1}( 01)}\left(( -1)^{\omega _{2}\left( g_{0} ,g_{0}^{-1} g_{1}\right)} c_{12}^{g_{1}}\right)^{n_{1}( 12)} X_{012}[ n_{0}] ,\label{eq4.5}\\
X_{012}[ n_{0}] =P_{012}[ n_{0}]\left(( -1)^{s_{1}( g_{0})} \gamma _{0B}^{g_{0}}\right)^{\mathrm{d} n_{1}( 012)} .\label{eq4.5.1}
\end{gather}
\end{widetext}
In the above expression, $|G_{b} |$ is the order of the group $G_{b}$, $|G_{b} |^{1/2}$ is just a normalization factor for the reduction of one lattice site; $\nu _{2}( g_{0} ,g_{1} ,g_{2})$ is a $U( 1)$ phase factor; the complex fermion terms $c^{\dagger } cc$ annihilate the possibly decorated (depend on $n_{1}$) complex fermions $c_{01}^{g_{0}}$ and $c_{12}^{g_{1}}$ on the right hand side of \eqref{eq4.4} and possibly create (depend on $n_{1}$) a new complex fermion $c_{02}^{g_{0}}$ on the left hand side. The $X_{012}[ n_{0}]$ term in \eqref{eq4.5.1} stands for the projection operator of Majorana chain decoration, it projects the Majorana chain configuration in the right hand side to the left hand side, which we will explain in detail later \footnote{See the explanation for \eqref{eq5.3} and \eqref{eq5.4} for the 2+1D case. The current 1+1D case is similar. Since the related calculation is not important in the 1+1D case, we will skip it here.}. A dangling Majorana mode $\gamma _{0B}^{g_{0}}$ is inserted to match the fermion parity of both sides. Finally the $(-1)$ phase factors come from the symmetry transformation stated in \eqref{eq4.18}.

It is important to note that the picture in \eqref{eq4.4} is merely an example used to illustrate the concept. In this example, $n_{0}( g_{0}) =n_{0}( g_{1}) =n_{0}( g_{2}) =0$ (indicating trivial Majorana pairs depicted by red solid arrows with red dots on both sides), $n_{1}( g_{0} ,g_{2}) =n_{1}( g_{1} ,g_{2}) =1$ (depicted by the black solid dots), and $n_{1}( g_{0} ,g_{1}) =0$ (depicted by the black hollow dot). In general, the $F$ move can accept any input configuration of $n_{0}$ and $n_{1}$. This argument applies to all figures in this paper, unless otherwise stated.

\subsubsection{\label{sec4.1.2}Symmetry transformation}

Since we are constructing FSPT states, the $F$ moves should be compatible with the symmetry action $U( g)$ defined as follows for any $g\in G_{b}$: 
\begin{enumerate}
    \item on the bosonic state, 
\begin{equation}
U( g)\ket{g_{i}} =\ket{gg_{i}} ;\label{eq4.6}
\end{equation}
    \item on the complex fermions, 
\begin{equation}
U( g) c_{ij\cdot \cdot \cdot k}^{\sigma } U^{\dagger }( g) =( -1)^{\omega _{2}( g,\sigma )} c_{ij\cdot \cdot \cdot k}^{g\sigma } ;\label{eq4.7}
\end{equation}
    \item on the Majorana fermions induced from the complex fermion $a_{ij\cdot \cdot \cdot k}^{\sigma } = \frac{1}{2}(\gamma _{ij\cdot \cdot \cdot k,A}^{\sigma } + i \gamma _{ij\cdot \cdot \cdot k,B}^{\sigma })$, 
\begin{align}
U( g) a_{ij\cdot \cdot \cdot k}^{\sigma } U^{\dagger }( g) &=( -1)^{\omega _{2}( g,\sigma )} a_{ij\cdot \cdot \cdot k}^{g\sigma } ,\label{eq4.8}\\
U( g) \gamma _{ij\cdot \cdot \cdot k,A}^{\sigma } U^{\dagger }( g) &=( -1)^{\omega _{2}( g,\sigma )} \gamma _{ij\cdot \cdot \cdot k,A}^{g\sigma } ,\label{eq4.9}\\
U( g) \gamma _{ij\cdot \cdot \cdot k,B}^{\sigma } U^{\dagger }( g) &=( -1)^{\omega _{2}( g,\sigma ) +s_{1}( g)} \gamma _{ij\cdot \cdot \cdot k,B}^{g\sigma } .\label{eq4.10}
\end{align}
\end{enumerate}

If we define the standard Majorana pairing by restricting the first Majorana fermion's label to $e$ (the identity element in group $G_b$) as 
\begin{equation}
-i\gamma _{i,C}^{e} \gamma _{j,D}^{g^{-1} h} =1 , \label{eq4.11}
\end{equation}
depicted by a red arrow from $\gamma _{i,C}^{e}$ at position $i$ (maybe a site or higher codimension domain wall depends on the dimension of the system) to $\gamma _{j,D}^{g^{-1} h}$ at position $j$ in the picture ($C,D$ take values in $\{A,B\}$), we get the generic Majorana pairing through the symmetry transformation 
\begin{equation}
-i\gamma _{i,C}^{g} \gamma _{j,D}^{h} =( -1)^{\omega _{2}\left( g,g^{-1} h\right) +s_{1}( g)( 1+\delta _{CB} +\delta _{DB})} \label{eq4.12}
\end{equation}
depicted by a blue arrow on a red solid line (as shown in Fig.~\ref{fig1}(c) and (d), for instance) to indicate the direction may be flipped after the symmetry transformation. Here, $\delta _{CB} =1$ when $C=B$, otherwise $\delta _{CB} =0$ ($\delta _{DB}$ is defined similarly). To simplify the description of the pairing, we introduce the projection operator of the Majorana fermion pairing as 
\begin{align}\label{eq4.13}
P_{iC,jD}^{g,h} &:=U( g) P_{iC,jD}^{e,g^{-1} h} U^{\dagger }( g)\\\nonumber
&=U( g)\frac{1}{2}\left( 1-i\gamma _{i,C}^{e} \gamma _{j,D}^{g^{-1} h}\right) U^{\dagger }( g)\\\nonumber
&=\frac{1}{2}\left[ 1-( -1)^{\omega _{2}\left( g,g^{-1} h\right) +s_{1}( g)( 1+\delta _{CB} +\delta _{DB})} i\gamma _{i,C}^{g} \gamma _{j,D}^{h}\right].
\end{align} 

Now, in order to be compatible with the symmetry transformation, the $F$ move must satisfy 
\begin{equation}
F(\{gg_{i}\}) =U( g) F(\{g_{i}\}) U^{\dagger }( g) .\label{eq4.14}
\end{equation}
That is, the follow diagram commutes 
\begin{equation}
\vcenter{\hbox{\includegraphics[scale=0.8]{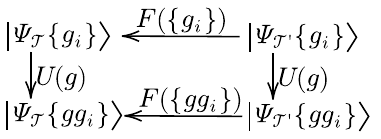}}}
,\label{eq4.15}
\end{equation}
where $\mathcal{T}$ and $\mathcal{T} '$ label different triangulations of the space.

The above $F$ move \eqref{eq4.5} is inconvenient in the derivation, so we often use 
a more convenient one called standard $F$ move with the first argument being the identity element $e\in G_{b}$
\begin{widetext}
\begin{gather}
\Psi \left( \adjincludegraphics[valign=c,scale=0.8]{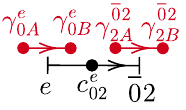} \right) =F( e,\overline{0} 1,\overline{1} 2) \Psi \left( \adjincludegraphics[valign=c,scale=0.8]{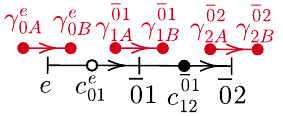} \right) ,\label{eq4.16}\\
F( e,\overline{0} 1,\overline{0} 2) =|G_{b} |^{1/2} \nu _{2}( e,\overline{0} 1,\overline{0} 2)\left( c_{02}^{e\dagger }\right)^{n_{1}( 02)}\left( c_{01}^{e}\right)^{n_{1}( 01)}\left( c_{12}^{\overline{0} 1}\right)^{n_{1}( 12)} X_{012}[ n_{0}] ,\label{eq4.17}\\
X_{012}[ n_{0}] =P_{012}[ n_{0}]\left( \gamma _{0B}^{e}\right)^{\mathrm{d} n_{1}( 012)} .\label{eq4.17.1}
\end{gather}
\end{widetext}
which is obtained from \eqref{eq4.4} and \eqref{eq4.5} by setting the three input group elements of $F$ move as $e$, $\overline{0} 1:=g_{0}^{-1} g_{1}$ and $\overline{0} 2:=g_{0}^{-1} g_{2}$ (we will constantly use the shorthand notation $i$ for $g_{i}$ and $\overline{i}$ for $g_{i}^{-1}$ from now on). We also used the shorthand notation for $n_{1}( ij) :=n_{1}( g_{i} ,g_{j}) =n_{1}\left( e,g_{i}^{-1} g_{j}\right) =n_{1}\left( g_{i}^{-1} g_{j}\right)$, where the former one with two variables is a homogeneous cocycle, and the later with only one variable is 
an inhomogeneous cocycle (we will interchangeably use both from now on, and the reader can distinguish them by the number of variables). See Appendix \ref{ap1} for more detail on the relation between homogeneous and inhomogeneous cocycles.

Once we have the standard $F$ move, using \eqref{eq4.14}, we can get the generic $F$ move as 
\begin{align}\label{eq4.18}\nonumber
F( g_{0} ,g_{1} ,g_{2}) &={}^{g_{0}} F\left( e,g_{0}^{-1} g_{1} ,g_{0}^{-1} g_{2}\right)\\
&:=U( g_{0}) F\left( e,g_{0}^{-1} g_{1} ,g_{0}^{-1} g_{2}\right) U^{\dagger }( g_{0}).
\end{align}
Additionally, the commutation relation \eqref{eq4.15} is automatically satisfied because the follow diagram commutes 
\begin{equation}
\adjincludegraphics[valign=c,scale=0.8]{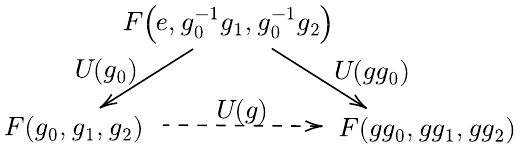} . \label{eq4.19}
\end{equation}
To simplify calculations, we consistently utilize the standard $F$ move after it has been defined.

\subsubsection{\label{sec4.1.3}Consistency equation}

The $F$ moves should satisfy that the following diagram
\begin{equation}
\adjincludegraphics[valign=c,scale=0.8]{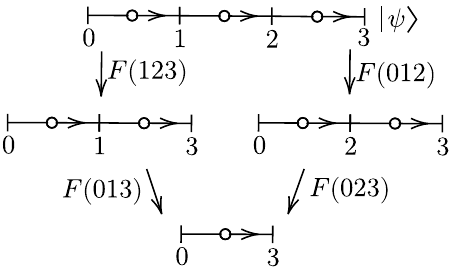}  \label{eq4.19.1}
\end{equation}
commutes, which is equivalent to 
\begin{equation}
F( e,\overline{0} 2,\overline{0} 3) \cdotp F( e,\overline{0} 1,\overline{0} 2) =F( e,\overline{0} 1,\overline{0} 3) \cdotp ^{\left( g_{0}^{-1} g_{1}\right)} F( e,\overline{1} 2,\overline{1} 3) .\label{eq4.19.2}
\end{equation}
Here the move $F(123)$ is non-standard. This commutative diagram \eqref{eq4.19.1} leads to the obstruction function \eqref{eq4.3} in three parts
\begin{equation}
\mathcal{O}_{3} [n_{1}] =\mathcal{O}_{3}^{c} \cdot \mathcal{O}_{3}^{c\gamma } \cdot \mathcal{O}_{3}^{\gamma } .\label{eq4.19.3}
\end{equation}

The firs part comes from the commutation relation between complex fermions and the symmetry transformation of complex fermions 
\begin{equation}
\begin{array}{ c c l }
\mathcal{O}_{3}^{c}( 0123) & = & ( -1)^{\omega _{2}( 012) n_{1}( 23)}\\
 & \times  & ( -1)^{n_{1}( 02) n_{1}( 23) +n_{1}( 23)[ n_{1}( 01) +n_{1}( 12)]}
\end{array} .\label{eq4.19.4}
\end{equation}
The second part comes from the commutation relation between complex fermions and Majorana fermions 
\begin{equation}
\mathcal{O}_{3}^{c\gamma }( 0123) =( -1)^{\mathrm{d} n_{1}( 023)\mathrm{d} n_{1}( 012) +\mathrm{d} n_{1}( 013)\mathrm{d} n_{1}( 123)} .\label{eq4.19.5}
\end{equation}
Finally the third part is the pure Majorana phase 
\begin{equation}
\mathcal{O}_{3}^{\gamma }( 0123) =( -1)^{s_{1}( 01)\mathrm{d} n_{1}( 123)} ,\label{eq4.19.6}
\end{equation}
which is calculated by 
\begin{equation}
\begin{array}{ c c l }
\mathcal{O}_{3}^{\gamma }( 0123) & = & \bra{\psi }\overline{P}_{123}\left(( -1)^{s_{1}( 01)}\textcolor[rgb]{0.82,0.01,0.11}{\gamma }\textcolor[rgb]{0.82,0.01,0.11}{_{1B}^{\overline{0} 1}}\right)^{\mathrm{d} n_{1}( 123)}\\
 & \times  & \overline{P}_{013}\left(\textcolor[rgb]{0.29,0.56,0.89}{\gamma }\textcolor[rgb]{0.29,0.56,0.89}{_{0B}^{e}}\right)^{\mathrm{d} n_{1}( 013)} P_{023}\left(\textcolor[rgb]{0.29,0.56,0.89}{\gamma }\textcolor[rgb]{0.29,0.56,0.89}{_{0B}^{e}}\right)^{\mathrm{d} n_{1}( 023)}\\
 & \times  & P_{012}\left(\textcolor[rgb]{0.29,0.56,0.89}{\gamma }\textcolor[rgb]{0.29,0.56,0.89}{_{0B}^{e}}\right)^{\mathrm{d} n_{1}( 012)}\ket{\psi }
\end{array} .\label{eq4.19.6}
\end{equation}

In the above formula of expectation value, the state $\ket{\psi }$ is taken to be the top figure of \eqref{eq4.19.1}. We project this state to the other states in this commutative diagram step by step in a counterclockwise order, and finally return to the original state $\ket{\psi }$ with potentially an extra Berry phase. We extract this Berry phase by this formula. The projector $\overline{P}$ represents the 'inverse' process of the projector $P$. However, they are not the real inverse operator of each other $\overline{P}P \neq \mathbb{I}$. We will explain how to calculate such an expectation value like \eqref{eq4.19.6} in detail in Sec. \ref{sec5.3}. In the current case, it turns out that this expectation value is just trivial, i.e., only the symmetry transformation factor $( -1)^{s_{1}( 01)\mathrm{d} n_{1}( 123)}$ is left.

\subsection{\label{sec4.2}Stacking of 1+1D FSPT}

Given two 1+1D FSPT phases $( n_{0} ,n_{1} ,\nu _{2})$ and $( n'_{0} ,n'_{1} ,\nu '_{2})$ with the same symmetry group $( G_{b} ,\omega _{2} ,s_{1})$, we stack these two phases to create a new 1+1D FSPT phase denoted by $( N_{0} ,N_{1} ,\mathcal{V}_{2})$. For this new phase, we aim to address the same question as we did in the 0+1D case. Due to the increasing length of the derivations, we first present the answers: 
\begin{align}
N_{0} &=n_{0} +n'_{0} \ (\bmod\ 2) ,\label{eq4.20}\\
N_{1} &=n_{1} +n'_{1} +m_{1} \ (\bmod\ 2) ,\label{eq4.21}\\
m_{1} &=s_{1} \cup n_{0} \cup n'_{0} ,\label{eq4.21.1}\\
\mathcal{V}_{2} &=\nu _{2} \nu '_{2}( -1)^{\epsilon _{2}} e^{2\pi i\theta _{2}[ n_{0} ,n'_{0}]} ,\label{eq4.22}\\
\epsilon _{2} &=m_{1} \cup ( n_{1} +n'_{1}) +n_{1} \cup n'_{1} +n_{1} \cup _{1}\mathrm{d} n'_{1} ,\label{eq4.22.1}\\
\theta _{2} &=\frac{3}{4} \omega _{2} \cup n_{0} \cup n'_{0} +\frac{1}{2} \omega _{2} \cup _{1} m_{1} .\label{eq4.22.2}
\end{align}
It is important to note that due to the obstruction of \eqref{eq4.2}: $\mathrm{d}n_{1} = \omega _{2} \cup n_{0}$, $\omega _{2}$ and $n_{0}$ can not be non-trivial simultaneously, otherwise $n_{1}$ will be obstructed. Therefore, the obstruction-free condition requires $\omega _{2}$ in \eqref{eq4.22.1} and \eqref{eq4.22.2} to be trivial, which means we can set $\omega_{2}=0$ \footnote{In general, $\omega _{2}$ could be a coboundary, and different choice of coboundaries may lead to different physical realizations. However, if we only consider the classification and extension problems, the results derived from different coboundaries should be isomorphic to each other.}. After this simplification, the stacking rule for the bosonic layer becomes rather simple as \eqref{eq1.5}.

Our results are equivalent to the previous results in Ref. \cite{Omer} up to a phase shift in the bosonic layer $\nu _{2}\rightarrow \nu _{2}(-1)^{s_{1}\cup n_{1}\cup n_{0}}$. Note that the bosonic phase $\nu _{2}$ itself is a torsor, so introducing a phase shift in the definition of bosonic layer does not change the physics. Subsequently, we proceed to derive each of these equations individually.

\subsubsection{\label{sec4.2.1}$\mathcal{P}$ move}

Since we are considering stacking two FSPT states into a single one, we need to define the $\mathcal{P}$ move, an FSLU transformation, which projects the two-layer system onto a one-layer system. We need to emphasize that although it maps two layers into a single layer, it does not change the number of degrees of freedom but only puts some degrees into the trivial vacuum state. Therefore, we can define an inverse operator to this $\mathcal{P}$. The explicit expression for it is
\begin{widetext}
\begin{gather}
\Psi \left( \adjincludegraphics[valign=c,scale=0.8]{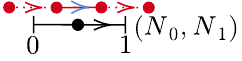} \right) =\mathcal{P}_{01} \Psi \left( \adjincludegraphics[valign=c,scale=0.8]{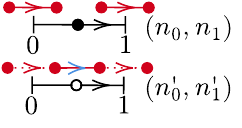} \right) ,\label{eq4.23}\\
\mathcal{P}_{01} =U( g_{0})\mathcal{P}_{01}^{\mathrm{std}} U^{\dagger }( g_{0}) =P_{01} \left(( -1)^{s_{1}( g_{0})} \gamma ^{\prime g_{0}}_{0B}\right)^{m_{1}( 01)} \left( C_{01}^{g_{0} \dagger }\right)^{N_{1}( 01)}\left( c^{\prime g_{0}}_{01}\right)^{n'_{1}( 01)}\left( c_{01}^{g_{0}}\right)^{n_{1}( 01)} ,\label{eq4.24}
\end{gather}
\end{widetext}
where $P_{01} =U( g_{0}) P_{01}^{\mathrm{std}} U^{\dagger }( g_{0})$ is the projection operator of Majorana chain configurations from the right hand side to the left hand side of the equation. A dangling Majorana mode $\gamma ^{\prime g_{0}}_{0B}$ is inserted for fermion parity matching. The standard $\mathcal{P}$ move is defined similar as the standard $F$ move 
\begin{equation}
\mathcal{P}_{01}^{\mathrm{std}} =P_{01}^{\mathrm{std}} \left( \gamma ^{\prime e}_{0B}\right)^{m_{1}( 01)} \left( C_{01}^{e\dagger }\right)^{N_{1}( 01)}\left( c^{\prime e}_{01}\right)^{n'_{1}( 01)}\left( c_{01}^{e}\right)^{n_{1}( 01)} \label{eq4.25}
\end{equation}
with the standard Majorana chain configuration projection operator $P_{01}^{\mathrm{std}}$. The same as the argument used in the $F$ move \eqref{eq4.4}, the illustration provided in \eqref{eq4.23} represents one possible configuration in which $n_{0}( g_{0}) =n_{0}( g_{1}) =0$, $n'_{0}( g_{0}) =n'_{0}( g_{1}) =1$, $n_{1}( 01) =1$ and $n'_{1}( 01) =0$ . In general, the $\mathcal{P}_{01}$ move is applicable to any possible input configuration of $n_{0}$ and $n_{1}$. This argument is always implicitly assumed, unless stated otherwise, and further emphasis will be omitted.

To ensure clarity, we will address several potential points of confusion about the $\mathcal{P}$ move. In this explanation, we will employ the 1+1D $\mathcal{P}$ move defined in \eqref{eq4.23} as an illustrative example, but it should be noted that the argument presented here is applicable to any dimension.

The first issue concerns the degrees of freedom associated with the $\mathcal{P}$ move. It is worth noting that the right-hand side of \eqref{eq4.23} includes two layers of FSPT phases identified as $(n _{0},n _{1},\nu _{2})$ and $(n '_{0},n '_{1},\nu '_{2})$, whereas the left-hand side comprises only one layer of FSPT phase denoted as $(N _{0},N _{1},\mathcal{V} _{2})$. At first glance, it may appear that the degrees of freedom are mismatched. However, this is not the case, as there are still two layers of degrees of freedom on the left-hand side. The upper layer, represented by $(N _{0},N _{1},\mathcal{V} _{2})$, is explicitly depicted in \eqref{eq4.23}, while the lower layer corresponds to a trivial FSPT phase that is implicitly assumed and therefore not represented in \eqref{eq4.23}.

The second issue concerns the definition of the $\mathcal{P}$ move. It has been observed that the triangulation of the upper and lower layers on the right-hand side of \eqref{eq4.23} is identical and the group element labels of the bosonic state for both the upper and lower layers are the same at each site. 
A compelling argument can be made for setting the triangulation and group element labels of the two layers to be the same in the definition of the $\mathcal{P}$ move, which we present as follows. 

\begin{figure}
    \centering
    \includegraphics[width=0.25\textwidth]{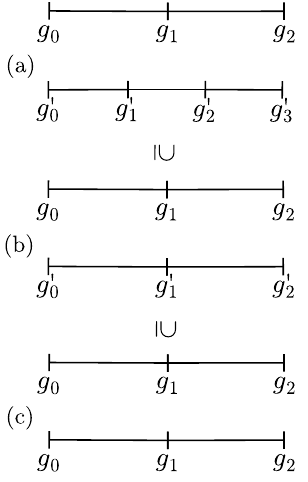}
    \caption{The wavefunctions of BSPT: (a) the triangulation and group element labels of the two layers are both different; (b) the triangulation is the same, while the group element labels are different; (c) the triangulation and group element labels are both the same, which is the setup for the $\mathcal{P}$ move in \eqref{eq4.23}. The wavefunction in (c) is a subset of (b), and the wavefunction in (b) is a subset of (a).} 
    \label{fig18}
\end{figure}

Let us consider the BSPT for simplicity and ignore the fermionic degrees of freedom. In general, there are three situations, as shown in Fig. \ref{fig18}. The first is the one in panel (a) where the triangulation and the group element labels for the two layers could be different as shown in Fig. \ref{fig18}(a). The second is panel (b) that depicts a scenario where the triangulation of the two layers is the same, but the group element labels are distinct. Thirdly,  panel (c) represents a case where the triangulation and group element labels in both layers are the same, corresponding to the setup of the $\mathcal{P}$ move in \eqref{eq4.23}.

It can be observed that the wavefunction shown in Fig. \ref{fig18}(c) is a special case of the wavefunction in (b), and the wavefunction in (b) is a special case of (a). It is important to note that the fixed point wavefunction of FSPT remains in the same class up on re-triangulation of the lattice (in general up to the renormalization of the wave-function). So starting from the case (a) where the upper and lower layers having different triangulations, we can also possess appropiate triangulation procedures such that the upper and lower layers have the same triangulation structure, which arrives at the case (b). 

To go from case (b) to case (c), we first note that in the decoupling limit of the two layers, in general the symmetry of the whole system is (temporally) enlarged to $G\times G$ whose elements is denoted as $U(g)\otimes U(g')$ where $g, g'\in G$. Importantly, we only need one $G$ symmetry so that we identify the diagonal one as our physical symmetry whose group elements are given by $U(g)\otimes U(g)$ where $g\in G$.
Now we turn on an ferromagnetic coupling 
\begin{equation}
    \mathcal{H} _{\mathrm{FC}} = -\sum\nolimits _{ \begin{array}{l}
g\in G\\
\forall \mathrm{site} \ i
\end{array}}\left(\ket{g}_{i} \otimes \ket{g} '_{i}\right)\left(\bra{g}_{i} \otimes \bra{g} '_{i}\right),
\end{equation}
which is allowed since it is symmetric under $U(g)\otimes U(g)$ and the final state is still in the same class. This coupling forces the degrees of freedom in both layers to be the same, i.e. the triangulation and the group element labels are forced to be the same as depicted in Fig. \ref{fig18}(c).
This is why we can define the $\mathcal{P}$ move as \eqref{eq4.23}, and this choice of definition will greatly simplify the problem.

\subsubsection{\label{sec4.2.2}Kitaev chain and complex fermion decorations}

The objective of this subsection is to derive the stacking rules \eqref{eq4.20} and \eqref{eq4.21} for the classification data of the Kitaev chain and the complex fermion respectively. The main idea is to design some FSLU transformations to convert the configuration of Kitaev chains from the two-layer system to an one-layer system, such that the lower layer in the original two-layer system becomes a trivial product state while the upper layer may still contain non-trivial Majorana pairings. The transformed system is a tensor product of the upper nontrivial FSPT state and a trivial product state in the lower layer. Therefore, the properties, especially the classification data, can be obtained by analyzing the upper layer.

The FSLU transformation of Kitaev chain is characterized by a transition loop (or transition graph). Starting from an initial Majorana pairing configuration, applying the FSLU transformation using certain projection operators in the form of \eqref{eq4.13} results in a new Majorana pairing configuration. By connecting the Majorana pairing configurations before and after the FSLU transformation, a transition loop is formed. This loop is Kasteleyn oriented if, when moving counterclockwise (or clockwise) along the loop, there is an odd number of arrows pointing in the opposite direction to our movement \cite{Kasteleyn1963}. In such cases, the fermion parity is conserved during the FSLU transformation. If the loop is not Kasteleyn oriented, the fermion parity will change.

\begin{figure*}
    \centering
    \includegraphics[width=0.65\textwidth]{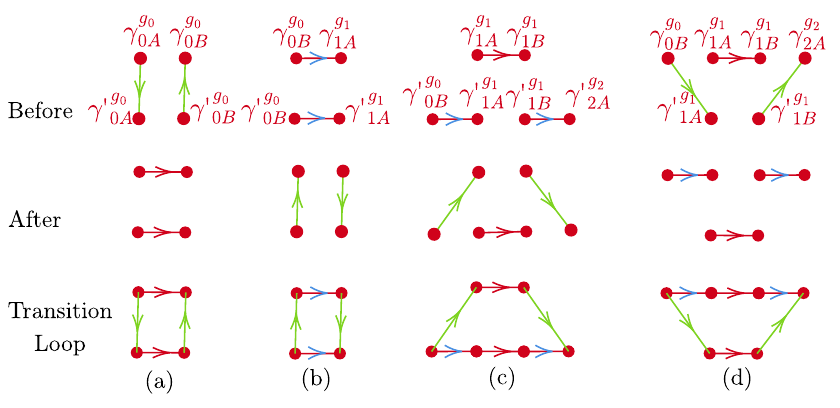}
    \caption{Four basic moves of FSLU transformation on Kitaev chains that are always Kasteleyn oriented.}
    \label{fig2}
\end{figure*}

In Fig.~\ref{fig2}, there are four basic moves of FSLU transformation on Kitaev chains, where (a), (c) and (d) are all Kasteleyn oriented. Note panel (b) may be non-Kasteleyn oriented depending on $s_{1}$. As before, the red arrow indicates that this Majorana pair will not change sign from the standard pairing
to the non-standard one. In contrast, the blue and green arrows indicate that the sign will be changed. Specifically, we use a green solid line with an arrow to indicate that this pair connects Majorana operators from different layers, also referred to as a 'bridge' pair. Our convention for the green arrow is as follows: if the line is vertical, then the arrow always points from $\gamma _{A}$ to $\gamma _{A} '$, or from $\gamma '_{B}$ to $\gamma _{B}$, such as $\gamma _{0A}^{g_{0}} \gamma ^{\prime g_{0}}_{0A}$ and $\gamma ^{\prime g_{0}}_{0B} \gamma _{0B}^{g_{0}}$ in Fig.~\ref{fig2}(a); if the line is not purely vertical, then the arrow points as the same direction as the horizontal one, such as $\gamma ^{\prime g_{0}}_{0B} \gamma _{1A}^{g_{1}}$ and $\gamma _{0B}^{g_{0}} \gamma _{1A}^{g_{1}}$ in Fig.~\ref{fig2}(c). These conventions ensure that all four basic moves are always Kasteleyn oriented when $s_{1}=0$.

\begin{figure*}
    \centering
    \includegraphics[width=0.65\textwidth]{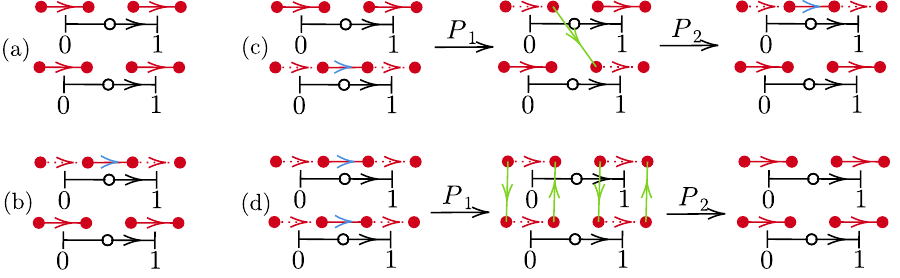}
    \caption{All possible FSLU transformations that reduce the two-layer system to one possibly non-trivial layer plus a trivial product state layer.}
    \label{fig3}
\end{figure*}

Now we explain the function of $P_{01}$ in \eqref{eq4.23} by exhaustive enumeration. In Fig.~\ref{fig3}, we have listed all the possible configurations for $n_{0}$ and $n'_{0}$ and demonstrate how to reduce the two-layer system to one layer, which may be non-trivial, and a trivial product state layer through FSLU transformations. In panel (a) and (b) of Fig.~\ref{fig3}, the Kitaev chain in the lower layer is already trivial, so no action is required. In Fig.~\ref{fig3}(c), the upper layer is trivial while the lower layer is non-trivial. Our objective is to shift the non-trivial Kitaev chain from the lower layer to the upper layer. This can be achieved by employing the basic move (c) in Fig.~\ref{fig2}, followed by the basic move (d) in Fig.~\ref{fig2}. Please note that for simplicity, only a portion of the transition loops are illustrated in Fig.~\ref{fig3}(c) as the complete loops involve more sites outside this picture. In Fig.~\ref{fig3}(d), the Kitaev chains in the two layers are all non-trivial, resulting in a trivial $N_{0}$. To achieve this, we first utilize the basic move Fig.~\ref{fig2}(b) (this step may potentially change the fermion parity), followed by the basic move Fig.~\ref{fig2}(a). Again, please note that only a portion of the transition loops are shown in the first step, and further emphasis will be omitted in similar situations. Ultimately, we obtain the stacking rule of the Majorana decoration as \eqref{eq4.20}.

Since only the three Kasteleyn oriented basic moves in Fig.~\ref{fig2} are used, the conservation of fermion parity is always maintained in the stacking processes Fig.~\ref{fig3}(a), (b) and (c) of Kitaev chains. The only process that may change the fermion parity is the first step in Fig.~\ref{fig3}(d). This fermion parity change can be summarized as $[s_{1}(g_{0})+s_{1}(g_{1})]n_{0}n'_{0}$. Using $\mathrm{d} s_{1} =0$, the above equation can be simplified as $s_{1}(g^{-1}_{0}g_{1})n_{0}n'_{0}:=m_{1}(01)$, which is \eqref{eq4.21.1}.

Therefore, the fermion parity of the Majorana chain decoration should be compensated by the fermion parity change of the complex fermion decoration, as the total fermion parity of the system must be conserved. This conservation immediately leads to the stacking rule \eqref{eq4.21} of complex fermion decoration. This is the first non-trivial stacking rule we get in this paper, since the $N_{1}$ decoration is not simply the addition of $n _{1} + n '_{1}$. Instead, it has an extra factor $m_{1}$ resulting from the fermion parity conservation.

\subsubsection{\label{sec4.2.3}Bosonic phase}

Finally, after projecting two layers into a single combined layer, we can define the total $F$ move of the combined system using the following commutation relation
\begin{equation}
\adjincludegraphics[valign=c,scale=0.8]{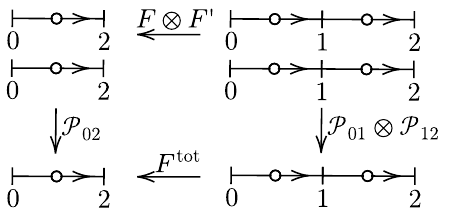},\label{eq4.26}
\end{equation}
where the $F:=F[ n_{0} ,n_{1}]$, $F':=F[ n'_{0} ,n'_{1}]$ and $F^{\mathrm{tot}} :=F[ N_{0} ,N_{1}]$. Since $F$ move depends on the $n_{0} ,n_{1}$ data, we use the shorthand notation $F[ n_{0} ,n_{1}]$ to indicate which layer is being referred to. Furthermore, note that since $\mathcal{P}_{01}$ and $\mathcal{P}_{12}$ act on different spaces, they naturally commute with each other. So the tensor product above is the same as act them one after another $\mathcal{P}_{01} \otimes \mathcal{P}_{12} =\mathcal{P}_{01}\mathcal{P}_{12} =\mathcal{P}_{12}\mathcal{P}_{01}$ (the same argument also works for $F\otimes F'=FF'=F'F$). In other words, the commutation relation \eqref{eq4.26} is equivalent to  
\begin{equation}
F\mathcal{^{\mathrm{tot}} P}_{01}\mathcal{P}_{12} =F^{\mathrm{tot}}(\mathcal{P}_{01} \otimes \mathcal{P}_{12}) =\mathcal{P}_{02}( F\otimes F') =\mathcal{P}_{02} FF',\label{eq4.27}
\end{equation}
or explicitly expand as: 
\begin{widetext}
\begin{equation}
F^{\mathrm{tot}}\left( e,g_{0}^{-1} g_{1} ,g_{0}^{-1} g_{2}\right) \cdotp \mathcal{P}_{01}^{\mathrm{std}} \cdotp \mathcal{P}_{12} =\mathcal{P}_{02}^{\mathrm{std}} \cdotp F\left(e,g_{0}^{-1} g_{1} ,g_{0}^{-1} g_{2}\right) \cdotp F'\left(e,g_{0}^{-1} g_{1} ,g_{0}^{-1} g_{2}\right) .\label{eq4.28}
\end{equation}
\end{widetext}
Note that only the operator $\mathcal{P}_{12}$ is non-standard: we already put $e$ at the site $0$, so we can only put $g_{0}^{-1} g_{1}$  instead of $e$ at the site $1$, making $\mathcal{P}_{12}$ non-standard. Using the definition \eqref{eq4.5} of the $F$ move and \eqref{eq4.24} of the $\mathcal{P}$ move, along with the anti-commutation relation of complex fermions and Majorana fermions and the symmetry transformation rules, we obtain the $(-1)^{\epsilon _{2}}$ part of the stacking rule 
\begin{equation}
\begin{array}{ c c l }
\epsilon _{2}( 012) & = & m_{1}( 01) m_{1}( 12) +N_{1}( 12) m_{1}( 01)\\
 &  & +\mathrm{d} n'_{1}( 012) n_{1}( 02) +n'_{1}( 12) n_{1}( 01)
\end{array} ,\label{eq4.29.1}
\end{equation}
which is exactly \eqref{eq4.22.1}. The total stacking rule for the bosonic layer contains two parts 
\begin{equation}
\mathcal{V}_{2}(012) =\nu _{2}(012) \nu '_{2}(012)( -1)^{\epsilon _{2}(012)} e^{2\pi i\theta _{2}(012)} ,\label{eq4.29}
\end{equation}
with the shorthand notation $\nu _{2}( 012) :=\nu _{2}(\overline{0} 1,\overline{0} 2) =\nu _{2}( e,\overline{0} 1,\overline{0} 2)$. In addition to the first part $\epsilon _{2}$, here's the second part $\theta _{2}$ called pure Majorana phase. This phase is calculated by the formula 
\begin{equation}
\begin{array}{ l }
e^{2\pi i\theta _{2}[ n_{0} ,n'_{0}]} =\bra{\psi } X_{5} X_{4} X_{3} X_{2} X_{1} X_{0}\ket{\psi }\\
=\bra{\psi }\overline{P}_{012}^{\mathrm{tot}}[ N_{0}]\textcolor[rgb]{0.74,0.06,0.88}{\gamma }\textcolor[rgb]{0.74,0.06,0.88}{_{0B}^{e,\mathrm{d} N_{1}( 012)}} P_{02}^{\mathrm{std}}\textcolor[rgb]{0.82,0.01,0.11}{\gamma '}\textcolor[rgb]{0.82,0.01,0.11}{_{0B}^{e,m_{1}( 02)}}\\
\times P_{012}[ n_{0}]\textcolor[rgb]{0.74,0.06,0.88}{\gamma }\textcolor[rgb]{0.74,0.06,0.88}{_{0B}^{e,\mathrm{d} n_{1}( 012)}} P'_{012}[ n'_{0}]\textcolor[rgb]{0.82,0.01,0.11}{\gamma '}\textcolor[rgb]{0.82,0.01,0.11}{_{0B}^{e,\mathrm{d} n_{1}( 012)}}\\
\times \overline{P}_{01}^{\mathrm{std}}\textcolor[rgb]{0.82,0.01,0.11}{\gamma '}\textcolor[rgb]{0.82,0.01,0.11}{_{0B}^{e,m_{1}( 01)}}\overline{P}_{01}^{\mathrm{std,b}} \overline{P}_{12}\left(( -1)^{s_{1}( 01)}\textcolor[rgb]{0.29,0.56,0.89}{\gamma '}\textcolor[rgb]{0.29,0.56,0.89}{_{1B}^{\overline{0} 1}}\right)^{m_{1}( 12)}\overline{P}_{12}^{\mathrm{b}}\ket{\psi }
\end{array}. \label{eq4.29.2}
\end{equation}
Here we used the abbreviated notation 
\begin{equation}
\begin{cases}
X_{0}  :=  \overline{P}_{12}\left(( -1)^{s_{1}( 01)}\textcolor[rgb]{0.29,0.56,0.89}{\gamma '}\textcolor[rgb]{0.29,0.56,0.89}{_{1B}^{\overline{0} 1}}\right)^{m_{1}( 12)}\overline{P}_{12}^{\mathrm{b}}\\
X_{1}  := \overline{P}_{01}^{\mathrm{std}}\textcolor[rgb]{0.82,0.01,0.11}{\gamma '}\textcolor[rgb]{0.82,0.01,0.11}{_{0B}^{e,m_{1}( 01)}}\overline{P}_{01}^{\mathrm{std,b}}\\
X_{2}  :=  P'_{012}[ n'_{0}]\textcolor[rgb]{0.82,0.01,0.11}{\gamma '}\textcolor[rgb]{0.82,0.01,0.11}{_{0B}^{e,\mathrm{d} n_{1}( 012)}}\\
X_{3}  :=  P_{012}[ n_{0}]\textcolor[rgb]{0.74,0.06,0.88}{\gamma }\textcolor[rgb]{0.74,0.06,0.88}{_{0B}^{e,\mathrm{d} n_{1}( 012)}}\\
X_{4}  := P_{02}^{\mathrm{std}}\textcolor[rgb]{0.82,0.01,0.11}{\gamma '}\textcolor[rgb]{0.82,0.01,0.11}{_{0B}^{e,m_{1}( 02)}}\\
X_{5}  :=  \overline{P}_{012}^{\mathrm{tot}}[ N_{0}]\textcolor[rgb]{0.74,0.06,0.88}{\gamma }\textcolor[rgb]{0.74,0.06,0.88}{_{0B}^{e,\mathrm{d} N_{1}( 012)}}
\end{cases}. \label{eq4.29.3}
\end{equation}
Here the $X_{0}$, $X_{1}$ and $X_{5}$ operators are the 'inverse' processes of $X_{12}=P_{12}\left(( -1)^{s_{1}( 01)} \gamma ^{\prime \overline{0} 1}_{1B}\right)^{m_{1}( 12)}$, $X_{01}^{\mathrm{std}}=P_{01}^{\mathrm{std}}\left( \gamma ^{\prime e}_{0B}\right)^{m_{1}( 01)}$ and $X_{012}^{\mathrm{tot}}=P_{012}^{\mathrm{tot}}\left( \gamma _{0B}^{e}\right)^{\mathrm{d} n_{1}( 012)}$, respectively. However, they are not the real inverse operators of each other, i.e. $X_{0}X_{12}\neq \mathbb{I}$,  $X_{1}X_{01}^{\mathrm{std}}\neq \mathbb{I}$ and $X_{5}X_{012}^{\mathrm{tot}}\neq \mathbb{I}$. For $X_{0}$ and $X_{1}$, the $\overline{P}^{\mathrm{b}}$ terms project the state to the second figures in Fig.~\ref{fig3}(c) and (d), which contain the 'bridge' Majorana pairs crossing two layers. The $\overline{P}$ terms project the state to the first figures (the starting states) in Fig.~\ref{fig3}(c) and (d), which are the final state of these 'inverse' processes. The dangling Majorana mode is inserted in between the $\overline{P}$ and $\overline{P}^{\mathrm{b}}$ operators, which is because the dangling Majorana mode should be inserted when fermion parity changes. This 'inverse' process is explained in detail around \eqref{eq5.23} for the 2+1D case. The current 1+1D case is similar.

In \eqref{eq4.29.2}, the state $\ket{\psi }$ represents the right bottom state in \eqref{eq4.26}. The \eqref{eq4.29.2} means that we start from the right bottom state $\ket{\psi }$ in \eqref{eq4.26}, project through this commutative diagram counterclockwise, and finally return to the same state with some extra Berry phase $e^{2\pi i\theta _{2}}\ket{\psi }$. This Berry phase is extracted by \eqref{eq4.29.2}.

The result of the pure Majorana phase $\theta _{2}$ is 
\begin{equation}
\theta _{2}(012) = \left[\frac{3}{4} \omega _{2}( 012) +\frac{1}{2} \omega _{2}( 012) s_{1}( 02)\right] \cdot n_{0} \cdot n'_{0} ,\label{eq4.29.4}
\end{equation}
which is exactly \eqref{eq4.22.2}. We will explain how to calculate such an expectation value like \eqref{eq4.29.2} in detail in Sec. \ref{sec5.3} for the 2+1D case. For the current 1+1D case, since this phase factor is not important \footnote{As we explained before, in \eqref{eq4.2}, non-trivial $\omega _{2}$ is always obstructed when $n_{0}=1$, so we can always set $\omega _{2}=0$. Then the phase in \eqref{eq4.29.4} always vanishes.} and the calculation is similar to the 2+1D case, we will skip it here.

\subsubsection{\label{sec4.2.4}Self-consistency condition}

Here is an important consistency condition for the stacking rules to satisfy. Given two layers of FSPT phase $( n_{0} ,n_{1} ,\nu _{2})$ and $( n'_{0} ,n'_{1} ,\nu '_{2})$ with the same symmetry $( G_{b} ,\omega _{2} ,s_{1})$, they are both solutions of the classification equations (\ref{eq4.1})-(\ref{eq4.3}). Now, if we stack them to get a new FSPT phase $( N_{0} ,N_{1} ,\mathcal{V}_{2})$ with the same symmetry according to stacking rules Eqs.~(\ref{eq4.20})-(\ref{eq4.22.2}), this new phase should also be a solution of the classification equations (\ref{eq4.1})-(\ref{eq4.3})
\begin{align}
\mathrm{d} N_{0} &=0\ (\bmod\ 2) ,\label{eq4.29.5}\\
\mathrm{d} N_{1} &=\omega _{2} \cup N_{0} ,\label{eq4.29.6}\\
\mathrm{d}_{s_1}\mathcal{V}_{2} &=\mathcal{O}_{3} [N_{1}] .\label{eq4.29.7}
\end{align}
The first two equations (\ref{eq4.29.5}) and (\ref{eq4.29.6}) are easy to verify, while the last one \eqref{eq4.29.7} is harder. Specifically, when combining the obstruction function \eqref{eq4.29.7} with the stacking rule \eqref{eq4.22}, the following equation should automatically hold 
\begin{equation}
\mathrm{d}_{s_1}\mathcal{E}_{2} = \frac{\mathrm{d}_{s_1}\mathcal{V} _{2}}{\mathrm{d}_{s_1}\mathcal{\nu} _{2}\cdot\mathrm{d}_{s_1}\mathcal{\nu} '_{2}} =\frac{\mathcal{O}_{3}[ N_{1}]}{\mathcal{O}_{3}[ n_{1}]\mathcal{O}_{3}[ n'_{1}]} :=\Delta \mathcal{O}_{3} .\label{eq4.29.8}
\end{equation}
Here $\mathcal{E}_{2} = ( -1)^{\epsilon _{2}} e^{2\pi i\theta _{2}}$ is the total phase factor of the stacking rule \eqref{eq4.22}.
Physically, this consistency condition signifies that after stacking two FSPT phases, the resulting phase should also be an FSPT phase.

We have checked that all our stacking rules are compatible with the classification equations, demonstrating the self-consistency of our construction. When checking this self-consistency condition, we found an useful equation 
\begin{equation}
\frac{1}{4}\mathrm{d}_{s_{1}} \omega _{2} =\frac{1}{2}[ s_{1} \cup \omega _{2} +\omega _{2} \cup _{1} \omega _{2}] \ (\bmod 1) .\label{eq4.29.9}
\end{equation}

\section{\label{sec5}Stacking group structure of 2+1D FSPT}

\subsection{\label{sec5.1}Classification of 2+1D FSPT}

After warming up in the lower dimensions, we are prepared to delve into the much more complex scenario in 2+1D spacetime, which constitutes the main findings of this paper. The degrees of freedom exhibit similarities to the 1+1D case with minor modifications:
\begin{enumerate}
    \item $|G_{b} |$ levels of bosonic (spin) states $\ket{g_{i}} \ ( g_{i} \in G_{b})$ on each vertex $i$ of the spatial lattice;
    \item $|G_{b} |$ species of complex fermions $c_{ijk}^{\sigma } \ ( \sigma \in G_{b})$ at the center of each triangle $\langle ijk\rangle $ of the spatial lattice;
    \item $|G_{b} |$ species of complex fermions (split to Majorana fermions) $a_{ij}^{\sigma } =\left( \gamma _{ij,A}^{\sigma } +i\gamma _{ij,B}^{\sigma }\right) /2\ ( \sigma \in G_{b})$ on each link $\langle ij\rangle $ of the spatial lattice.
\end{enumerate}

An FSPT phase is classified by a triplet $( n_{1} ,n_{2} ,\nu _{3})$, where $n_{1} \in H^{1}( G_{b} ,\mathbb{Z}_{2})$ represents the Majorana chain decoration, $n_{2} \in C^{2}( G_{b} ,\mathbb{Z}_{2})$ denotes the complex fermion decoration, and $\nu _{3} \in C^{3}_{s_1}( G_{b} ,U( 1))$ indicates the bosonic layer. If the $n_{1}( ij) =0$, the Majorana operator $\gamma _{ij,A}^{\sigma }$ and $\gamma _{ij,B}^{\sigma }$ are all in trivial pairing $-i\gamma _{ij,A}^{\sigma } \gamma _{ij,B}^{\sigma }=1$ for all $\sigma \in G_{b}$. On the other hand, if $n_{1}( ij) \neq 0$, only the Majorana pair $\gamma _{ij,A}^{g_{i}}$ and $\gamma _{ij,B}^{g_{i}}$ are in non-trivial paring, while the remaining Majorana pairs $\gamma _{ij,A}^{\sigma }$ and $\gamma _{ij,B}^{\sigma }$ for $\sigma \neq g_{i}$ are all in trivial paring. If $n_{2}( ijk) =0$, there is no complex fermion at the center of triangle $\langle ijk\rangle $. Conversely, if $n_{2}( ijk) =1$, a complex fermion $c_{ijk}^{g_{i}}$ is decorated at this center. The complex fermions $c_{ijk}^{g}$ with $g\neq g_i$ are all in vacuum state.

\begin{figure*}
    \centering
    \includegraphics[width=0.7\textwidth]{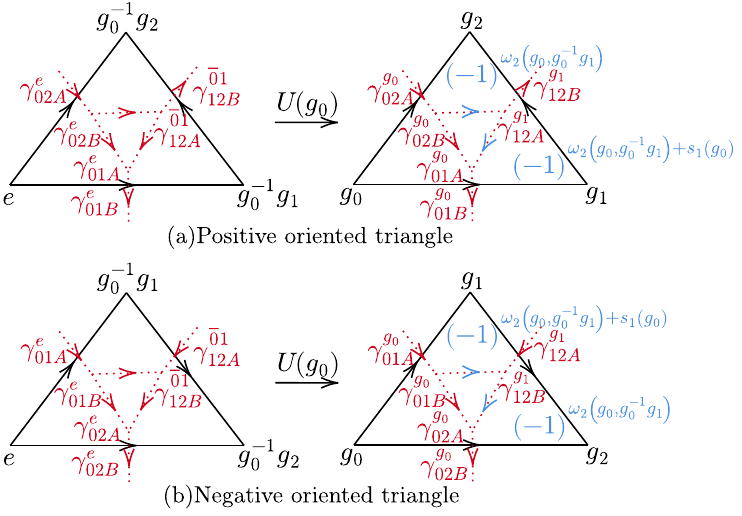}
    \caption{Kitaev chain decoration convention.}
    \label{fig4}
\end{figure*}

Given a spatial manifold, we begin by constructing a triangulation and assigning a branching structure. This can be achieved by labeling each vertex with a distinct natural number $i\in \mathbb{N}$. Subsequently, for each link $\langle ij\rangle $ ($i< j$) in the triangulation, we assign an arrow pointing from $i$ to $j$. The decoration convention for Kitaev chain is depicted in Fig.~\ref{fig4}. In the resolved dual lattice, the presence of red arrows signifies that if two Majorana fermions are paired at a resolved dual link, their pairing should align with the direction indicated by the red arrow. Initially, this convention is established within the standard triangle. Then, a symmetry transformation is applied to yield a generic (non-standard) triangle, wherein the blue arrow on the resolved dual lattice indicates that the direction of this Majorana pair may change [based on $\omega _{2}$ and $s_{1}$ according to \eqref{eq4.13}] subsequent to the symmetry transformation.

The standard $F$ move is defined as 
\begin{widetext}
\begin{gather}
\Psi \left( \adjincludegraphics[valign=c,scale=0.65]{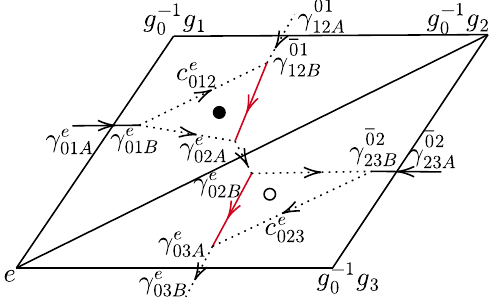} \right) = F\left( e,g_{0}^{-1} g_{1} ,g_{0}^{-1} g_{2} ,g_{0}^{-1} g_{3}\right) \Psi \left( 
\adjincludegraphics[valign=c,scale=0.65]{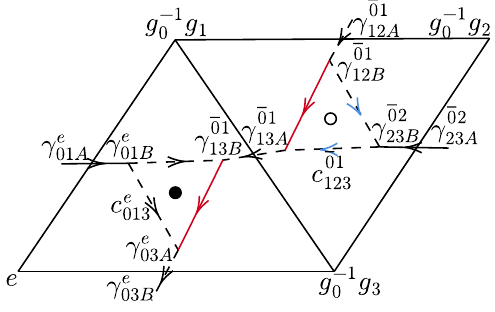} \right) ,\label{eq5.1}\\
F\left( e,g_{0}^{-1} g_{1} ,g_{0}^{-1} g_{2} ,g_{0}^{-1} g_{3}\right) =\nu _{3}(\overline{0} 1,\overline{1} 2,\overline{2} 3)\left( c_{012}^{e\dagger }\right)^{n_{2}( 012)}\left( c_{023}^{e\dagger }\right)^{n_{2}( 023)}\left( c_{013}^{e}\right)^{n_{2}( 013)}\left( c_{123}^{\overline{0} 1}\right)^{n_{2}( 123)} X_{0123}[ n_{1}] .\label{eq5.2}
\end{gather}
\end{widetext}
Here a red solid line with arrow indicates non-trivial Majorana pairing, while a dashed line in the resolved dual lattice represents no nontrivial pairing. A solid black dot indicates a decorated complex fermion, while a hollow black dot means no complex fermion. As noted before in \ref{sec3.1}, only the potentially non-trivial species of Majorana fermions and complex fermions are labeled in the picture; all other unlabeled species are considered trivial by default. The non-standard $F$ move can be obtained via symmetry transformation \eqref{eq4.18}.

The operator $X_{0123}[ n_{1}]$ projects the Kitaev chain decoration from the right-hand side of \eqref{eq5.2} to the left-hand side. In terms of Majorana fermions pairing projection operators \eqref{eq4.13}, it's general expression is 
\begin{gather}
X_{0123}[ n_{1}] =P_{0123}[ n_{1}] \cdotp \left( \gamma _{23B}^{\overline{0} 2}\right)^{\mathrm{d} n_{2}( 0123)} ,\label{eq5.3}\\
\begin{array}{ c c l }
P_{0123}[ n_{1}] & = & \left(\prod _{\mathrm{loop} \ i} 2^{( L_{i} -1) /2}\right)\\
 &  & \times \left(\prod _{\mathrm{Majorana\ pairs} \ \langle a,b\rangle \ \mathrm{in} \ \mathcal{T}} P_{a,b}^{g_{a} ,g_{b}}\right)\\
 &  & \times \left(\prod _{\mathrm{link} \ \langle ij\rangle \notin \mathcal{T}}\prod _{\sigma \in G_{b}} P_{ijA.ijB}^{\sigma ,\sigma }\right)
\end{array} .\label{eq5.4}
\end{gather}
The Majorana fermion $\gamma _{23B}^{\overline{0} 2}$ is introduced to ensure the matching of fermion parities. If a transition loop of the Kitaev chain is not Kasteleyn oriented, the fermion parity changes after the transition, which is not allowed physically. Therefore, in such cases, we first modify the fermion parity of the original Kitaev chain decoration by applying $\gamma _{23B}^{\overline{0} 2}$. Pictorially, this operation means flipping the direction of the Majorana pair containing this $\gamma _{23B}^{\overline{0} 2}$. Then the transition loop between the modified and final Kitaev chain decorations is Kasteleyn oriented, so that the fermion parities on both sides match. Now, we can safely use a sequence of Majorana fermions pairing projection operators \eqref{eq4.13} represented by $P_{0123}[ n_{1}]$ for the projection. Otherwise, $P_{0123}[n_{1}]$ would project to zero without the presence of the $\gamma_{23B}^{\overline{0}2}$ operator.

The term $\mathrm{d} n_{2}( 0123)$ in \eqref{eq5.3} counts the fermion parity change of the complex fermion, which must be the same as the fermion parity change of the Kitaev chain to ensure fermion parity conservation of the entire system. The later has been calculated in Ref.~\cite{WangGu2020} as \eqref{eq5.6}. The pairing projection operator $P_{0123}[ n_{1}]$ in \eqref{eq5.3} consists of three terms. The first term is a normalization factor, where $2L_{i}$ is the length of the $i$-th transition loop in the transition graph of Majorana pairing dimer configurations on the right triangulation lattice $\mathcal{T}$ and left lattice $\mathcal{T} '$. The second term projects the right state to the Majorana pairing configuration state in the left figure. The third term is the product of the vacuum projection operators for those Majorana fermions that do not appear explicitly in the left figure.

The $F$ moves should satisfy fermionic pentagon equation which leads the consistency condition \eqref{eq5.7}. Therefore, the classification of 2+1D FSPT phase corresponds to the solutions of the following set of equations
\begin{align}
\mathrm{d} n_{1} &=0\ (\bmod\ 2) ,\label{eq5.5}\\
\mathrm{d} n_{2} &=\omega _{2} \cup n_{1} +s_{1} \cup n_{1} \cup n_{1} \ (\bmod\ 2) ,\label{eq5.6}\\
\mathrm{d}_{s_1} \nu _{3} &=\mathcal{O}_{4}[ n_{2}]( 01234)\nonumber\\\nonumber
&=( -1)^{[ \omega _{2} \cup n_{2} +n_{2} \cup n_{2} +n_{2} \cup _{1}\mathrm{d} n_{2} +\mathrm{d}( s_{1} \cup n_{2} +n_{2} \cup _{2}\mathrm{d} n_{2})]( 01234)}\\\nonumber
&\quad\times ( -1)^{\omega _{2}( 013)\mathrm{d} n_{2}( 1234) +\mathrm{d} n_{2}( 0124)\mathrm{d} n_{2}( 0234)}\\
&\quad\times ( -i)^{\mathrm{d} n_{2}( 0123)[ 1-\mathrm{d} n_{2}( 0124)] (\mathrm{mod}\ 2)}
.\label{eq5.7}
\end{align}
It is important to note that there is a coboundary term $( -1)^{\mathrm{d}( s_{1} \cup n_{2} +n_{2} \cup _{2}\mathrm{d} n_{2})}$ in \eqref{eq5.7}, which was stated to be omitted in Ref.~\cite{WangGu2020} for the sake of simplifying the obstruction formula. However, this coboundary term is essential for the consistency condition \eqref{eq5.41}, so we will keep it throughout this paper.

\subsection{\label{sec5.2}Stacking of 2+1D FSPT}

\subsubsection{\label{sec5.2.1}Summary of results}

Given two 2+1D FSPT states $( n_{1} ,n_{2} ,\nu _{3})$ and $( n'_{1} ,n'_{2} ,\nu '_{3})$ with the same symmetry group $( G_{b} ,\omega _{2} ,s_{1})$, we stack these two states to form a new 2+1D FSPT phase denoted by $( N_{1} ,N_{2} ,\mathcal{V}_{3})$. The stacking rules are summarized as: 
\begin{enumerate}
    \item Kitaev chain decoration
\begin{equation}
N_{1} =n_{1} +n'_{1} ;\label{eq5.8}
\end{equation}
    \item complex fermion decoration
\begin{align}
N_{2} &=n_{2} +n'_{2} +m_{2} ,\label{eq5.9}\\
m_{2} &=n_{1} \cup n'_{1} +s_{1} \cup ( n_{1} \cup _{1} n'_{1}) ;\label{eq5.10}
\end{align}
    \item bosonic phase
\begin{align}
\mathcal{V}_{3} &=\nu _{3} \nu '_{3}\mathcal{E}_{3} ,\label{eq5.11}\\
\mathcal{E}_{3} &=( -1)^{\epsilon_{3}} e^{2\pi i\theta _{3}[ n_{1} ,n'_{1}]} ,\label{eq5.12}
\end{align}
where $\epsilon _{3}$ comes from the reordering sign of the complex fermions and Majorana fermions, while $\theta _{3}$ is an additional Majorana phase.
\end{enumerate}

The calculation of the bosonic phase is the most complicated aspect. The part involving complex fermions ($n_2$ and $n_2'$) in \eqref{eq5.12} is 
\begin{gather}
\begin{array}{ c c l }
\epsilon_{3} & = & n_{2} \cup _{1} n'_{2} +\mathrm{d} n_{2} \cup _{2} n_{2} +n'_{2} \cup _{2}\mathrm{d} n_{2} +\mathrm{d} n'_{2} \cup _{2} n'_{2}\\
 &  &+(\mathrm{d} n_{2} +\mathrm{d} n'_{2}) \cup _{2} N_{2} +m_{2} \cup _{1} N_{2} +m_{2} \cup _{2}\mathrm{d} N_{2}
\end{array} .\label{eq5.13}
\end{gather}
On the other hand, the pure Majorana fermion phase $\theta _{3}$ can be divided into four components as \eqref{eq5.14}. Here, $\theta _{3}^{0}$ represents the scenario when both $\omega_{2}$ and $s_{1}$ are trivial (consistent with the spin bordism results in Ref.~\cite{Morgan2018}), $\theta _{3}^{\omega _{2}}$ represents the scenario when $s_{1}$ is trivial but $\omega_{2}$ is non-trivial (consistent with the conjecture in Ref.~\cite{Maissam2022}), $\theta _{3}^{s_{1}}$ represents the scenario when $\omega_{2}$ is trivial but $s_{1}$ is non-trivial, and $\theta _{3}^{\omega _{2} ,s_{1}}$ represents the scenario when both $\omega_{2}$ and $s_{1}$ are non-trivial. The explicit expressions of them are
\begin{widetext}
\begin{align}
\theta _{3}[ n_{1} ,n'_{1}] &=\theta _{3}^{0} +\theta _{3}^{\omega _{2}} +\theta _{3}^{s_{1}} +\theta _{3}^{\omega _{2} ,s_{1}},\label{eq5.14}\\
\theta _{3}^{0} &=\frac{1}{2} n_{1} \cup ( n_{1} \cup _{1} n'_{1}) \cup n'_{1} +\frac{1}{4} n_{1} \cup n'_{1} \cup n'_{1} ,\label{eq5.15.1}\\
\theta _{3}^{\omega _{2}} &=\frac{1}{2}( \omega _{2} \cup N_{1}) \cup _{2}( n_{1} \cup n'_{1}) +\frac{3}{4} \omega _{2} \cup ( n_{1} \cup _{1} n'_{1}) ,\label{eq5.15.2}\\\nonumber
\theta _{3}^{s_{1}} & =  \frac{1}{2}[s_{1} \cup s_{1} \cup ( n_{1} \cup _{1} n'_{1}) +s_{1} \cup ( s_{1} \cup _{1} n_{1} \cup _{1} n'_{1}) \cup ( n_{1} \cup _{1} n'_{1}) +n_{1} \cup ( s_{1} \cup _{1} n'_{1}) \cup ( n_{1} \cup _{1} n'_{1})\\\nonumber
&\quad   +( s_{1} \cup _{1} n_{1}) \cup N_{1} \cup ( n_{1} \cup _{1} n'_{1}) +( s_{1} \cup _{1} n_{1}) \cup ( n_{1} \cup _{1} n'_{1}) \cup n_{1} +s_{1} \cup ( n_{1} \cup _{1} n'_{1}) \cup n'_{1}]\\
&\quad   +\frac{1}{4}[s_{1} \cup n'_{1} \cup n_{1} +s_{1} \cup ( n_{1} \cup _{1} n'_{1}) \cup N_{1}] +\frac{5}{8} s_{1} \cup n_{1} \cup n'_{1},\label{eq5.15.3}\\\nonumber
\theta _{3}^{\omega _{2} ,s_{1}} & =  \frac{1}{2}\{( \omega _{2} \cup _{1} s_{1}) \cup ( n_{1} \cup _{1} n'_{1}) +[ \omega _{2} \cup _{2}( s_{1} \cup ( n_{1} \cup _{1} n'_{1}))] \cup N_{1}\\
&\quad   +[ \omega _{2} \cup _{2}( s_{1} \cup n'_{1})] \cup n_{1} +[ \omega _{2} \cup _{2}( s_{1} \cup n_{1})] \cup ( n'_{1} +n_{1} \cup _{1} n'_{1})\}. \label{eq5.15.4}
\end{align}
\end{widetext}
To the best of our knowledge, the expressions for $\theta _{3}^{s_{1}}$ and $\theta _{3}^{\omega _{2} ,s_{1}}$ are completely new in the literature. The dependence on $s_{1}$ in these expressions is exceedingly complex and virtually impossible to predict intuitively~\footnote{Observe that the consistency condition \eqref{eq5.41} is linear for $\omega _{2}$ but non-linear for $s_{1}$. Consequently, it is considerably simpler to predict or calculate formulas for $\omega _{2}$, whereas it is significantly more challenging to obtain formulas for $s_{1}$}. It is also remarkable to discover that $\theta _{3}^{s_{1}}$ in \eqref{eq5.15.3} actually includes a $\mathbb{Z}_{8}$ phase instead of solely $\mathbb{Z}_{4}$ phase in the obstruction functions Eqs.~(\ref{eq5.5})-(\ref{eq5.7}) (see the discussions in the paragraph following \eqref{eq5.41} to understand the necessity of introducing a $\mathbb Z_8$ phase). We will derive these results step by step in the subsequent sections.

Note that the equations \eqref{eq5.15.1}-\eqref{eq5.15.4} we listed here is not exactly the same with the results of the formula \eqref{eq5.25}. Instead, we added a coboundary to simplify the results. Please see Appendix \ref{ap4} for more details.

\subsubsection{\label{sec5.2.2}$\mathcal{P}$ move}

Similar to the 1+1D case, we first define the standard $\mathcal{P}^{\mathrm{std}}$ move which projects two layers to one layer 
\begin{widetext}
\begin{gather}
\Psi \left( \adjincludegraphics[valign=c,scale=0.7]{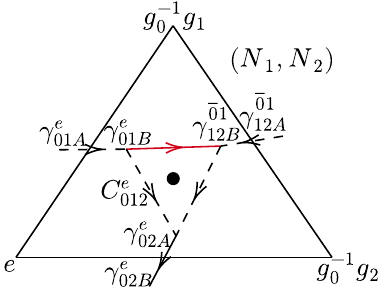} \right) =\mathcal{P}_{012}^{\mathrm{std}}[ n_{1} ,n'_{1} ,n_{2} ,n'_{2}] \Psi \left( \adjincludegraphics[valign=c,scale=0.7]{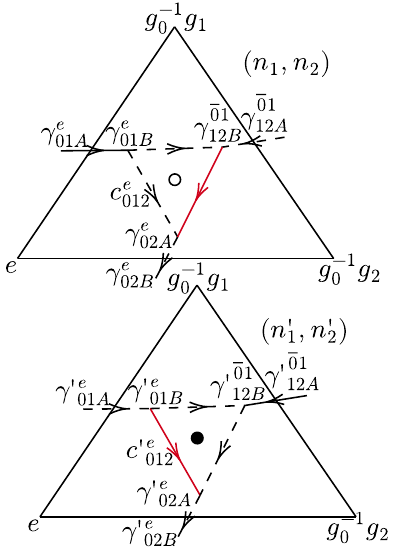} \right) ,\label{eq5.16}\\
\mathcal{P}_{012}^{\mathrm{std}}[ n_{1} ,n'_{1} ,n_{2} ,n'_{2}] =P_{012}^{\mathrm{std}}[ n_{1} ,n'_{1}]\left( \gamma ^{\prime \overline{0} 1}_{12B}\right)^{m_{2}( 012)}\left( C_{012}^{e\dagger }\right)^{N_{2}( 012)}\left( c^{\prime e}_{012}\right)^{n'_{2}( 012)}\left( c_{012}^{e}\right)^{n_{2}( 012)} .\label{eq5.17}
\end{gather}
\end{widetext}
The $P_{012}^{\mathrm{std}}$ has exactly the same meaning as \eqref{eq5.4}, representing products of a series Majorana pairing projection operators \eqref{eq4.13} (up to an unimportant normalization factor) that project to the left hand side Kitaev chain configuration. We add an operator $\gamma ^{\prime \overline{0} 1}_{12B}$ for the same reason as \eqref{eq5.3}, which is to match the fermion parity of both sides. The $m_{2}( 012)$ term counts the fermion parity difference during the projection of Kitaev chains. The Majorana operator $\gamma ^{\prime \overline{0} 1}_{12B}$ is chosen because all terms in $m_{2}( 012)$ contains $n'_{1}( 12)$. Since the total fermion parity of the system should conserve, the change of fermion parity in the Kitaev chains must be compensated by the fermion parity change of the complex fermions, leading to the stacking rule \eqref{eq5.9}. The derivation of the expression \eqref{eq5.10} of $m_{2}$ will be postponed until the next subsection.

The non-standard $\mathcal{P}$ move is obtained through a $U( g_{0})$ symmetry action 
\begin{widetext}
\begin{equation}
\mathcal{P}_{012} =U(g_0) \mathcal{P}_{012}^\mathrm{std} U(g_0)^\dagger
=P_{012}\left[( -1)^{\omega _{2}\left( g_{0} ,g_{0}^{-1} g_{1}\right) +s_{1}( g_{0})} \gamma ^{\prime g_{1}}_{12B}\right]^{m_{2}( 012)}\left( C_{012}^{g_{0} \dagger }\right)^{N_{2}( 012)}\left( c^{\prime g_{0}}_{012}\right)^{n'_{2}( 012)}\left( c_{012}^{g_{0}}\right)^{n_{2}( 012)} ,\label{eq5.18}
\end{equation}
\end{widetext}
where the phase $( -1)^{\omega _{2}\left( g_{0} ,g_{0}^{-1} g_{1}\right) +s_{1}( g_{0})}$ comes from \eqref{eq4.10}.

\subsubsection{\label{sec5.2.3}Kitaev chain and complex fermion decorations}

The primary objective of our construction is to formulate a procedure for projecting one Kitaev chain configuration onto another. Given that the Kitaev chain, as an invertible topological order, is classified by $\mathbb Z_2$, the stacking of two Kitaev chains results in a trivial configuration. Hence, the stacking rule for Kitaev chain decoration data can be expressed as:
\begin{equation}
N_{1} =n_{1} +n'_{1}. 
\end{equation}
The challenge then lies in designing an FSLU transformation that effectively transforms one layer into a trivial state while configuring the other to a nontrivial state with $N_1$ decoration. The construction is initially designed for individual triangles, and subsequently, the total transformation is assembled by combining these individual transformations.

\begin{figure}
    \centering
    \includegraphics[width=0.47\textwidth]{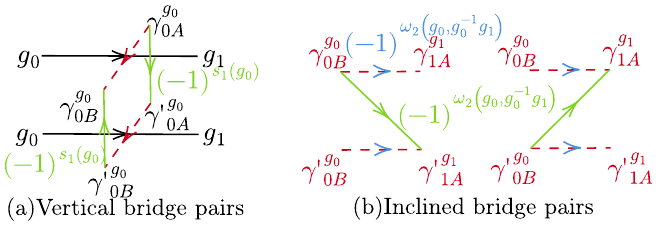}
    \caption{Convention of the bridge Majorana pairs.}
    \label{fig5}
\end{figure}

In order to establish a connection between the Kitaev chain decoration in the two layers, it is necessary to introduce “bridge" Majorana pairs that span across the layers. There are two types of bridge pairs (depicted by the green arrows), the first type is vertical shown in Fig.~\ref{fig5}(a), where the standard vertical pairing direction is always from $A$ to $A'$ and $B'$ to $B$, such as $\gamma _{0A}^{g_{0}} \gamma ^{\prime g_{0}}_{0A}$; the second type is inclined shown in Fig.~\ref{fig5}(b), where the inclined pairing direction is always the same as the corresponding horizontal one, such as $\gamma _{0B}^{g_{0}} \gamma ^{\prime g_{1}}_{1A}$ is decided by $\gamma _{0B}^{g_{0}} \gamma _{1A}^{g_{1}}$ (or equivalently $\gamma ^{\prime g_{0}}_{0B} \gamma ^{\prime g_{1}}_{1A}$). The phase factor after the symmetry transformation, as determined by the rule \eqref{eq4.13}, consistently results in a $( -1)^{s_{1}}$ phase for the vertical pairing, while the inclined pairing maintains the same phase factor as the corresponding horizontal pairing. We note the prototype of inclined bridge pair have already appeared 1+1D case, such as the one in Fig.\ref{fig3}.

After establishing the pairing conventions for the bridge pairs, we can define one more basic move of FSLU transformation on Kitaev chains except those in Fig.\ref{fig2}. The new one basic move is shown in Fig.~\ref{fig6} where the transition loop contains both a vertical bridge pair and an inclined bridge pair. Whether it is Kasteleyn oriented or not depends on the values of $\omega _{2}$ and $s_{1}$. In the example depicted in this figure, it is Kasteleyn oriented because the standard transition loop is Kasteleyn oriented, and the phases resulting from symmetry transformation nullify each other.

\begin{widetext}
\begin{figure*}
    \centering
    \includegraphics[width=0.8\textwidth]{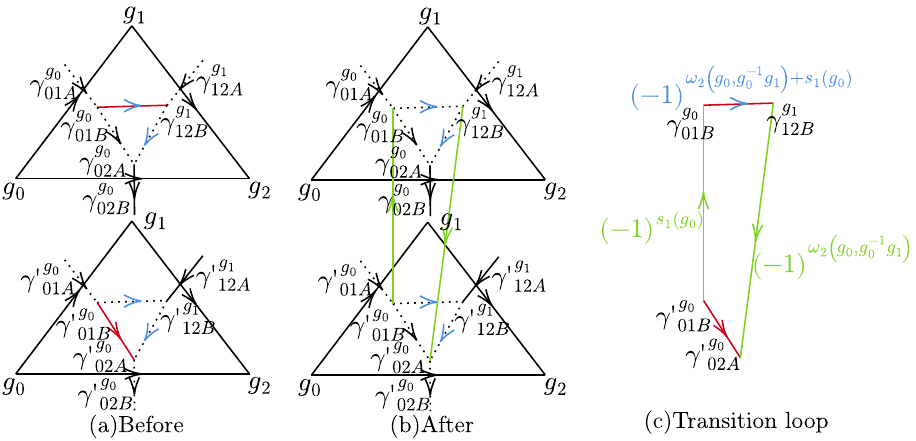}
    \caption{The 5th basic move of FSLU transformation on Kitaev chains, which may not be Kasteleyn oriented.}
    \label{fig6}
\end{figure*}
Now, with the assistance of the four basic moves in Fig.~\ref{fig2} ((a), (c) and (d) are Kasteleyn oriented, while (b) may be non-Kasteleyn oriented) and one possibly non-Kasteleyn oriented basic move in Fig.~\ref{fig6}, we can now devise a projection scheme to implement the operator $P_{012}$ and determine the expression for $m_{2}( 012)$ as outlined in \eqref{eq5.17} and \eqref{eq5.10}, which would help us to map all the nontrivial decoration into the first layer successfully. This scheme depends on two scenarios, one of which involves two nontrivial but non-parallel initial pairs in two layers and is new for 2+1D FSPT as compared to 1+1D (discussed below in details).  The other scenario which involves one or two but parallel initial pairs in two layers is similar to that in 1+1D FSPT.

The new situation of the projection scheme, where $n_{1}$ and $n'_{1}$ are both non-trivial and not parallel, involves three steps:
\begin{enumerate}
    \item Begin with two separate layers and use the basic move in Fig.~\ref{fig6} to connect the two layers. This generates one vertical bridge pair and one inclined bridge pair. This projection process is denoted as $P_{1}$. (We note that this is the only move that may have non-Kasteleyn transition loop and then change the fermion parity.)
    \item Convert the inclined bridge pair to the corresponding horizontal pair in the upper layer using the basic move Fig.~\ref{fig2}(c) or (d). This projection process is denoted as $P_{2}$.
    \item Transform all vertical bridge pairs into trivial horizontal pairs using Fig.~\ref{fig2}(a). This projection process is denoted as $P_{3}$.
\end{enumerate}
An example is drawn in Fig.~\ref{fig7}, where the solid lines represent current Majorana pairings (black for trivial paring, red for nontrivial one, green for bridge pairs), while dotted lines are unpairs drawn as references (the red dotted lines are meant to remind the reader the initial non-trivial horizontal pairs before the transition). The labels of vertices and Majorana pairs are omitted for clarity, but they should be easily identified by the reader. Although only one triangle is depicted, steps 2 and 3 can only be performed along neighboring triangles, so some parts of the transition loop are incomplete in this figure. For instance, the bridge pair $\gamma _{01A}^{g_{0}} \gamma ^{\prime g_{0}}_{01A}$ does not belong to this triangle $\langle 012\rangle $ but rather comes from the first projection step of the nearby triangle. It is also important to keep in mind that only the first step $P_{1}$ may change the fermion parity.

\begin{figure*}
    \centering
    \includegraphics[width=0.9\textwidth]{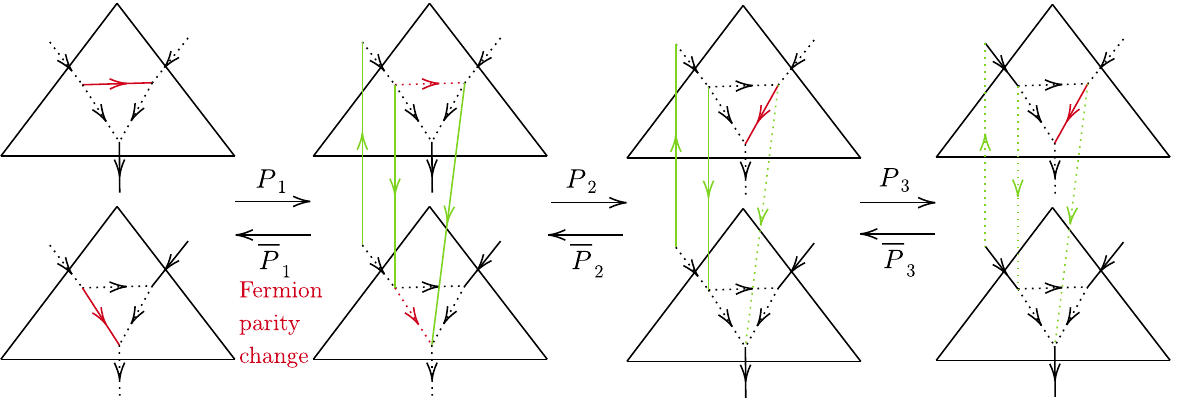}
    \caption{An example of the projection scheme for $n_{1}$ and $n'_{1}$ are nontrivial and parallel.}
    \label{fig7}
\end{figure*}

Above, we have described the most challenging scenario, in which $n_{1}$ and $n'_{1}$ are both non-trivial and not parallel. However, there are also simpler cases to consider that we have met similarly in 1+1D FSPT. When both $n_{1}$ and $n'_{1}$ are no-trivial but parallel with each other, we can proceed with just two steps:
\begin{enumerate}
    \item Since the two horizontal pairs are parallel, we can utilize the basic move shown in Fig.~\ref{fig2}(b). This will result in two vertical bridge pairs. This projection process is also denoted as $P_{1}$.
    \item We further use Fig.~\ref{fig2}(a) to convert all the vertical bridge pairs into trivial horizontal pairs. This projection process is also denoted as $P_{3}$.
\end{enumerate}
An example illustrating this case is presented in Fig.~\ref{fig8}.

\begin{figure*}
    \centering
    \includegraphics[width=0.75\textwidth]{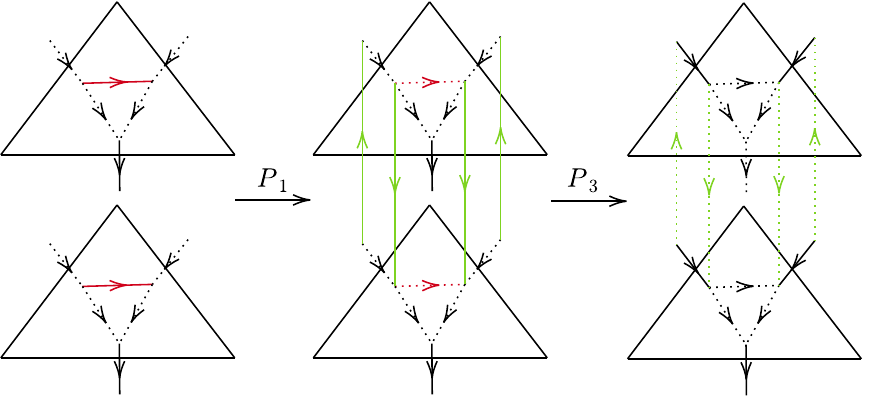}
    \caption{An example of the projection scheme for $n_{1}$ and $n'_{1}$ are nontrivial but parallel.}
    \label{fig8}
\end{figure*}

When $n_{1}$ is trivial but $n'_{1}$ is non-trivial,  a slightly complex situation arises. An example is shown in Fig.~\ref{fig9}. This situation resembles the second step $P_{2}$ in the first case, but it needs to be divided into two smaller steps:
\begin{enumerate}
    \item Utilize one of the basic moves illustrated in Fig.~\ref{fig2}(c) and (d), based on the nearby configurations of the Kitaev chain, in order to obtain a single inclined bridge pair. This projection process is referred to as $P'_{2}$.
    \item Apply the other basic move from Fig.~\ref{fig2}(c) and (d) to eliminate the inclined bridge pair and generate a horizontal non-trivial pair in the upper layer. This projection process is denoted as $P''_{2}$.
\end{enumerate}

\begin{figure*}
    \centering
    \includegraphics[width=0.75\textwidth]{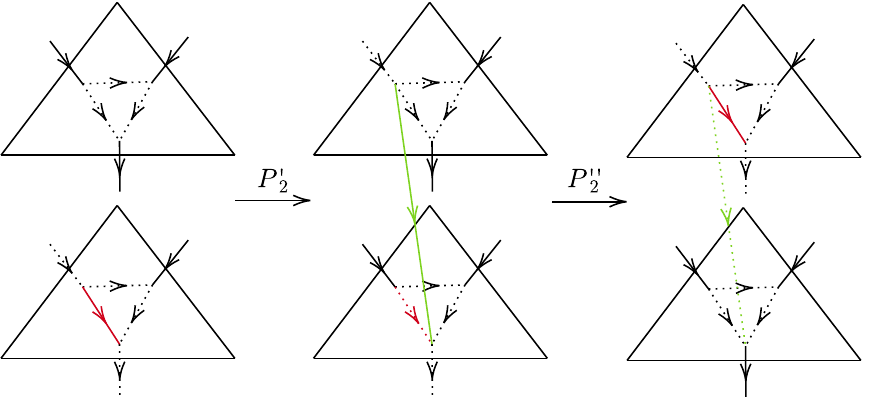}
    \caption{An example of the projection scheme for $n_{1}$ trivial but $n'_{1}$ nontrivial.}
    \label{fig9}
\end{figure*}

Finally, if the $n'_{1}$ is trivial in the lower layer, we do not need to take any action because the lower layer is already a trivial product state that we desire.

Now, we have developed a comprehensive projection scheme for all situations within a single triangle. This scheme enables us to accurately calculate the change in fermion parity at each step. Remember that the fermion parity only changes in the 1st step $P_{1}$, when $n_{1}$ and $n'_{1}$ are both non-trivial and not parallel in our construction. The detailed calculation for counting the fermion parity change $m_{2}$ is summarized in the Appendix \ref{ap2}. The expression for $m_{2}$ is given by \eqref{eq5.10}, which corresponds to the stacking rule of the complex fermion decoration. Through direct observation, we obtain the stacking rule \eqref{eq5.8} of the Kitaev chain decoration.

\subsubsection{\label{sec5.2.4}Bosonic phase}

Similar to the 1+1D case, we can define the total $F$ move of the combined system using the following commutation relation
\begin{equation}
\adjincludegraphics[valign=c,scale=0.6]{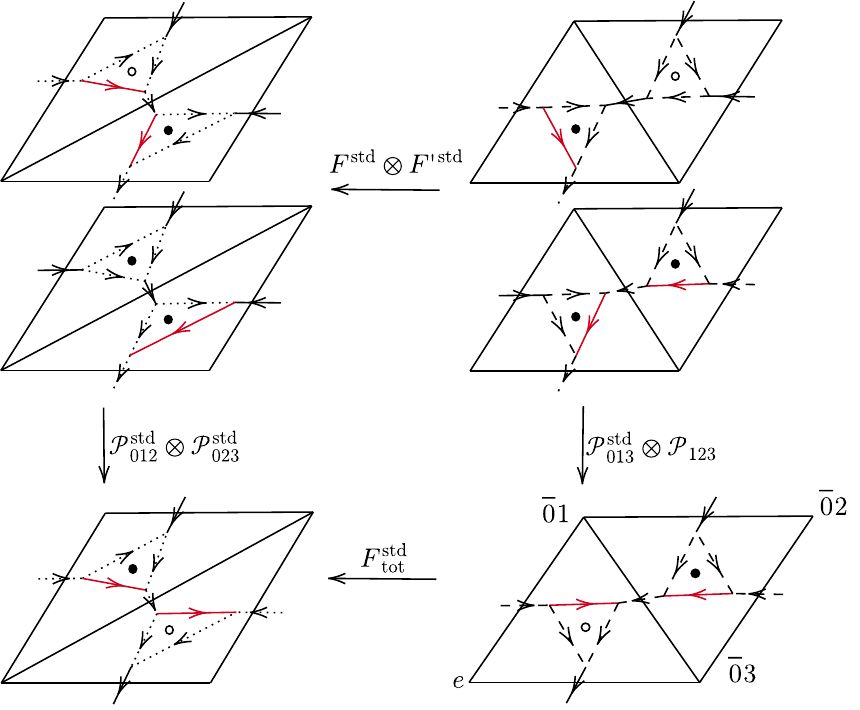}.\label{eq5.19}
\end{equation}
This means that 
\begin{equation}
\left(\mathcal{P}_{012}^{\mathrm{std}} \otimes \mathcal{P}_{023}^{\mathrm{std}}\right)\left( F^{\mathrm{std}} \otimes F^{\prime \mathrm{std}}\right) =F_{\mathrm{tot}}^{\mathrm{std}}\left(\mathcal{P}_{013}^{\mathrm{std}} \otimes \mathcal{P}_{123}\right) ,\label{eq5.20}
\end{equation}
where only the $\mathcal{P}_{123}$ operator is non-standard. As we mentioned in the 1+1D case, the two operators in the tensor product always commute. For example, we have $\mathcal{P}_{012}^{\mathrm{std}} \otimes \mathcal{P}_{023}^{\mathrm{std}} =\mathcal{P}_{012}^{\mathrm{std}}\mathcal{P}_{023}^{\mathrm{std}} =\mathcal{P}_{023}^{\mathrm{std}}\mathcal{P}_{012}^{\mathrm{std}}$. Therefore, \eqref{eq5.20} can be written as 
\begin{equation}
\mathcal{P}_{012}^{\mathrm{std}}\mathcal{P}_{023}^{\mathrm{std}} F^{\mathrm{std}} F^{\prime \mathrm{std}} =F\mathcal{_{\mathrm{tot}}^{\mathrm{std}} P}_{013}^{\mathrm{std}}\mathcal{P}_{123} .\label{eq5.21}
\end{equation}
This can further be rewritten as
\begin{equation}
F\mathcal{_{\mathrm{tot}}^{\mathrm{std}}} =\mathcal{P}_{012}^{\mathrm{std}}\mathcal{P}_{023}^{\mathrm{std}} F^{\mathrm{std}} F^{\prime \mathrm{std}}\overline{\mathcal{P}}_{013}^{\mathrm{std}}\overline{\mathcal{P}}_{123}\label{eq5.22}
\end{equation}
by introducing the `inverse' operator $\overline{\mathcal{P}}$ of the $\mathcal{P}$ move 
\begin{align}
\overline{\mathcal{P}}_{012}^{\mathrm{std}} &=\left( c_{012}^{e\dagger }\right)^{n_{2}( 012)}\left( c^{\prime e\dagger }_{012}\right)^{n'_{2}( 012)}\left( C_{012}^{e}\right)^{N_{2}( 012)}\overline{P}_{012}^{\mathrm{std}}\left( \gamma ^{\prime \overline{0} 1}_{12B}\right)^{m_{2}( 012)}\overline{P}_{012}^{\mathrm{std,b}} ,\label{eq5.23}\\
\overline{\mathcal{P}}_{012}
&=U(g_0) \overline{\mathcal{P}}_{012}^{\mathrm{std}} U(g_0)^\dagger
=\left( c_{012}^{g_{0} \dagger }\right)^{n_{2}( 012)}\left( c^{\prime g_{0} \dagger }_{012}\right)^{n'_{2}( 012)}\left( C_{012}^{g_{0}}\right)^{N_{2}( 012)}\overline{P}_{012}\left[( -1)^{\omega _{2}\left( g_{0} ,g_{0}^{-1} g_{1}\right) +s_{1}( g_{0})} \gamma ^{\prime g_{1}}_{12B}\right]^{m_{2}( 012)}\overline{P}_{012}^{\mathrm{b}} .\label{eq5.24}
\end{align}

The $\overline{P} \gamma \overline{P}^{\mathrm{b}}$ part in above definition is a bit tricky; it represents the `inverse' process of the $P\gamma $ part. The reason why $\gamma $  is sandwiched between the $\overline{P}$ and $\overline{P}^{\mathrm{b}}$ operators is as explained in the following. Since the dangling Majorana mode $\gamma $ is inserted to ensure fermion parity conservation before and after the FSLU transformation, it must be inserted at the point where fermion parity changes (always inserted before the projection $P_{i}$ as $P_{i} \gamma $ if this step changes the fermion parity). As previously mentioned in Fig.~\ref{fig7}, only the first step $P_{1}$ may change the fermion parity. Thus, the the projection of Kitaev chain in $\mathcal{P}_{012}$ is given by $P_{012} \gamma '_{12B} =P_{3} P_{2} P_{1} \gamma '_{12B}$. Accordingly, the inverse process in $\overline{\mathcal{P}}_{012}$ is $\overline{P}_{012} \gamma '_{12B}\overline{P}_{012}^{\mathrm{b}} =\overline{P}_{1} \gamma '_{12B}\overline{P}_{2}\overline{P}_{3}$, where $\overline{P}_{i}$ is defined as projecting the final state of $P_{i}$ to the initial state of $P_{i}$, i.e. reverse the projection process of $P_{i}$. We can observe that the $\overline{P}_{2}\overline{P}_{3}$ part is related to the bridge Majorana pairs, so we label this part as $\overline{P}^{\mathrm{b}}$, and we denote the other part $\overline{P}_{1}$ as $\overline{P}$.

In Fig.~\ref{fig8} and Fig.~\ref{fig9} , the fermion parity remains unchanged, so there is no inserted dangling Majorana mode $\gamma $. The $\overline{P}\overline{P}^{\mathrm{b}}$ part as a whole represents the `inverse' of $P$, where $\overline{P}^{\mathrm{b}}$ also represents the part related to bridge pairs. We provide detailed examples in Appendix \ref{ap3} to illustrate this process.

By utilizing the standard $F$ move \eqref{eq5.2} [alongside \eqref{eq5.3} and \eqref{eq5.4}], as well as the $\mathcal{P}$ move \eqref{eq5.17} and \eqref{eq5.18}, in conjunction with the `inverse' move $\overline{\mathcal{P}}$ defined above, we can rearrange all the complex fermion and Majorana fermion operators according to the anti-commutation relation. This reordering process yields the stacking rule \eqref{eq5.13}.

The extra phase $\mathcal{E}_{3}$ in \eqref{eq5.11} consists of two components. The first component $\epsilon _{3}$ arises from the anti-commutation relation of complex fermions themselves and complex fermions with Majorana fermions, as discussed above. The second component $\theta _{3}$ is a pure phase factor associated with the Majorana fermion, which poses the greatest challenge in our calculation. The detailed computation of this pure Majorana fermion phase will be presented in the subsequent section.

\subsection{\label{sec5.3}Calculation of the Majorana phase $\theta _{3}$}

\subsubsection{\label{sec5.3.1}Formula}

\begin{figure}
    \centering
    \includegraphics[width=0.45\textwidth]{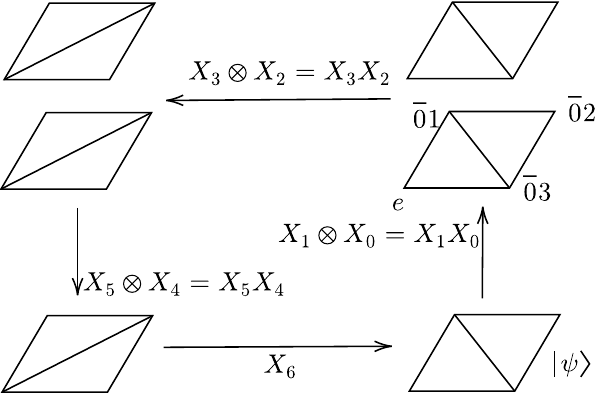}
    \caption{Extract the pure Majorana phase $\theta_{3}$.}
    \label{fig10}
\end{figure}

The calculation of the pure Majorana phase $\theta _{3}$ is illustrated in Fig.~\ref{fig10}. Starting from a wave function $\ket{\psi }$ representing the lower right figure in Fig.~\ref{fig10}, we follow the counterclockwise projection processes and eventually return to the lower right figure with a possible additional phase $\ket{\psi '} =e^{2\pi i\theta _{3}}\ket{\psi }$. This represents the desired pure Majorana phase, which can be extracted using the formula  
\begin{equation}
\begin{array}{ c c l }
e^{i\theta _{3}[ n_{1} ,n'_{1}]} & = & \bra{\psi } X_{6} X_{5} X_{4} X_{3} X_{2} X_{1} X_{0}\ket{\psi }\\
 & = & \bra{\psi }\overline{P}_{0123}^{\mathrm{tot}}[ N_{1}]\textcolor[rgb]{0.29,0.56,0.89}{\gamma }\textcolor[rgb]{0.29,0.56,0.89}{_{23B}^{\overline{0} 2,\mathrm{d} N_{2}( 0123)}} P_{012}^{\mathrm{std}}\textcolor[rgb]{0.49,0.83,0.13}{\gamma '}\textcolor[rgb]{0.49,0.83,0.13}{_{12B}^{\overline{0} 1,m_{2}( 012)}} P_{023}^{\mathrm{std}}\textcolor[rgb]{0.82,0.01,0.11}{\gamma '}\textcolor[rgb]{0.82,0.01,0.11}{_{23B}^{\overline{0} 2,m_{2}( 023)}} P_{0123}[ n_{1}]\textcolor[rgb]{0.29,0.56,0.89}{\gamma }\textcolor[rgb]{0.29,0.56,0.89}{_{23B}^{\overline{0} 2,\mathrm{d} n_{2}( 0123)}}\\
 &\qquad &\times   P'_{0123}[ n'_{1}]\textcolor[rgb]{0.82,0.01,0.11}{\gamma '}\textcolor[rgb]{0.82,0.01,0.11}{_{23B}^{\overline{0} 2,\mathrm{d} n'_{2}( 0123)}}\overline{P}_{013}^{\mathrm{std}}\textcolor[rgb]{0.74,0.06,0.88}{\gamma '}\textcolor[rgb]{0.74,0.06,0.88}{_{13B}^{\overline{0} 1,m_{2}( 013)}}\overline{P}_{013}^{\mathrm{std,b}}\overline{P}_{123}\left(( -1)^{\omega _{2}(\overline{0} 1,\overline{1} 2) +s_{1}(\overline{0} 1)}\textcolor[rgb]{0.82,0.01,0.11}{\gamma '}\textcolor[rgb]{0.82,0.01,0.11}{_{23B}^{\overline{0} 2}}\right)^{m_{2}( 123)}\overline{P}_{123}^{\mathrm{b}}\ket{\psi }
\end{array} .\label{eq5.25}
\end{equation}
Here we used the abbreviated notation 
\begin{equation}
\begin{cases}
X_{0} := \overline{P}_{123}\left(( -1)^{\omega _{2}(\overline{0} 1,\overline{1} 2) +s_{1}(\overline{0} 1)}\textcolor[rgb]{0.82,0.01,0.11}{\gamma '}\textcolor[rgb]{0.82,0.01,0.11}{_{23B}^{\overline{0} 2}}\right)^{m_{2}( 123)}\overline{P}_{123}^{\mathrm{b}}\\
X_{1} := \overline{P}_{013}^{\mathrm{std}}\textcolor[rgb]{0.74,0.06,0.88}{\gamma '}\textcolor[rgb]{0.74,0.06,0.88}{_{13B}^{\overline{0} 1,m_{2}( 013)}}\overline{P}_{013}^{\mathrm{std,b}}\\
X_{2} := P'_{0123}[ n'_{1}]\textcolor[rgb]{0.82,0.01,0.11}{\gamma '}\textcolor[rgb]{0.82,0.01,0.11}{_{23B}^{\overline{0} 2,\mathrm{d} n'_{2}( 0123)}}\\
X_{3} := P_{0123}[ n_{1}]\textcolor[rgb]{0.29,0.56,0.89}{\gamma }\textcolor[rgb]{0.29,0.56,0.89}{_{23B}^{\overline{0} 2,\mathrm{d} n_{2}( 0123)}}\\
X_{4} := P_{023}^{\mathrm{std}}\textcolor[rgb]{0.82,0.01,0.11}{\gamma '}\textcolor[rgb]{0.82,0.01,0.11}{_{23B}^{\overline{0} 2,m_{2}( 023)}}\\
X_{5} := P_{012}^{\mathrm{std}}\textcolor[rgb]{0.49,0.83,0.13}{\gamma '}\textcolor[rgb]{0.49,0.83,0.13}{_{12B}^{\overline{0} 1,m_{2}( 012)}}\\
X_{6} := \overline{P}_{0123}^{\mathrm{tot}}[ N_{1}]\textcolor[rgb]{0.29,0.56,0.89}{\gamma }\textcolor[rgb]{0.29,0.56,0.89}{_{23B}^{\overline{0} 2,\mathrm{d} N_{2}( 0123)}}
\end{cases} .\label{eq5.26}
\end{equation}
Different colors are used above to denote different inserted dangling Majorana modes, and it is observed that there are only four distinct dangling Majorana modes 
\begin{equation}
\left(\textcolor[rgb]{0.29,0.56,0.89}{\gamma }\textcolor[rgb]{0.29,0.56,0.89}{_{23B}^{\overline{0} 2,(\mathrm{d} N_{2} +\mathrm{d} n_{2})( 0123)}} ,\textcolor[rgb]{0.49,0.83,0.13}{\gamma '}\textcolor[rgb]{0.49,0.83,0.13}{_{12B}^{\overline{0} 1,m_{2}( 012)}} ,\textcolor[rgb]{0.74,0.06,0.88}{\gamma '}\textcolor[rgb]{0.74,0.06,0.88}{_{13B}^{\overline{0} 1,m_{2}( 013)}} ,\textcolor[rgb]{0.82,0.01,0.11}{\gamma '}\textcolor[rgb]{0.82,0.01,0.11}{_{23B}^{\overline{0} 2,\mathrm{d} n'_{2}( 0123) +m_{2}( 023) +m_{2}( 123)}}\right) .\label{eq5.27}
\end{equation}
We use the quadruplet of their number 
\begin{equation}
[\textcolor[rgb]{0.29,0.56,0.89}{(}\mathrm{\textcolor[rgb]{0.29,0.56,0.89}{d}}\textcolor[rgb]{0.29,0.56,0.89}{N}\textcolor[rgb]{0.29,0.56,0.89}{_{2}}\textcolor[rgb]{0.29,0.56,0.89}{+}\mathrm{\textcolor[rgb]{0.29,0.56,0.89}{d}}\textcolor[rgb]{0.29,0.56,0.89}{n}\textcolor[rgb]{0.29,0.56,0.89}{_{2}}\textcolor[rgb]{0.29,0.56,0.89}{)}\textcolor[rgb]{0.29,0.56,0.89}{(}\textcolor[rgb]{0.29,0.56,0.89}{0123}\textcolor[rgb]{0.29,0.56,0.89}{)} ,\textcolor[rgb]{0.49,0.83,0.13}{m}\textcolor[rgb]{0.49,0.83,0.13}{_{2}}\textcolor[rgb]{0.49,0.83,0.13}{(}\textcolor[rgb]{0.49,0.83,0.13}{012}\textcolor[rgb]{0.49,0.83,0.13}{)} ,\textcolor[rgb]{0.74,0.06,0.88}{m}\textcolor[rgb]{0.74,0.06,0.88}{_{2}}\textcolor[rgb]{0.74,0.06,0.88}{(}\textcolor[rgb]{0.74,0.06,0.88}{013}\textcolor[rgb]{0.74,0.06,0.88}{)} ,\mathrm{\textcolor[rgb]{0.82,0.01,0.11}{d}}\textcolor[rgb]{0.82,0.01,0.11}{n'}\textcolor[rgb]{0.82,0.01,0.11}{_{2}}\textcolor[rgb]{0.82,0.01,0.11}{(}\textcolor[rgb]{0.82,0.01,0.11}{0123}\textcolor[rgb]{0.82,0.01,0.11}{)}\textcolor[rgb]{0.82,0.01,0.11}{+m}\textcolor[rgb]{0.82,0.01,0.11}{_{2}}\textcolor[rgb]{0.82,0.01,0.11}{(}\textcolor[rgb]{0.82,0.01,0.11}{023}\textcolor[rgb]{0.82,0.01,0.11}{)}\textcolor[rgb]{0.82,0.01,0.11}{+m}\textcolor[rgb]{0.82,0.01,0.11}{_{2}}\textcolor[rgb]{0.82,0.01,0.11}{(}\textcolor[rgb]{0.82,0.01,0.11}{123}\textcolor[rgb]{0.82,0.01,0.11}{)}] \quad(\bmod\ 2) \label{eq5.28}
\end{equation}
\end{widetext}
to indicate the presence or absence (mod 2) of these four Majorana fermions in \eqref{eq5.25}.

\subsubsection{\label{sec5.3.2}General computation approach}

Let us focus on \eqref{eq5.25} and consider how to evaluate this phase. Firstly, we note that every $P$ operators in \eqref{eq5.25} is in the form of \eqref{eq5.4}, and we will not care about the normalization factor $2^{( L_{i} -1) /2}$ in \eqref{eq5.4} during our calculation \footnote{Since $\theta _{3}$ in \eqref{eq5.25} is simply an $U( 1)$ phase, the normalization factor will always ensure this factor to be $|e^{2\pi i \theta _{3}} | =1$. That's why this normalization factor is not important and is ignored throughout this paper}. Secondly, $\theta _{3}$ in \eqref{eq5.25} involves many $P$ terms as \eqref{eq5.4}. Additionally, \eqref{eq5.4} is also a product of the projection operators of the Majorana fermion pairing \eqref{eq4.13}. Therefore, we can imagine that \eqref{eq5.25} expands as 
\begin{widetext}
\begin{align}
e^{i\theta _{3}} & \sim  \bra{\psi }\left( 1-( -1)^{\alpha _{1}} i\gamma _{i_{1}} \gamma _{i_{2}}\right) \cdots \left( 1-( -1)^{\alpha _{2}} i\gamma _{i_{3}} \gamma _{i_{4}}\right) \gamma _{j_{1}}\left( 1-( -1)^{\alpha _{3}} i\gamma _{i_{4}} \gamma _{i_{5}}\right)\\ 
&\quad \times   \cdots \left( 1-( -1)^{\alpha _{3}} i\gamma _{i_{4}} \gamma _{i_{5}}\right) \gamma _{j_{2}}\left( 1-( -1)^{\alpha _{4}} i\gamma _{i_{6}} \gamma _{i_{7}}\right) \cdots \left( 1-( -1)^{\alpha _{5}} i\gamma _{i_{8}} \gamma _{i_{9}}\right)\ket{\psi },\label{eq5.29}
\end{align}
\end{widetext}
where $\left( 1-( -1)^{\alpha _{1}} i\gamma _{i_{a}} \gamma _{i_{b}}\right)$ represents a particular projection operator of the Majorana fermion pairing \eqref{eq4.13}, and $\gamma _{j_{a}}$ is an inserted dangling Majorana mode. Naively, we can further expand each $\left( 1-( -1)^{\alpha _{1}} i\gamma _{i_{a}} \gamma _{i_{b}}\right)$ term to get a polynomial-like summation, as shown in an simplest example 
\begin{align}\label{eq5.30}
&\quad\bra{\psi }( 1-i\gamma _{1} \gamma _{2}) \gamma _{5}( 1-i\gamma _{3} \gamma _{4}) \gamma _{6}\ket{\psi }\\\nonumber
&=\bra{\psi }( \gamma _{5} \gamma _{6} -i\gamma _{1} \gamma _{2} \gamma _{5} \gamma _{6} -i\gamma _{5} \gamma _{3} \gamma _{4} \gamma _{6} -\gamma _{1} \gamma _{2} \gamma _{5} \gamma _{3} \gamma _{4} \gamma _{6})\ket{\psi },
\end{align}
After expanding all $\left( 1-( -1)^{\alpha _{1}} i\gamma _{i_{a}} \gamma _{i_{b}}\right)$ terms, we aim to evaluate these polynomial-like terms. The Majorana fermions satisfy the anti-commutation relation 
\begin{gather}
\{\gamma _{i} ,\gamma _{j}\} \equiv \gamma _{i} \gamma _{j} +\gamma _{j} \gamma _{i} = 2\delta _{ij} ,\label{eq5.31}
\end{gather}
So for a product $\gamma _{1} \cdots \gamma _{n}$ of several Majorana operators, we can use \eqref{eq5.31} to reorder the Majorana operators and eliminate two neighboring identical ones. This will give us a reduced product $\overline{\gamma }_{1} \cdots \overline{\gamma }_{m}$, where $\overline{\gamma }_{i} \neq \overline{\gamma }_{j}$ if $i\neq j$ for all $i,j\in \{1,2,\cdots ,m\}$.

Assuming we have obtained a reduced product $\overline{\gamma }_{1} \cdots \overline{\gamma }_{m}$, there are three cases to consider when evaluating the expectation value: 
\begin{enumerate}
\item If $m$ is even and we can reorder the product using \eqref{eq5.31} to get $\overline{\gamma }_{i_{1}} \cdots \overline{\gamma }_{i_{m}}$ such that every pairing $\overline{\gamma }_{i_{2l}}\overline{\gamma }_{i_{2l+1}}$ is contained in the Majorana configuration of the state $\ket{\psi }$, according to \eqref{eq4.12} 
\begin{equation}
-i\overline{\gamma }_{i_{2l-1}}\overline{\gamma }_{i_{2l}}\ket{\psi } =\ket{\psi } \ (\mathrm{for} \ l=1,\cdots ,m/2) ,\label{eq5.32}
\end{equation} the expectation value will be non-zero 
\begin{equation}
\begin{array}{ l }
\quad\bra{\psi }\overline{\gamma }_{1} \cdots \overline{\gamma }_{m}\ket{\psi }\\
=\bra{\psi }(\overline{\gamma }_{i_{1}}\overline{\gamma }_{i_{2}}) \cdots (\overline{\gamma }_{i_{2l-1}}\overline{\gamma }_{i_{2l}}) \cdots (\overline{\gamma }_{i_{m-1}}\overline{\gamma }_{i_{m}})\ket{\psi }\\
\neq 0\ (\mathrm{even} \ m).
\end{array} \label{eq5.33}
\end{equation}
\item If $m$ is even but we can not reorder the product using \eqref{eq5.31} to get $\overline{\gamma }_{i_{1}} \cdots \overline{\gamma }_{i_{m}}$ such that every pairing $\overline{\gamma }_{i_{2l}}\overline{\gamma }_{i_{2l+1}}$ is contained in the Majorana configuration of the state $\ket{\psi }$, meaning that at least one pair $\overline{\gamma }_{i_{2l-1}}\overline{\gamma }_{i_{2l}}$ is not contained in the state $\ket{\psi }$ 
\begin{equation}
-i\overline{\gamma }_{i_{2l-1}}\overline{\gamma }_{i_{2l}}\ket{\psi } =0,\label{eq5.34}
\end{equation}
the expectation value will be zero
\begin{equation}
\bra{\psi }\overline{\gamma }_{1} \cdots \overline{\gamma }_{m}\ket{\psi } =0\ (\mathrm{even} \ m) .\label{eq5.35}
\end{equation}
\item If $m$ is odd, even if we can reorder the product and eliminate the pairs avoiding the last case, there will always be a single Majorana operator remained. Since the single Majorana expectation is always zero 
\begin{equation}
\bra{\psi } \gamma \ket{\psi } =0,\label{eq5.36}
\end{equation}
the expectation will always be zero 
\begin{equation}
\bra{\psi }\overline{\gamma }_{1} \cdots \overline{\gamma }_{m}\ket{\psi } =0\ (\mathrm{odd} \ m) .\label{eq5.37}
\end{equation}
\end{enumerate}

Knowing the rules outlined above for evaluating all possible terms in the polynomial-like summation, we can calculate the pure Majorana phase $\theta _{3}$ directly using a computer. This involves expanding \eqref{eq5.25}, similar to what was done in \eqref{eq5.30}, evaluating each term, summing over all terms in the polynomial-like summation, and finally normalizing the result to a U(1) phase factor.

\subsubsection{\label{sec5.3.3}Graphical computation approach}

Now, we can represent the aforementioned observation in a ``graphical" manner. Let's revisit \eqref{eq5.30}. We gather all the Majorana pairings inserted inside the bracket to form a set $\mathrm{Par}_{\mathrm{inserted}} =\{\gamma _{1} \gamma _{2} ,\gamma _{3} \gamma _{4}\}$. Similarly, we can also gather all Majorana pairings contained in state $\ket{\psi }$ to from a set $\mathrm{Par}_{\ket{\psi }} =\left\{\gamma _{i} \gamma _{j} |\ \gamma _{i} \gamma _{j}\ket{\psi } =\pm i \ket{\psi }\right\}$. Let's consider the total Majorana pairing ``pool" $\mathrm{Par}$ as the union of the above two sets $\mathrm{Par} =\mathrm{Par}_{\mathrm{inserted}} \bigcup \mathrm{Par}_{\ket{\psi }}$, which encompasses all the Majorana pairings that are inserted in \eqref{eq5.30} and are contained in $\ket{\psi }$. Then, a term in the polynomial-like summation is non-zero if and only if when there is a ``path" comprised of Majorana pairings in the Majorana pairing pool $\mathrm{Par}$, which can connect the inserted dangling Majorana operators $\gamma _{5}$ and $\gamma _{6}$. In this context, we only concern ourselves with the presence of the non-zero terms, rather than its value (the value will be considered later when evaluating a specific path). Hence, the direction of a pair is insignificant, and we will view $\gamma _{i} \gamma _{j}$ and $\gamma _{j} \gamma _{i}$ as the same element in the set $\mathrm{Par}$.

For example, if there is pair $\gamma _{5} \gamma _{6} \in \mathrm{Par}$ (a path of length one), there must be a nonzero term in the polynomial-like summation corresponding to $\bra{\psi } \gamma _{5}( \gamma _{5} \gamma _{6}) \gamma _{6}\ket{\psi }$ (up to reordering). If there are two pairs $\gamma _{6} \gamma _{3} ,\gamma _{5} \gamma _{3} \in \mathrm{Par}$ (a path of length two), there must be a nonzero term in the polynomial-like summation corresponding to $\bra{\psi } \gamma _{5}( \gamma _{6} \gamma _{3})( \gamma _{5} \gamma _{3}) \gamma _{6}\ket{\psi }$ (up to reordering).

Using this graphical method of calculation, we can easily evaluate the general form of $\theta _{3}$ in \eqref{eq5.29}. After expanding \eqref{eq5.29}, most terms in the polynomial-like summation are zero; while the remaining non-zero terms can be found as follows. If there are only two types of dangling Majorana modes in \eqref{eq5.27}, then each ``path'' consisting of Majorana pairings in the Majorana pairing pool $\mathrm{Par}$ that connects the inserted dangling Majorana operators corresponds to a series of non-zero terms subject to reordering of Majorana operators. A path tells the Majorana pairs that should be inserted in \eqref{eq5.25} to produce a non-zero term. However, it does not indicate the specific order of these inserted pairs in \eqref{eq5.25}. Each different order of Majorana pairs corresponds to a distinct term (further explanation is provided in Appendix \ref{ap3}).

If there are four types of dangling Majorana modes in \eqref{eq5.27}, we must first divide them into two pairs, and connecting each pair by a path in $\mathrm{Par}$, so the different divisions and paths corresponds to different non-zero terms in the polynomial-like summation (fortunately, this complex situation does not arise in this study). It should be noted that two paths can be combined to form a composite path: if a path already corresponds to a non-zero term, anther non-zero term can be obtained by adding a ``loop'' to this path. Finally, $\theta _{3}$ is obtained by summing over all these non-zero terms in \eqref{eq5.29} and normalize the result to a $U( 1)$ phase.

Note that one can just adopt the general computation approach introduced in Sec.\ref{sec5.3.2} to calculate the pure Majorana phase $\theta_{3}$ by directly expanding \eqref{eq5.25} and evaluating all the terms one by one. The graphical computation approach introduced here is just a way to simplify the computation and make the process more tractable by humans. In Appendix \ref{ap3}, several examples are provided to illustrate the practical application of the graphical computation approach.

\subsubsection{\label{sec5.3.4}Self-consistency condition with $\mathbb{Z}_{8}$ solution}

As we noted before in Sec. \ref{sec4.2.4}, here is an important consistency condition for the stacking rules to satisfy. Given two layers of FSPT phase $( n_{1} ,n_{2} ,\nu _{3})$ and $( n'_{1} ,n'_{2} ,\nu '_{3})$ with the same symmetry $( G_{b} ,\omega _{2} ,s_{1})$, they are both solutions of the classification equations (\ref{eq5.5})-(\ref{eq5.7}). Now, if we stack them to get a new FSPT phase $( N_{1} ,N_{2} ,\mathcal{V}_{3})$ with the same symmetry according to stacking rules Eqs.~(\ref{eq5.8})-(\ref{eq5.15.4}), this new phase should also be a solution of the classification equations (\ref{eq5.5})-(\ref{eq5.7})
\begin{align}
\mathrm{d} N_{1} &=0\ (\bmod\ 2) ,\label{eq5.38}\\
\mathrm{d} N_{2} &=\omega _{2} \cup N_{1} +s_{1} \cup N_{1} \cup N_{1} \ (\bmod\ 2) ,\label{eq5.39}\\
\mathrm{d}_{s_1}\mathcal{V}_{3} &=\mathcal{O}_{4}[ N_{2}] .\label{eq5.40}
\end{align}
The first two equations (\ref{eq5.38}) and (\ref{eq5.39}) are easy to verify, while the last one \eqref{eq5.40} is tricky. Specifically, when combining the obstruction function \eqref{eq5.40} with the stacking rule \eqref{eq5.11}, the following equation should automatically hold 
\begin{equation}
\mathrm{d}_{s_1}\mathcal{E}_{3} = \frac{\mathrm{d}_{s_1}\mathcal{V} _{3}}{\mathrm{d}_{s_1}\mathcal{\nu} _{3}\cdot\mathrm{d}_{s_1}\mathcal{\nu} '_{3}} =\frac{\mathcal{O}_{4}[ N_{2}]}{\mathcal{O}_{4}[ n_{2}]\mathcal{O}_{4}[ n'_{2}]} :=\Delta \mathcal{O}_{4} .\label{eq5.41}
\end{equation}
Physically, this consistency condition signifies that after stacking two FSPT phases, the resulting phase should also be an FSPT phase.

We have checked that all our stacking rules are compatible with the classification equations, demonstrating the self-consistency of our construction. Additionally, we attempted to directly solve \eqref{eq5.41} as a system of integral linear equations numerically using Hermite normal form. It turns out that when nontrivial $s_{1}$ is considered, there is no solution in $\mathbb{Z}_{4}$ phase (i.e. $\mathcal{E}_{3} \in \{\pm 1,\pm i\}$), but rather are solutions in $\mathbb{Z}_{8}$ phase (i.e. $\mathcal{E}_{3} \in \left\{\pm 1,\pm i,( \pm 1\pm i) /\sqrt{2}\right\}$). This observation aligns with our formula \eqref{eq5.15.3}, which incorporates a $\mathbb{Z}_{8}$ term $\frac{5}{8} s_{1} \cup n_{1} \cup n'_{1}$ in $\theta _{3}^{s_{1}}$. This observation parallels the 3+1D FSPT stacking group structure with trivial $\omega_2$ and $s_1$, as detailed in Supplementary Material IV of Ref.~\cite{WangGu2018}.

\section{\label{sec6}Examples}



\subsection{\label{sec6.3}2d wallpaper groups}

According to the crystalline equivalence principle~\cite{Else2018} for bosonic SPT, crystalline topological liquids with symmetry group $G$ are in one-to-one correspondence with topological phases protected by the same symmetry $G$, but acting
internally, where if an element of $G$ is orientation reversing, it is realized as an anti-unitary symmetry in the internal symmetry group.

However, the fermionic crystalline equivalence principle is somewhat more intricate~\cite{Debray2021,Manjunath2023}. Here, we follow the conjecture proposed in Ref.~\cite{Manjunath2023}, which is supported by the mathematical background provided in Ref.~\cite{Debray2021}.

This conjecture asserts that there exists a one-to-one correspondence between the classification of invertible fermionic topological phases with spatial symmetry $G_{f}$,
defined by the data $(G_{b}, \omega _{2}, s_{1})$, and the classification of invertible fermionic topological phases with an effective internal symmetry $G_{f} ^{\mathrm{eff}}$, defined by data $(G_{b}, \omega _{2} ^{\mathrm{eff}}, s_{1} ^{\mathrm{eff}})$, where $G_{b}$ acts trivially on space. The internal symmetry data can be fully determined by the crystalline symmetry data according to 
\begin{align}
s_{1} ^{\mathrm{eff}} &=s_{1} +w_{1} ,\label{eq6.1}\\
\omega _{2} ^{\mathrm{eff}} &=\omega _{2} + w_{2} + w_{1} \cup ( s_{1} + w_{1}) .\label{eq6.2}
\end{align}
In these equations, the $w_{1}$ and $w_{2}$ are obtained by pulling back the Stiefel-Whitney classes. If there are no reflections in $G_{f}$, $w_{1}=0$; if there are no rotations, $w_{2}=0$. As the formulation may appear intricate, we highly recommend interested readers to refer to Ref.~\cite{Manjunath2023} for a more detailed explanation.

Based on this fermionic crystalline equivalence principle, we compute the stacking groups of interacting FSPT protected by 2d wallpaper groups (viewed as onsite symmetries) through the computation method \cite{Ouyang2021, QiGithub}, which turn out to be isomorphic to the stacking groups of corresponding crystalline topological superconductors calculated by another approach called real-space construction~\cite{Zhang2020,Zhang2022,zhang202204,Zhang2023}. 

In Tab.~\ref{tab7}, we listed the stacking group structure of interacting spinless (spin-0) FSPT protected by 2d wallpaper groups (viewed as onsite symmetries). The first column $G_{b}$ represents the standard names of the corresponding wallpaper groups. The solutions of the FSPT classification are listed in terms of the Majorana chain (MC), complex fermion (FC), and bosonic (B) layer. The last column (Extension) represents the overall stacking group structure, i.e. answers the group extension problem. Similarly, the stacking group structure of spin-1/2 interacting FSPT is listed in Tab.~\ref{tab8}.

It is important to note that the crystalline equivalence principle maps a \textbf{spinless internal} FSPT protected by onsite symmetry to a \textbf{spin-1/2 crystalline} FSPT protected by the corresponding crystalline symmetry, while maps a \textbf{spin-1/2 internal} FSPT protected by onsite symmetry to a \textbf{spinless crystalline} FSPT protected by the corresponding crystalline symmetry. Therefore, the results of spinless FSPT in Tab.~\ref{tab7} should be compared with the results of spin-1/2 crystalline FSPT in Tab.~\uppercase\expandafter{\romannumeral2} of Ref.~\cite{Zhang2022}. On the other hand, the results of spin-1/2 FSPT in Tab.~\ref{tab8} should be compared with the results of spinless crystalline FSPT in Tab.~\uppercase\expandafter{\romannumeral1} of Ref.~\cite{Zhang2022}. This rule applies to all the results in Tab.~\ref{tab7}-\ref{tab8}. This rule applies to all the results in Tab.~\ref{tab7}-\ref{tab10}.

\subsection{\label{sec6.4}The combination of 2d wallpaper groups and onsite time-reversal symmetry}

In the last subsection, we calculated the stacking groups of interacting FSPT protected by 2d wallpaper groups (viewed as onsite symmetries), which are isomorphic to the stacking groups of interacting crystalline SPT protected by the 2d wallpaper groups (viewed as space symmetries) according to the crystalline equivalence principle. However, these results are already known. In this subsection, we take a further step to obtain some new results.

The motivation behind our work is to determine the stacking groups of interacting crystalline SPT protected by a combination of onsite symmetries and space symmetries. The bosonic part of the total symmetry group can be expressed as the direct product of the form $G_{b} = G_{\mathrm{wp}} \times \mathbb{Z}_{2} ^{T}$, where $G_{\mathrm{wp}}$ is a 2d wallpaper group (viewed as space symmetries) and $\mathbb{Z}_{2} ^{T}$ is the onsite time-reversal symmetry.

Using the above formulas \eqref{eq6.1} and \eqref{eq6.2}, we compute the stacking group structure of spinless interacting FSPT protected by internal symmetries in Tab.~\ref{tab9}. Similarly, we compute the stacking group structure of spin-1/2 interacting FSPT protected by internal symmetries in Tab.~\ref{tab10}. Note that we omit the results of the first wallpaper group $G_{\mathrm{wp}}=\mathrm{p1}$ here due to its trivial nature.

\section{\label{sec7}Conclusions and Discussions}


In this work, we have derived the stacking group structure of FSPT states in dimensions ranging from 0+1 to 2+1. To achieve this, the key is to design an fermionic local unitary transformation that allows us to convey the nontrivial classifying data (such as $(n_1,n_2,\nu_3)$ for 2D) in two layers of FSPT to a single layer while leaving another layer completely trivial. By utilizing these stacking rules, we have obtained the group structure of classification of the 2D wallpaper groups with the assistance of the fermionic crystalline equivalence principle via the computation method developed in Ref. \cite{Ouyang2021,QiGithub}.

The main technical challenge in deriving the stacking group structure of FSPT states is calculating the pure Majorana phase $\theta_3$ using equation \eqref{eq5.25} which is obtained in assistance of computer program. 
Below we list some future directions:
\begin{enumerate}
    \item Generalize to study the stacking group extension problem for 3D and higher FSPT.
    \item Utilize the stacking group structure obtained in this paper to calculate more groups, such as those involve both spacial and internal group and also the spin space group  classified recently \cite{Xiao2023,CF2023}.
\end{enumerate}


\begin{acknowledgments}
We would like to thank Weicheng Ye for pointing out our errors in the previous version.
X.-Y.R. is grateful to Tian Yuan and Tian Lan for invaluable discussions. Z.-C.G. is supported by Direct Grant No. 4053578 from The Chinese University of Hong Kong and funding from Hong Kong’s Research Grants Council (GRF No. 14306420, ANR/RGC Joint Research Scheme No. A-CUHK402/18). Q.-R.W. is supported by the National Natural Science Foundation of China (Grant No. 12274250). Y.Q. is supported by the National Natural Science Foundation of China (Grant No. 11874115).
\end{acknowledgments}

\appendix

\section{\label{ap1}Group cohomology}

\subsection{\label{ap1.1}$G$-module}

Given a group $G$ and an Abelian group $M$, the $G$-module is defined by a $G$ action on $M$, denoted as $g\triangleright a\in M$ for $g\in G$ and $a,b\in M$, such that the $G$ action is compatible with the group multiplication of $M$  
\begin{equation}
g\triangleright ( ab) =( g\triangleright a)( g\triangleright b) .\label{eq.ap1.1}
\end{equation}
In this paper, we always consider $M$ to be the $U( 1)$ phase. If $G$ contains only unitary operation, then we set the $G$ action to be trivial $g\triangleright a=a$; while if $G$ contains anti-unitary operation such as time-reversal transformation, we define the $G$ action as $g\triangleright a=a^{s_{1}( g)}$, where $s_{1}( g) =-1$ if group element $g$ is anti-unitary, otherwise $s_{1}( g) =1$.

\subsection{\label{ap1.2}An algebraic definition of inhomogeneous cochain}

An $n$-cochain $\omega _{n}( g_{1} ,g_{2} ,\cdots ,g_{n})$ is a function of $n$ group elements with values in the $G$-module $M$. In other words, an $n$-cochain is a map $\omega _{n} :G^{n} \rightarrow M$ from the Cartesian product $G\times G\times \cdots \times G$ to $M$. The set of all $n$-cochains is denoted as $\mathcal{C}^{n}( G,M)$.

The coboundary homomorphism $\mathrm{d}_{n} :\mathcal{C}^{n}( G,M) \rightarrow \mathcal{C}^{n+1}( G,M)$, for $n\in \mathbb{Z}^{+}$, is a map from the set of all $n$-cochains $\mathcal{C}^{n}( G,M)$ to the set of all $( n+1)$-cochains $\mathcal{C}^{n+1}( G,M)$. Given an $n$-cochain $\omega _{n}( g_{1} ,g_{2} ,\cdots ,g_{n})$, we apply the coboundary homomorphism $\mathrm{d}_{n}$ to obtain an $( n+1)$-cochain $(\mathrm{d}_{n} \omega _{n})( g_{1} ,g_{2} ,\cdots ,g_{n} ,g_{n+1})$ 
\begin{equation}
\begin{array}{ l }
(\mathrm{d}_{n} \omega _{n})( g_{1} ,\cdots ,g_{n+1})\\
=[ g_{1}\triangleright \omega _{n}( g_{2} ,\cdots ,g_{n+1})] \omega _{n}^{( -1)^{n+1}}( g_{1} ,\cdots ,g_{n}) \\
\times \prod _{i=1}^{n} \omega _{n}^{( -1)^{i}}( g_{1} ,\cdots ,g_{i-1} ,g_{i} g_{i+1} ,g_{i+2} ,\cdots ,g_{n+1})
\end{array} .\label{eq.ap1.2}
\end{equation}
In some cases, we may omit the subscript $n$ of $\mathrm{d}_{n}$ if there is no ambiguity. Using \eqref{eq.ap1.2}, we can explicitly verify that the composition map of $\mathrm{d}_{n}$ and $\mathrm{d}_{n+1}$ satisfies 
\begin{equation}
\mathrm{d}_{n+1} \circ \mathrm{d}_{n} =0, \label{eq.ap1.3}
\end{equation}
where $0$ is the identity element of $M$. This implies that the image $\mathcal{B}^{n}$ of map $\mathrm{d}_{n-1}$ 
\begin{equation}
\begin{array}{ l }
\mathcal{B}^{n}( G,M) =\\
\left\{\omega _{n} \in \mathcal{C}^{n}( G,M) |\omega _{n} =\mathrm{d}_{n-1} \omega _{n-1} |\omega _{n-1} \in \mathcal{C}^{n-1}( G,M)\right\}
\end{array} \label{eq.ap1.4}
\end{equation}
is contained in the kernel $\mathcal{Z}^{n}$ of map $\mathrm{d}_{n}$
\begin{equation}
\mathcal{Z}^{n}( G,M) =\left\{\omega _{n} \in \mathcal{C}^{n}( G,M) |\mathrm{d}_{n} \omega _{n} =0\right\} . \label{eq.ap1.5}
\end{equation}
This allows us to define the cohomology group $\mathcal{H}^{n}( G,M)$ as the quotient 
\begin{equation}
\mathcal{H}^{n}( G,M) =\mathcal{Z}^{n}( G,M) /\mathcal{B}^{n}( G,M) . \label{eq.ap1.6}
\end{equation}
An element in $\mathcal{B}^{n}( G,M)$ is called an $n$-coboundary, while an element in $\mathcal{Z}^{n}( G,M)$ is called an $n$-cocycle.

\subsection{\label{ap1.3}Homogeneous cochain}

What we introduced above is referred to as an inhomogeneous cochain $\omega _{n} :G^{n} \rightarrow M$. This inhomogeneous cochain can be transformed into a homogeneous cochain $\nu _{n} :G^{n+1} \rightarrow M$ by the correspondence 
\begin{equation}
\begin{array}{ l }
\omega _{n}( g_{1} ,g_{2} ,\cdots ,g_{n})\\
=\nu _{n}( 1,g_{1} ,g_{1} g_{2} ,g_{1} g_{2} g_{3} ,\cdots ,g_{1} g_{2} \cdots g_{n})\\
\equiv \nu _{n}( 1,\tilde{g}_{1} ,\tilde{g}_{2} ,\cdots ,\tilde{g}_{n})
\end{array} \label{eq.ap1.7}
\end{equation}
with $\tilde{g}_{i} =g_{1} g_{2} \cdots g_{i}$. The $G$-action on the homogeneous cochain satisfies 
\begin{equation}
g\triangleright \nu _{n}( g_{0} ,g_{1} ,\cdots ,g_{n}) =\nu _{n}( gg_{0} ,gg_{1} ,\cdots ,gg_{n}) . \label{eq.ap1.8}
\end{equation}
Using \eqref{eq.ap1.2}, we can check that the coboundary homomorphism $\mathrm{d}_{n} :\nu _{n} \mapsto \nu _{n+1}$ can be reduced to a simplified form 
\begin{equation}
\begin{array}{ l }
(\mathrm{d}_{n} \nu _{n})( g_{0} ,g_{1} ,\cdots ,g_{n+1})\\
=\prod\nolimits _{i=0}^{n+1} \nu _{n}^{( -1)^{i}}( g_{0} ,\cdots ,\hat{g}_{i} ,\cdots ,g_{n+1})
\end{array} , \label{eq.ap1.9}
\end{equation}
where $\hat{g}_{i}$ denotes omit $g_{i}$. The inhomogeneous $n$-cochain $\omega _{n}( g_{1} ,\cdots ,g_{n})$ and the corresponding homogeneous $n$-cochain $\nu _{n}( g_{0} ,g_{1} ,\cdots ,g_{n})$ can be easily distinguished: the former has $n$ input group element, while the latter has $( n+1)$. Therefore, in the main text, we may not always explicitly indicate whether a cochain is inhomogeneous or homogeneous, but the reader should be able to distinguish them easily.

\subsection{\label{ap1.4}Steenrod's higher cup product}

The Steenrod's higher cup product is defined in Ref.~\cite{Steen1974}. Since the definition of higher cup is more involved, here we only list the cup-0 and cup-1, with the assumption of trivial $G$-action on the coefficient $M$. Given two cochain $f_{m} \in \mathcal{C}^{m}( G,M)$ and $h_{n} \in \mathcal{C}^{n}( G,M)$, the cup product (cup-0) is a map $\cup \ :\mathcal{C}^{m}( G,M) \times \mathcal{C}^{n}( G,M) \rightarrow \mathcal{C}^{m+n}( G,M)$ that is defined as 
\begin{equation}
( f_{m} \cup h_{n})( g_{1} ,\cdots ,g_{m+n}) =f_{m}( g_{1} ,\cdots ,g_{m}) h_{n}( g_{1} ,\cdots ,g_{n}) . \label{eq.ap1.10}
\end{equation}
The cup-$1$ product is a map $\cup_{1} :\mathcal{C}^{m}( G,M) \times \mathcal{C}^{n}( G,M) \rightarrow \mathcal{C}^{m+n-1}( G,M)$ that is defined as 
\begin{equation}
\begin{array}{ l }
( f_{m} \cup _{1} h_{n})( g_{1} ,\cdots ,g_{m+n-1})\\
=\sum _{j=0}^{m-1}( -1)^{( m-j)( n+1)} f_{m}( g_{1} ,\cdots ,g_{j} ,g_{j+n} ,\cdots ,g_{m})\\
\times h_{n}( g_{j} ,\cdots ,g_{j+n-1})
\end{array} . \label{eq.ap1.11}
\end{equation}
The higher cup products satisfy the following Leibniz's rule 
\begin{equation}
\begin{array}{ l }
\mathrm{d}( f_{m} \cup _{i} h_{n}) =\mathrm{d} f_{m} \cup _{i} h_{n} +( -1)^{m} f_{m} \cup _{i}\mathrm{d} h_{n}\\
+( -1)^{m+n-i} f_{m} \cup _{i-1} h_{n} +( -1)^{mn+m+n} h_{n} \cup _{i-1} f_{m}
\end{array} . \label{eq.ap1.12}
\end{equation}

\section{\label{ap2}Counting fermion parity change}

In this appendix, we exhaustively enumerate the stacking rules for the Kitaev chain \eqref{eq5.8} and complex fermion \eqref{eq5.9} and \eqref{eq5.10}, deriving them in the process.

\begin{figure*}
    \centering
    \includegraphics[width=0.85\textwidth]{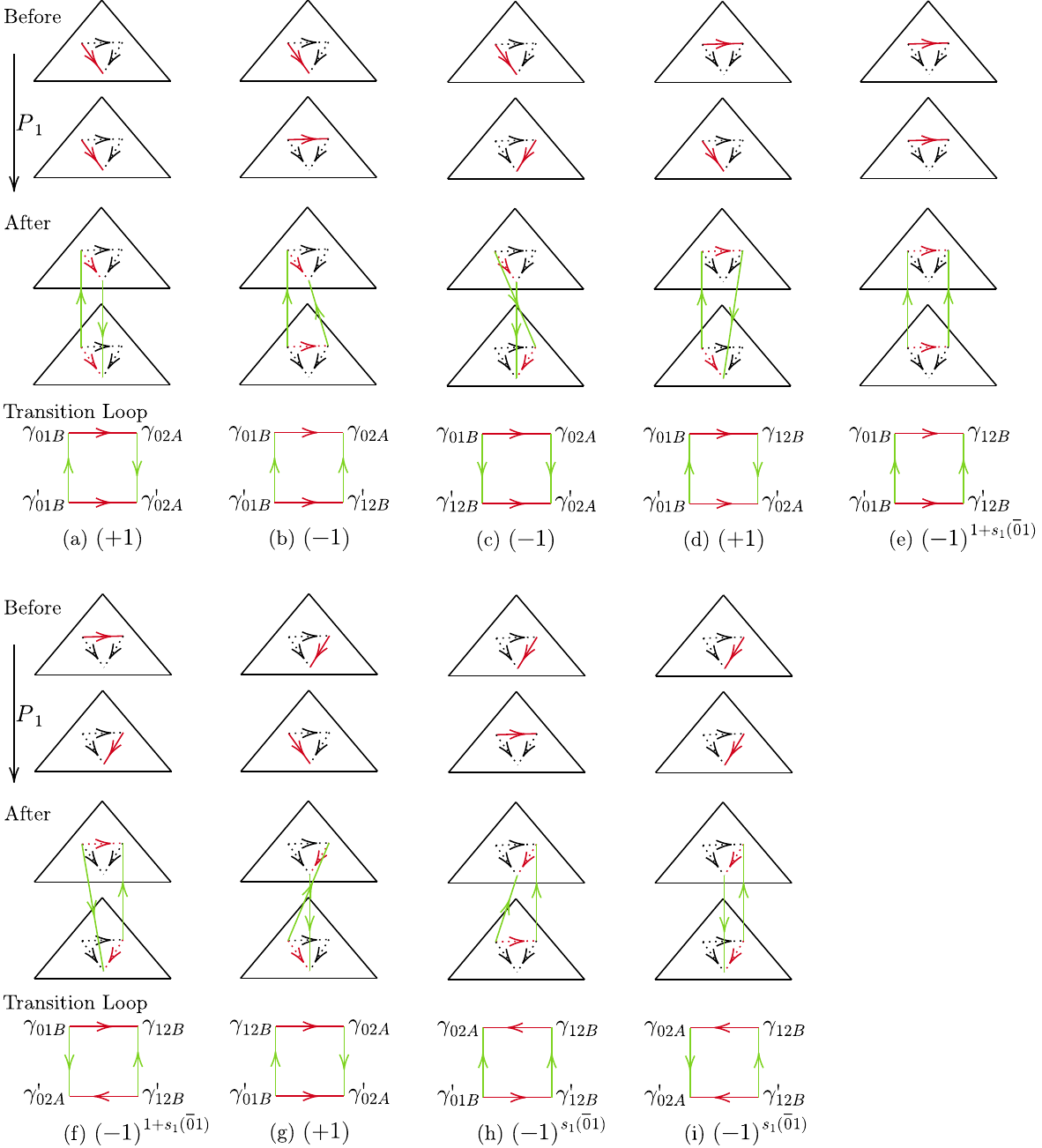}
    \caption{Fermion parity changes of all possible Kitaev chain decorations.}
    \label{fig11}
\end{figure*}

In Fig.~\ref{fig11}, we provide a comprehensive list of all possible configurations of the Kitaev chain under stacking. According to the projection rules outlined in Sec.\ref{sec5.2.3}, only the first step in Fig.~\ref{fig7} may possibly change the Kasteleyn orientation and consequently change the fermion parity of Kitaev chain decoration. Therefore, we solely display the first step of Fig.~\ref{fig7} in Fig.~\ref{fig11}, 
illustrating the corresponding initial and final states with transition loops and phase factors ($-1$ representing fermion parity change, while $+1$ indicating no change). The phase factor $(-1)^{m_{2}(012)}$ (representing the fermion parity change of the Kitaev chain decoration) of all cases can be succinctly summarized as $m_{2}(012)$ in \eqref{eq5.10}.

By simple observation, we can deduce the stacking rule for the Kitaev chain \eqref{eq5.8}. Since the fermion parity of the whole system should conserve, any change in the fermion parity of the Kitaev chain decoration must be counterbalanced by a corresponding change in the fermion parity of the complex fermion, which leads to the stacking rule \eqref{eq5.9}.

\section{\label{ap3}Examples of graphical approach}

We present multiple examples of the calculation of the pure Majorana phase $\theta_{3}$ in \eqref{eq5.25} using the graphical approach introduced in Sec.\ref{sec5.3.3}. In each of the following examples, we specify a particular set of input data for $n_{1}$, $n'_{1}$, $\omega_{2}$ and $s_{1}$, which fully determines the Majorana decoration.

\subsection{\label{ap3.1}Example 1}

The configurations of the Kitaev chain are determined by setting $n_{1}( 01) =n_{1}( 23) =n'_{1}( 12) =0$ and $n_{1}( 12) =n'_{1}( 01) =n'_{1}( 23) =1$ (all other $n_{1} ,n'_{1}$ are fixed by the cocycle condition \eqref{eq5.5}), with $\omega _{2}( 012) =1$ (other $\omega _{2}$ are irrelevant) and trivial $s_{1}=0$. The quadruplet \eqref{eq5.28} of the total dangling Majorana modes is $[ 1,0,0,1]$, there are one $\gamma _{23B}$ operator in $X_{6}$ and three $\gamma '_{23B}$ operators inserted in $X_{4}$, $X_{2}$ and $X_{0}$.

\begin{figure*}
    \centering
    \includegraphics[width=0.8\textwidth]{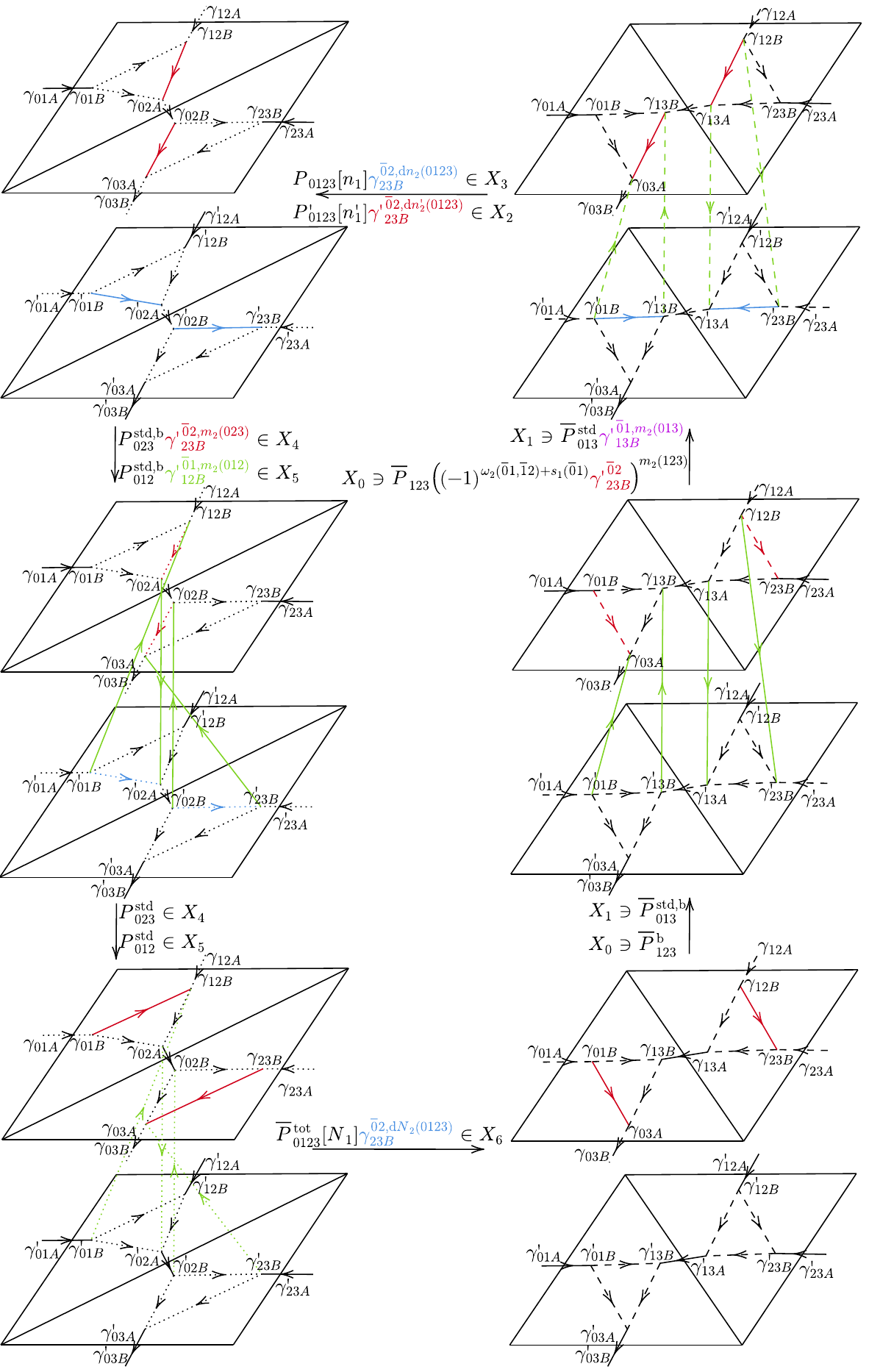}
    \caption{Example 1: detailed projection process of Fig.~\ref{fig10}.}
    \label{fig12}
\end{figure*}

The entire projection process is depicted in Fig.~\ref{fig12}. The solid lines with arrows represent Majorana pairings: black for trivial pairings, red for non-trivial pairings in the upper layer, blue for non-trivial pairings in the lower layer. The dotted (left column) or dashed (right column) lines with arrows indicate no pairing for the species we are working with, and the colors serve as a reminder of the non-trivial pairings that appeared in previous processes. All the other Majorana pairs not shown in the figures are in trivial pairing. We only draw the bridge pairs that are inside the tetrahedron $\langle 0123\rangle$.

\begin{figure*}
    \centering
    \includegraphics[width=0.55\textwidth]{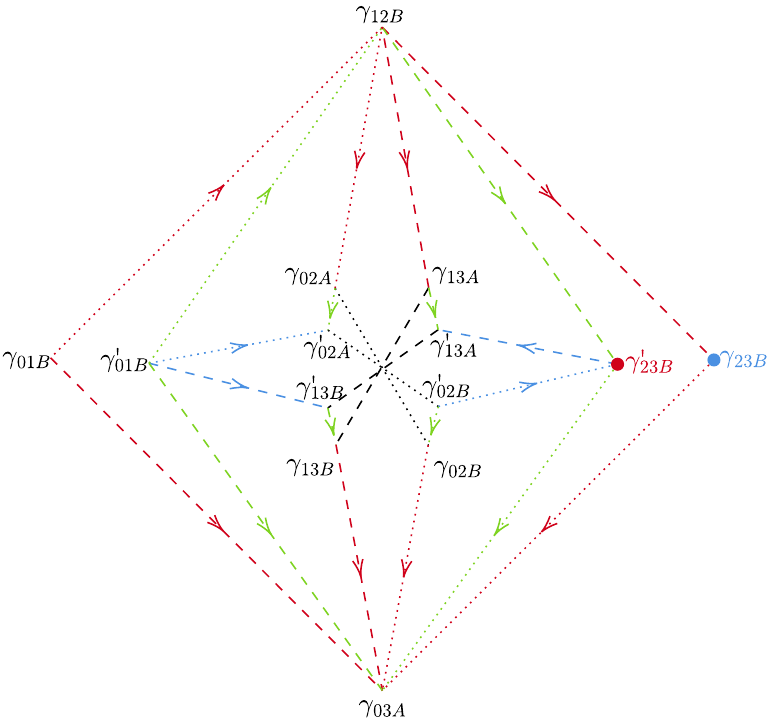}
    \caption{Example 1: the total Majorana pairing pool $\mathrm{Par}$ introduced in Sec.\ref{sec5.3.3}. The inserted dangling Majorana modes are $\gamma _{23B}$ and $\gamma '_{23B}$, marked by blue and red solid circles. Note though there are there black lines crossing at the center of picture, there is no Majorana operator at the center; we also omit the arrows of the three black line to avoid overcrowd picture.}
    \label{fig13}
\end{figure*}

After determining the transition process, we illustrate the Majorana pairing pool $\mathrm{Par}$ in Fig.~\ref{fig13}. This pool encompasses all the pairings that may be utilized when calculating \eqref{eq5.25} that arises in the transition process illustrated in Fig.~\ref{fig12}.  The color convention of the pairings depicted in Fig.~\ref{fig13} remains consistent with that of Fig.~\ref{fig12}. Additionally, the dotted lines correspond to the left column of Fig.~\ref{fig12}, while the dashed lines correspond to the right column. 

Then, we select four paths that connect the two inserted dangling Majorana modes $\gamma _{23B}$ and $\gamma '_{23B}$ in the Majorana pairing pool $\mathrm{Par}$. These paths are illustrated in Fig.~\ref{fig13}. We use these paths as examples to demonstrate how to obtain all the nonzero terms in the polynomial-like summation of the expansion of \eqref{eq5.25} and evaluate their values.

\begin{figure*}
    \centering
    \includegraphics[width=0.6\textwidth]{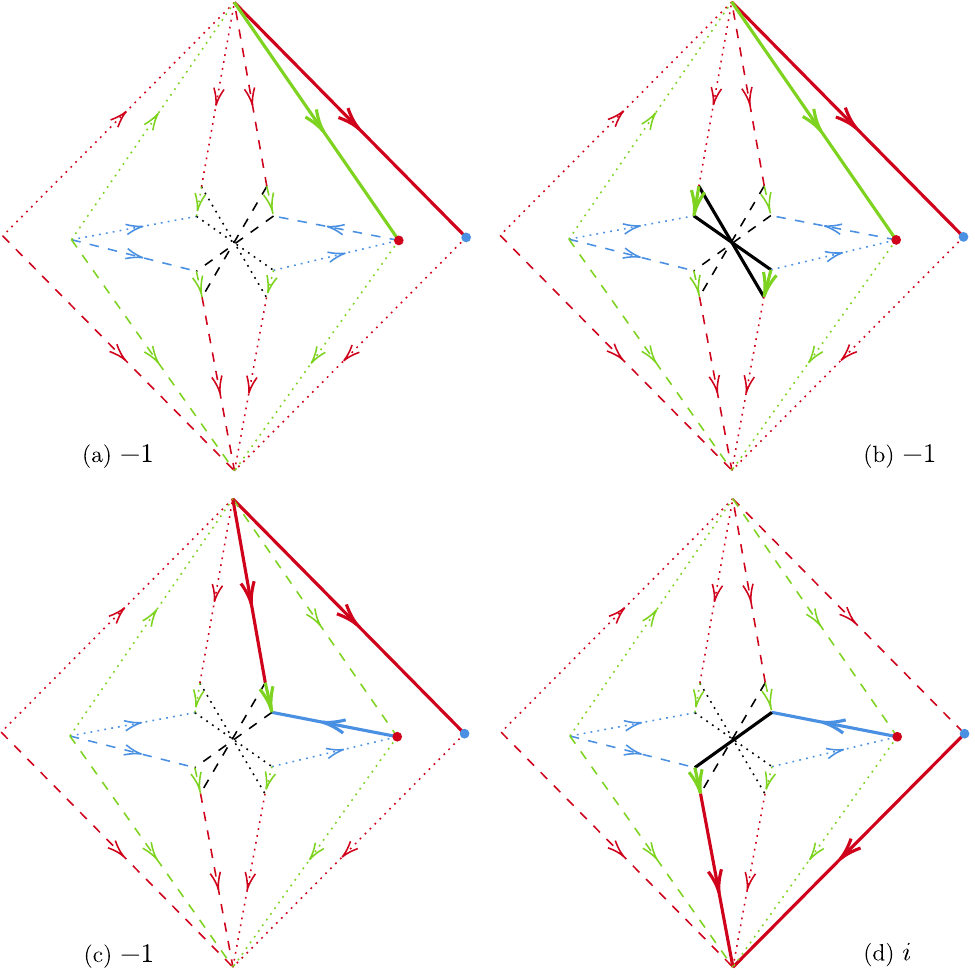}
    \caption{Example 1: different paths (represented by the solid bold lines) in the graphical computation approach contributing to the nonzero terms in the polynomial-like summation of the expansion of \eqref{eq5.25}.}
    \label{fig14}
\end{figure*}

\begin{figure*}
    \centering
    \includegraphics[width=0.8\textwidth]{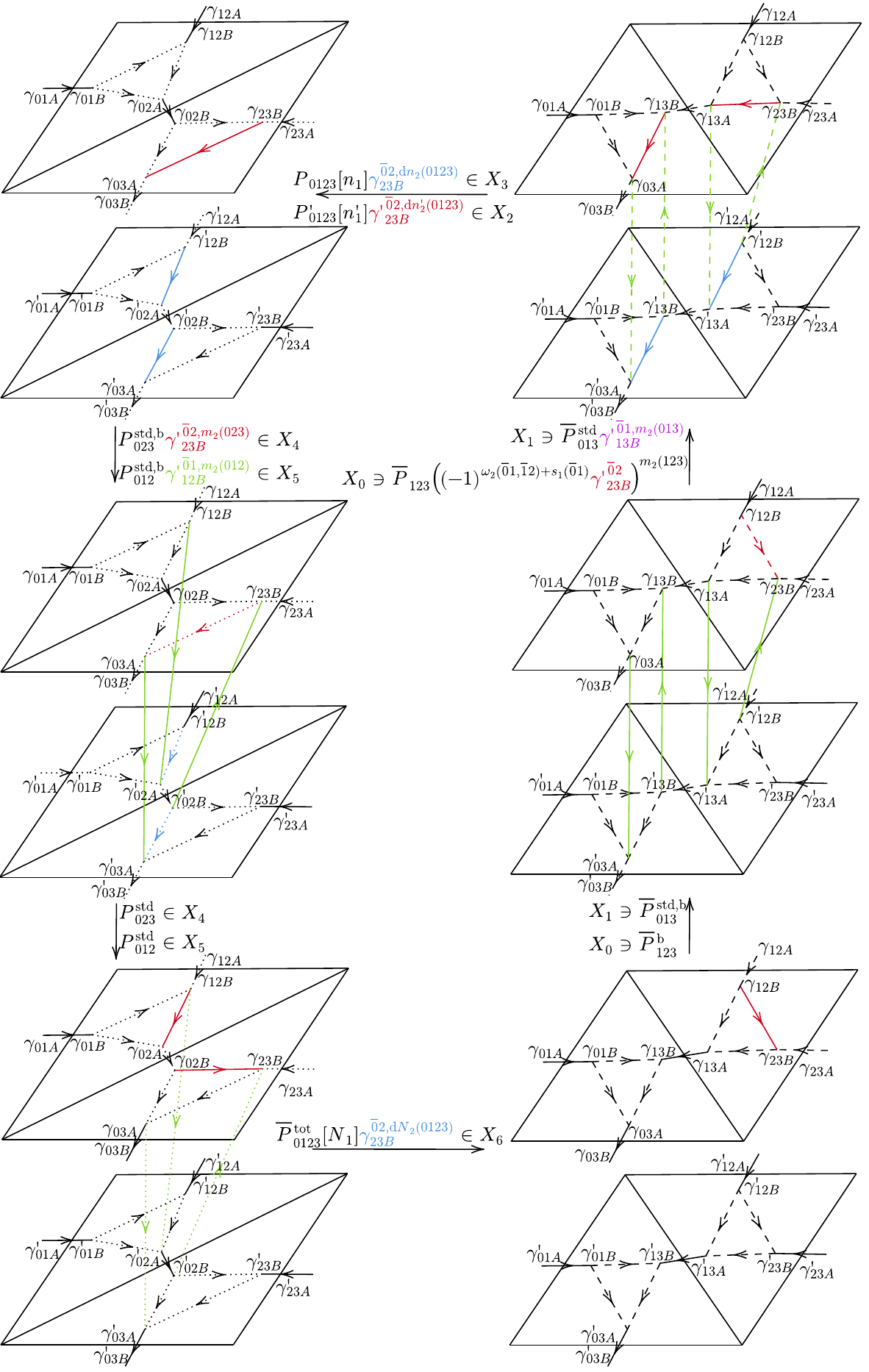}
    \caption{Example 3: detailed projection process of Fig.~\ref{fig10}.}
    \label{fig15}
\end{figure*}

\begin{figure}
    \centering
    \includegraphics[width=0.35\textwidth]{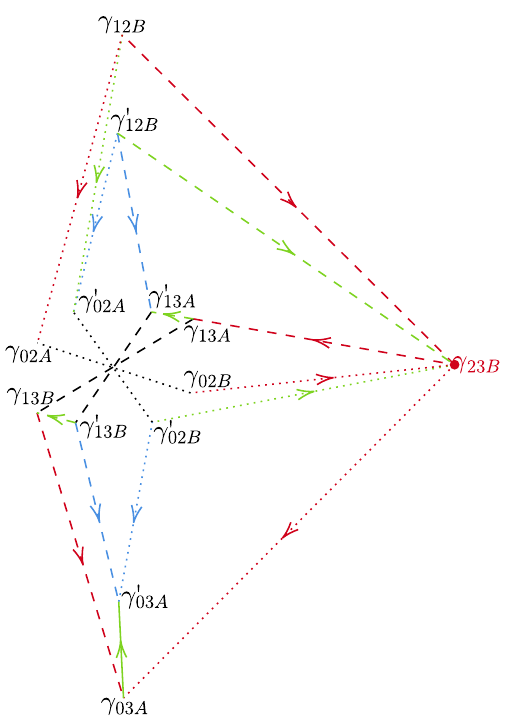}
    \caption{Example 3: the total Majorana pairing pool $\mathrm{Par}$ introduced in Sec.\ref{sec5.3.3}. The inserted dangling Majorana modes are two $\gamma '_{23B}$, marked by red solid circles. The solid green arrow means that this pair appears both in the left column and right column of Fig.~\ref{fig15}.}
    \label{fig16}
\end{figure}

\begin{figure}
    \centering
    \includegraphics[width=0.47\textwidth]{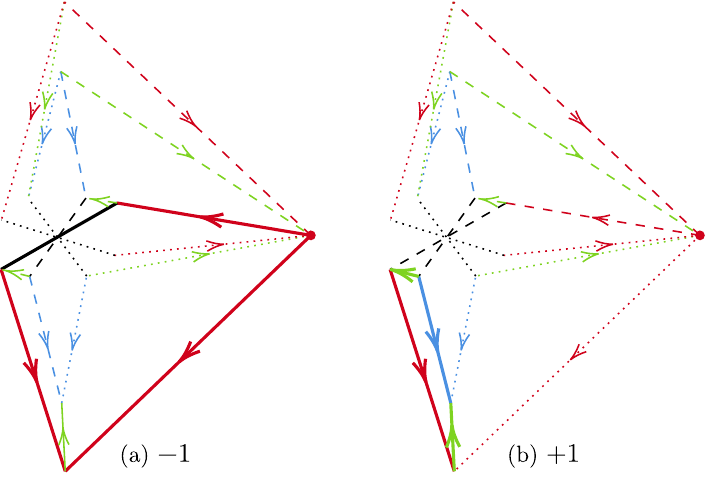}
    \caption{Example 3: different paths (represented by the solid bold lines) in the graphical computation approach contributing to the nonzero terms in the polynomial-like summation of the expansion of \eqref{eq5.25}.}
    \label{fig17}
\end{figure}

The first path is illustrated in Fig.~\ref{fig14}(a) and evaluated in Tab.~\ref{tab1}. This path contains of two pairs, $( +i\gamma _{12B} \gamma _{23B})$ and $( +i\gamma _{12B} \gamma '_{23B})$, shown as bold solid lines in Fig.~\ref{fig14}(a). Upon returning to the projection process in Fig.~\ref{fig13}, we observe that they belong to $X_{6}$ and $X_{0}$ respectively. By substituting these two pairs back into \eqref{eq5.25} in place of $X_{6}$ and $X_{0}$ respectively, we get a nonzero term in the polynomial-like summation of the expansion of \eqref{eq5.25} as 
\begin{equation}
\begin{array}{ l }
\bra{\psi }( +i\gamma _{12B} \gamma _{23B})\textcolor[rgb]{0.29,0.56,0.89}{\gamma }\textcolor[rgb]{0.29,0.56,0.89}{_{23B}}\textcolor[rgb]{0.82,0.01,0.11}{\gamma '}\textcolor[rgb]{0.82,0.01,0.11}{_{23B}}\textcolor[rgb]{0.82,0.01,0.11}{\gamma '}\textcolor[rgb]{0.82,0.01,0.11}{_{23B}}\textcolor[rgb]{0.82,0.01,0.11}{(}\textcolor[rgb]{0.82,0.01,0.11}{-\gamma '}\textcolor[rgb]{0.82,0.01,0.11}{_{23B}}\textcolor[rgb]{0.82,0.01,0.11}{)}\\
\times ( +i\gamma _{12B} \gamma '_{23B})\ket{\psi } =-1
\end{array}. \label{eq.ap3.1}
\end{equation}
The evaluation is performed using the anti-commutation relation \eqref{eq5.31} of Majorana fermions. All the Majorana operators mentioned above are listed in Tab.~\ref{tab1}, in the same order as in \eqref{eq.ap3.1}. Therefore, each table listed below represents the evaluation of the corresponding non-zero term.

\begin{table*}
\caption{Evaluation of the paths in the examples.}
\label{tab0}
\centering
\subtable[Example 1: evaluation of path (a) in Fig.~\ref{fig14}.]{
\begin{ruledtabular}
\begin{tabular}{cccccccc}
$X_{6}$ & $X_{5}$ & $X_{4}$ & $X_{3}$ & $X_{2}$ & $X_{1}$ & $X_{0}$ & $e^{2\pi i\theta}$\\ \hline
$( +i\gamma _{12B} \gamma _{23B}) \textcolor[rgb]{0.29,0.56,0.89}{\gamma }\textcolor[rgb]{0.29,0.56,0.89}{_{23B}}$ &  & $\textcolor[rgb]{0.82,0.01,0.11}{\gamma '}\textcolor[rgb]{0.82,0.01,0.11}{_{23B}}$ &  & $\textcolor[rgb]{0.82,0.01,0.11}{\gamma '}\textcolor[rgb]{0.82,0.01,0.11}{_{23B}}$ &  & $\textcolor[rgb]{0.82,0.01,0.11}{(}\textcolor[rgb]{0.82,0.01,0.11}{-\gamma '}\textcolor[rgb]{0.82,0.01,0.11}{_{23B}}\textcolor[rgb]{0.82,0.01,0.11}{)} ( +i\gamma _{12B} \gamma '_{23B})$ & $-1$\\
\end{tabular}
\end{ruledtabular}
\label{tab1}
}

\subtable[Example 1: evaluation of path (b) in Fig.~\ref{fig14}.]{
\begin{ruledtabular}
\begin{tabular}{cccccccc}
$X_{6}$ & $X_{5}$ & $X_{4}$ & $X_{3}$ & $X_{2}$ & $X_{1}$ & $X_{0}$ & $e^{2\pi i\theta}$\\ \hline
$( +i\gamma _{12B} \gamma _{23B}) \textcolor[rgb]{0.29,0.56,0.89}{\gamma }\textcolor[rgb]{0.29,0.56,0.89}{_{23B}}$ & $( -i\gamma _{02A} \gamma _{02B})( -i\gamma '_{02A} \gamma '_{02B})( -i\gamma _{02A} \gamma '_{02A})$ & $( -i\gamma '_{02B} \gamma _{02B})\textcolor[rgb]{0.82,0.01,0.11}{\gamma '}\textcolor[rgb]{0.82,0.01,0.11}{_{23B}}$ &  & $\textcolor[rgb]{0.82,0.01,0.11}{\gamma '}\textcolor[rgb]{0.82,0.01,0.11}{_{23B}}$ &  & $\textcolor[rgb]{0.82,0.01,0.11}{(}\textcolor[rgb]{0.82,0.01,0.11}{-\gamma '}\textcolor[rgb]{0.82,0.01,0.11}{_{23B}}\textcolor[rgb]{0.82,0.01,0.11}{)}( +i\gamma _{12B} \gamma '_{23B})$ & $-1$\\
\end{tabular}
\end{ruledtabular}
\label{tab2}
}

\subtable[Example 1: evaluation of path (c) in Fig.~\ref{fig14}.]{
\begin{ruledtabular}
\begin{tabular}{cccccccc}
$X_{6}$ & $X_{5}$ & $X_{4}$ & $X_{3}$ & $X_{2}$ & $X_{1}$ & $X_{0}$ & $e^{2\pi i\theta}$\\ \hline
$( +i\gamma _{12B} \gamma _{23B})\textcolor[rgb]{0.29,0.56,0.89}{\gamma }\textcolor[rgb]{0.29,0.56,0.89}{_{23B}}$ &  & $\textcolor[rgb]{0.82,0.01,0.11}{\gamma '}\textcolor[rgb]{0.82,0.01,0.11}{_{23B}}$ &  & $\textcolor[rgb]{0.82,0.01,0.11}{\gamma '}\textcolor[rgb]{0.82,0.01,0.11}{_{23B}}$ &  & $( -i\gamma _{12B} \gamma _{13A})( +i\gamma '_{23B} \gamma '_{13A})\textcolor[rgb]{0.82,0.01,0.11}{(}\textcolor[rgb]{0.82,0.01,0.11}{-\gamma '}\textcolor[rgb]{0.82,0.01,0.11}{_{23B}}\textcolor[rgb]{0.82,0.01,0.11}{)}( -i\gamma _{13A} \gamma '_{13A})$ & $-1$\\
\end{tabular}
\end{ruledtabular}
\label{tab3}
}

\subtable[Example 1: evaluation of path (d) in Fig.~\ref{fig14}.]{
\begin{ruledtabular}
\begin{tabular}{cccccccc}
$X_{6}$ & $X_{5}$ & $X_{4}$ & $X_{3}$ & $X_{2}$ & $X_{1}$ & $X_{0}$ & $e^{2\pi i\theta}$\\ \hline
$( -i\gamma '_{13A} \gamma '_{13B})\textcolor[rgb]{0.29,0.56,0.89}{\gamma }\textcolor[rgb]{0.29,0.56,0.89}{_{23B}}$ &  & $( -i\gamma _{23B} \gamma _{03A})\textcolor[rgb]{0.82,0.01,0.11}{\gamma '}\textcolor[rgb]{0.82,0.01,0.11}{_{23B}}$ &  & $\textcolor[rgb]{0.82,0.01,0.11}{\gamma '}\textcolor[rgb]{0.82,0.01,0.11}{_{23B}}$ & $( -i\gamma _{13B} \gamma _{03A})( -i\gamma '_{13B} \gamma _{13B})$ & $( +i\gamma '_{23B} \gamma '_{13A})\textcolor[rgb]{0.82,0.01,0.11}{(}\textcolor[rgb]{0.82,0.01,0.11}{-\gamma '}\textcolor[rgb]{0.82,0.01,0.11}{_{23B}}\textcolor[rgb]{0.82,0.01,0.11}{)}$ & $i$\\
\end{tabular}
\end{ruledtabular}
\label{tab4}
}

\subtable[Example 3: evaluation of path (a) in Fig.~\ref{fig17}.]{
\begin{ruledtabular}
\begin{tabular}{cccccccc}
$X_{6}$ & $X_{5}$ & $X_{4}$ & $X_{3}$ & $X_{2}$ & $X_{1}$ & $X_{0}$ & $e^{2\pi i\theta}$\\ \hline
$( -i\gamma _{13A} \gamma _{13B})$ &  &  & $( -i\gamma _{23B} \gamma _{03A})$ & $\textcolor[rgb]{0.82,0.01,0.11}{\gamma '}\textcolor[rgb]{0.82,0.01,0.11}{_{23B}}$ & $( -i\gamma _{13B} \gamma _{03A})$ & $( -i\gamma _{23B} \gamma _{13A}) \textcolor[rgb]{0.82,0.01,0.11}{(}\textcolor[rgb]{0.82,0.01,0.11}{-\gamma '}\textcolor[rgb]{0.82,0.01,0.11}{_{23B}}\textcolor[rgb]{0.82,0.01,0.11}{)}$ & $-1$\\
\end{tabular}
\end{ruledtabular}
\label{tab5}
}

\subtable[Example 3: evaluation of path (b) in Fig.~\ref{fig17}.]{
\begin{ruledtabular}
\begin{tabular}{cccccccc}
$X_{6}$ & $X_{5}$ & $X_{4}$ & $X_{3}$ & $X_{2}$ & $X_{1}$ & $X_{0}$ & $e^{2\pi i\theta}$\\ \hline
 &  &  &  & $\textcolor[rgb]{0.82,0.01,0.11}{\gamma '}\textcolor[rgb]{0.82,0.01,0.11}{_{23B}}$ & $( -i\gamma '_{13B} \gamma '_{03A})( -i\gamma _{13B} \gamma _{03A})( -i\gamma _{03A} \gamma '_{03A})( +i\gamma '_{13B} \gamma _{13B})$ & $\textcolor[rgb]{0.82,0.01,0.11}{(}\textcolor[rgb]{0.82,0.01,0.11}{-\gamma '}\textcolor[rgb]{0.82,0.01,0.11}{_{23B}}\textcolor[rgb]{0.82,0.01,0.11}{)}$ & $1$\\
\end{tabular}
\end{ruledtabular}
\label{tab6}
}
\end{table*}


The second path is illustrated in Fig.~\ref{fig14}(b) and evaluated in Tab.~\ref{tab2}. This path is constructed by adding an independent loop to the first path. The loop consists of the following pairs: $( -i\gamma _{02A} \gamma _{02B})( -i\gamma '_{02A} \gamma '_{02B})( -i\gamma _{02A} \gamma '_{02A})\in X_{5}$ and $( -i\gamma '_{02B} \gamma _{02B})\in X_{4}$. This example demonstrates that by adding any independent loop $\Omega$ to a legitimate path $\Gamma$ ($\Omega$ and $\Gamma$ have no common vertex), we obtain another legitimate (composite) path $\Gamma '$. It is worth noting that some Majorana pairs may appear in multiple $X_{i}$s. For instance, the pair $( -i\gamma '_{02A} \gamma '_{02B})$ exists in all $X_{i}$s except $X_{0}$ and $X_{1}$. We have the option to insert this pair at the place of $X_{3}$ or at the place of $X_{6}$ (or any other $X_{i}$s except $X_{0}$ and $X_{1}$). Different choices correspond to potentially different non-zero terms (possibly with different values) in principle. However, empirical observations reveal that these different choices yield the same value in most cases. This is not a coincidence but rather a general structure applicable to all situations. Despite this subtlety, we can represent them all using the same path shown in Fig.~\ref{fig14}(b). Therefore, we consistently represent them with a single picture and include one choice in the table.


The third path is illustrated in Fig.~\ref{fig14}(c) and evaluated in Tab.~\ref{tab3}. Although this path appears unrelated to the first path in Fig.~\ref{fig14}(a), it can be regarded as a deformation of the first path through a loop that shares some vertices and edges with the first path. By considering these three cases, we demonstrate that by adding a loop $\Omega$ to a legitimate path $\Gamma$, or deforming $\Gamma$ with the loop $\Omega$, we obtain another legitimate path $\Gamma '$.


The fourth path, presented in Fig.~\ref{fig14}(d) and evaluated in Tab.~\ref{tab4}, is distinguished by its value of $i$, which differs from the value of $(-1)$ observed in the first three paths.


The aforementioned four paths serve as examples to illustrate the computation of the pure Majorana phase $\theta _{3}$.  In principle, we should identify all the non-zero terms in the expansion of \eqref{eq5.25}, summing them within a polynomial-like summation, and ultimately normalizing the resulting norm to 1. However, this process is prohibitively intricate for manual computation. Therefore, we have developed a computer program to perform the calculations. Although we can employ either the general computation approach outlined in Sec.\ref{sec5.3.2} or the graphical computation approach described in Sec.\ref{sec5.3.3}, we opt for the latter due to its faster speed and greater ease of generalization to more complex scenarios in the future.

Upon calculating all the potential non-zero terms in this scenario, we observe that there exist only two distinct values: $(-1)$ and $i$. Remarkably, the number of non-zero terms in each class is equal, thereby resulting in a final outcome of a $\mathbb{Z}_{8}$ phase, specifically $e^{2\pi i\theta_{3}} =(-1+i)/\sqrt{2}$.

The origin of the $\mathbb{Z}_{8}$ phase can be attributed to the existence of odd-length loops in Fig.~\ref{fig13}. Given that each Majorana pair incorporates an $\pm i$ factor, when beginning with a path having a value of $\pm 1$, incorporating an odd-length loop to this path or deforming the path through an odd-length loop will yield a new path with a value of $\pm i$, and vise versa.

\subsection{\label{ap3.2}Example 2}

If we modify the last example in Appendix \ref{ap3.1} by keeping $n_{1}$, $n'_{1}$ and $s_{1}$ unchanged, but changing $\omega _{2}( 012) =0$ to trivial, we find that the quadruplet \eqref{eq5.28} of the total dangling Majorana modes becomes trivial as well, with the form $[ 0,0,0,0]$. Additionally, we observe the insertion of two $\gamma '_{23B}$ operators inserted in $X_{4}$ and $X_{0}$, respectively.

The projection process and the Majorana pairing pool $\mathrm{Par}$ are the same as Fig.~\ref{fig12} and Fig.~\ref{fig13} respectively. Although the inserted dangling Majorana modes differ, the only distinction lies in the sign of certain pairings, which is determined by $\omega _{2}$ as discussed in Appendix \ref{ap3.1}.

The two inserted dangling Majorana modes are now identical. Therefore, any loop present in Fig.~\ref{fig13} contributes to a non-zero term in the polynomial-like summation. By evaluating and summing over all loops in Fig.~\ref{fig13}, we obtain a $\mathbb{Z}_{8}$ phase $e^{2\pi i\theta_{3}} =(1-i)/\sqrt{2}$.

This example serves to demonstrate that the emergence of the $\mathbb{Z}_{8}$ phase solely depends on the values of $n_{1}$ and $n'_{1}$. As emphasized in Appendix \ref{ap3.1}, the presence of odd-length loops in the Majorana pairing pool $\mathrm{Par}$ is the reason behind the $\mathbb{Z}_{8}$ phase. The variables $\omega _{2}$ and $s_{1}$ only influence the direction of a Majorana pair, thus having no impact on the appearance of the $\mathbb{Z}_{8}$ phase.

\subsection{\label{ap3.3}Example 3}

The configurations of the Kitaev chain are determined by fixing $n_{1}( 01) =n_{1}( 12) =n'_{1}( 01) =n'_{1}( 23) =0$ and $n_{1}( 23) =n'_{1}( 12) =1$ (all other $n_{1} ,n'_{1}$ are fixed by the cocycle condition \eqref{eq5.5}), with $\omega _{2}( 012) =0$ (other $\omega _{2}$ values are irrelevant) and $s_{1}(01) =s_{1}(12) =1$ (other $s_{1}$ values are irrelevant or fixed by the cocycle condition). The quadruplet \eqref{eq5.28} representing the total dangling Majorana modes is $[ 0,0,0,0]$. Two $\gamma '_{23B}$ operators are inserted in $X_{2}$ and $X_{0}$, respectively.

We illustrate the entire projection process in Fig.~\ref{fig15} and the Majorana pairing pool $\mathrm{Par}$ in Fig.~\ref{fig16}, following the same conventions discussed in Appendix \ref{ap3.1}.

In this example, it is important to note that the two inserted dangling Majorana modes are identical. Therefore, any loop shown in Fig.~\ref{fig16} will contribute to a non-zero term in the polynomial-like summation. For instance, Fig.~\ref{fig17} illustrates two loops: the first one in Fig.~\ref{fig17}(a), computed in Tab.~\ref{tab5} and evaluated as $-1$, and the second one in Fig.~\ref{fig17}(b), computed in Tab.~\ref{tab6} and evaluated as $+1$.



Upon considering all possible loops and evaluating each of them, it becomes evident that there are only two distinct outcomes: $-1$ and $+1$. However, the presence of $-1$ loops is more prevalent, whereas the occurrence of $+1$ loops is relatively scarce. Consequently, by summing over all non-zero terms, the resulting value becomes $e^{2\pi i\theta_{3}} =-1$.

\section{\label{ap4}Simplification on the formula of $\theta _{3}$}

The original result of Majorana phase in \eqref{eq5.25} differs from the formulas listed in \eqref{eq5.14}-\eqref{eq5.15.4}. However, the original result in \eqref{eq5.25} can be obtained by adding a coboundary $\mathrm{d} x_{2}$ to \eqref{eq5.14}, where the 2-cochain $x_{2}$ is 
\begin{equation}
\begin{array}{ c c l }
x_{2} & = & \frac{15}{16} n_{1} n'_{1} +\frac{1}{8}( n_{1} \cup _{1} n'_{1}) n'_{1}\\
 &  & +\frac{1}{8} n_{1}( n_{1} \cup _{1} n'_{1}) +\frac{7}{8}( n_{1} \cup _{1} n'_{1})( n_{1} \cup _{1} n'_{1})
\end{array}. \label{eq.ap4.1}
\end{equation}
Note the coboundary operator here is defined as 
\begin{equation}
\begin{array}{ c c l }
\mathrm{d}_{s_1} x_{2}( 0123) & = & -x_{2}( 012) +x_{2}( 013) -x_{2}( 023)\\
 &  & +[ 1-2s_{1}( 01)] x_{2}( 123)
\end{array}, \label{eq.ap4.2}
\end{equation}
which explicitly involves $s_{1}$.

The original formula is much longer than \eqref{eq5.14}. Additionally, shifting the bosonic phase by a coboundary does not change the physics and is equivalent to the original formula. Thus, we adopt the simpler formula \eqref{eq5.14}.

The original result reveals the appearance of the $\mathbb{Z}_{8}$ phase in certain special geometric configurations of decorations. However, the general theory of this phenomenon remains unknown. Further investigations in our future research will address this question.

\nocite{*}

\bibliography{apssamp}

\end{document}